\documentclass[a4paper,11pt]{article}
\pdfoutput=1 % if your are submitting a pdflatex (i.e. if you have
\usepackage{jheppub} % for details on the use of the package, please
\usepackage{xcolor}
\usepackage{pagecolor,lipsum}% http://ctan.org/pkg/{pagecolor,lipsum}
\usepackage{mdframed}
\usepackage{verbatim}
\usepackage{subcaption,graphicx}
\usepackage[T1]{fontenc} % if needed
\usepackage{mathrsfs}
\usepackage{arydshln}
\usepackage{color}
\usepackage{slashed}
\usepackage{epsfig}
\usepackage{enumerate}
\usepackage{color}
\usepackage{wrapfig}
\usepackage{amsfonts}
\usepackage{multirow}
\usepackage{amsthm}
\usepackage{amsmath}
\usepackage{tikz}
\usepackage{float}
\usepackage[numbers]{natbib}

 \usepackage[normalem]{ulem}

 \definecolor{darkgreen}{cmyk}{1,0,1,0.4}
 \definecolor{pink}{cmyk}{0.4,1,0.3,0}

\newcommand{\ba}{\begin{array}}
\newcommand{\ea}{\end{array}}
\newcommand{\bd}{\begin{displaymath}}
\newcommand{\ed}{\end{displaymath}}
\def\be{\begin{equation}}
\def\ee{\end{equation}}
\def\bsube{\begin{subequation}}
\def\esube{\end{subequation}}
\def\bea{\begin{eqnarray}}
\def\eea{\end{eqnarray}}
\def\bal{\begin{align}}
\def\ealign{\end{align}}
\def\eal{\end{align}}
\def\ben{\begin{enumerate}}
\def\een{\end{enumerate}}

\def\beq{\begin{equation}\ba{rcl}}
\def\eeq{\ea\end{equation}}

\title{Spin-$0^\pm$ portal induced Dark Matter}
\author[a,\,\# ]{Sukanta Dutta,}
\author[b,\,\,\$]{Ashok Goyal,}
\author[a,b,\,\,\dagger]{Lalit Kumar Saini}
\affiliation[a]{SGTB Khalsa College, University of Delhi, Delhi, India.}
\affiliation[b]{Department of Physics $\&$ Astrophysics, University of Delhi, Delhi, India.}
\emailAdd{${}^\#  $ Sukanta.Dutta@gmail.com}
\emailAdd{${}^\$$agoyal45@yahoo.com}
\emailAdd{${}^{\dagger}$Corresponding~Author: sainikrlalit@gmail.com}

\abstract{Standard model (SM) spin-zero singlets are constrained through their di-Bosonic decay channels {\it via} an effective coupling induced by a vector-like quark (VLQ) loop  at the LHC for $\sqrt{s}$ = 13 TeV. These spin-zero  resonances are then considered as portals for scalar, vector or fermionic dark matter  particle interactions with SM gauge bosons. We find that the model is validated with respect to the observations from  LHC data and from cosmology, indirect and direct detection experiments for an appreciable range of scalar, vector and fermionic DM masses greater than 300 GeV and  VLQ masses $\ge$ 400 GeV, corresponding to the three  choice of portal masses 270  GeV, 500 GeV and 750 GeV respectively.}
\keywords{dark matter model, scalar portal, vector like quark,  indirect and direct detection}
\arxivnumber{1709.00720}
\begin{document} 
\maketitle
\flushbottom
\section{Introduction}
 ATLAS and CMS \cite{Aaboud:2016tru,Aad:2015zqe, Aaboud:2016trl,Khachatryan:2016txa} have been assiduously searching for di-Boson production in the semileptonic, fully leptonic and di-Bosonic channels at $\sqrt{s}$ = 13 TeV with the integrated luminosity of 3.2 $fb^{-1}$. They have looked for a scalar spin zero resonance of mass > 200 GeV and a spin 2 Randall-Sundrum graviton state as benchmark model of mass > 500 GeV. Assuming a scaling of cross-section for an s-channel resonance produced by  gluon fusion (light quark-antiquark annihilations) the consistency between the 13 TeV data and the data collected at the 8 TeV is found at the level of 1.2 (2.1) standard deviation. An excess of di-photon events at a mass of 750 GeV reported by the LHC's ATLAS and CMS experiments in 2015 had led to a flurry of activity resulting in more than 500 papers in a short span of time (see for example \cite{Staub:2016dxq} and references therein). The LHC phenomenology of the  750 GeV di-photon  resonance was also studied in the framework of the effective field theory (EFT) and extended to include  this new found resonance induced interactions of  the standard model (SM) singlet fermionic and/ or scalar DM  with the SM gauge Bosons of the visible world  \cite{DEramo:2016aee,Chao:2015ttq,Han:2015yjk,Mambrini:2015wyu,Morgante:2016cfv}.  The excess reported in 2015 however did not show up in 2016 data. ATLAS and CMS results of run 1 at LHC also saw the enhanced  production of  SM Higgs Boson in association with a top quark. A possible explanation put forward in \cite{vonBuddenbrock:2015ema} was to interpret the data due to the existence of another heavier scalar with the best fit mass of $272^{+12}_{-9}$ GeV. This particle unlike the SM Higgs was supposed to interact with dark matter. The most promising mechanism for the production of di-photon resonance  discussed in the literature is through gluon-gluon fusion and its subsequent decay into SM di-Bosons {\it viz} $gg\rightarrow$ $\phi^0$/$A^0$ $\rightarrow$ $\gamma\gamma$. We will examine the possibility of this resonance to act as a portal between the dark matter particle (DM) of any spin (0, 1/2, 1) with the SM particles and examine the constraints on the model parameters from the observed relic density, direct and indirect observations while satisfying the constraints from ATLAS and CMS results \cite{Aaboud:2016tru,Aad:2015zqe, Aaboud:2016trl,Khachatryan:2016txa}.
 \par In section \ref{TheModel} we describe a simple renormalizable model by augmenting the particle content of the SM to include an $SU(2)_L$  singlet scalar/pseudo-scalar particle and vector-like SM colour-triplet fermions of exotic charge $Q_\psi$. The DM particles in this model are neutral SM $SU(2)_L\times U(1)_Y$ singlets which are odd under a discrete $Z_2$ symmetry and can be scalars, fermions or vectors. These particles interact with the SM gauge Bosons through the scalar/pseudo-scalar portal.  We compute  the partial decay-widths of the scalar and pseudo-scalar  in the subsection \ref{partialdkywidths}. In the subsection  \ref{PortalSection} we analyse the di-Boson production cross-section observed by the ATLAS and the CMS experiments \cite{Aaboud:2016tru} in $p-p$ collision in the context of the model discussed here and obtain constraints on the coupling of the di-Boson resonance with vector-like fermions. With these constrained couplings of the portal scalar and pseudo-scalar, we compute the relic density contribution of the viable DM candidates through their interactions with the visible world  in the subsection \ref{RelicDensity}. The indirect detection of the DM candidates through the emission of monochromatic $\gamma$-rays by Fermi-LAT \cite{Fermi-LAT:2016uux,Ackermann:2015lka}, a satelite based $\gamma$ ray observatory and the ground based Cherenkov telescope  H.E.S.S. \cite{Abramowski:2011hc, Abramowski:2013ax}   is discussed in the subsection \ref{Indirect}. Further, we investigate  the possibility of direct detection of such DM particles in the elastic DM - nucleon scattering experiments in Dark-Side50 (2016) \cite{Agnes:2015ftt}, LUX \cite{Akerib:2016vxi,Savage:2015xta}, XENON \cite{Aprile:2017iyp,
Aprile:2015uzo} and PANDA \cite{Cui:2017nnn} collaborations  in the subsection  \ref{Direct}. Section \ref{conclusion} summaries our analysis and results through composite figures, where all experimental constraints are used to look out for allowed region of the model.

\section{The Model}
We consider a portal induced dark  matter model in which the di-Boson resonance is either a CP even  scalar ($\Phi$) or a CP odd (P) scalar. The di-Boson coupling is introduced through vector-like $SU(3)_C$  triplet fermion  with exotic charge $Q_\psi$ = +5/3. In addition we propose the dark matter to be a real scalar, a real vector or a spin 1/2 Dirac fermion. In order to avoid  mixing of the VLQ's  with SM quarks and to stabilise the DM particles we invoke an Abelian $U(1)_d$ gauge symmetry. The $U(1)_d$ sector gauge Lagrangian contains
\label{TheModel} 
\begin{eqnarray}
{\cal L}_d \subset -\,\frac{1}{4} V^0_{\mu\nu}\,{V^0}^{\mu\nu} +\left( {\cal D}_\mu\varphi\right)^\dagger \, \left( {\cal D}^\mu\varphi\right)- V(\varphi)
\end{eqnarray}
 \noindent where $\varphi$ is a charged scalar, $V^0_{\mu\nu}$ is the $U(1)_d$ field strength tensor of the gauge field $V^0_\mu$ and $V\left(\varphi\right)$ is the scalar potential. The charged scalar field in the dark sector allows the spontaneous breaking of $U(1)_d$ gauge symmetry to a $Z_2$ subgroup after $\varphi$ develops a non-zero VEV $ v_\varphi$.  The imaginary part of $\varphi$  is eaten up by $V^0_\mu$ to give it a mass $m_{V^0} = \lambda_\varphi v_\varphi/2$, where $\lambda_\varphi$ is the gauge coupling \cite{Arcadi:2016qoz,Ko:2016wce}. The usual charge conjugation  $Z_2$  symmetry  $V^0_\mu\to -\,V^0_\mu$  makes this massive gauge field a viable stable DM candidate.
 
The Lagrangian of the model is given as follows:
\begin{equation}
 \mathcal{L} = \mathcal{L_{SM}} + \mathcal{L_\psi}^{VLQ} - V(H,\Phi,P) + \mathcal{L_{DM}}
\end{equation}
where
\begin{eqnarray}V (H,\Phi,P) &=& \mu^2\, \left\vert H\right\vert^2\, +\, \lambda\, \left\vert H\right\vert^4\, +\, \mu_\Phi^2\, \left\vert \Phi\right\vert^2\, +\, \mu^2_P\, \left\vert P\right\vert^2\, +\, \lambda_\Phi\, \left\vert \Phi\right\vert^4\, +\, \lambda_{P}\, \left\vert P\right\vert^4\, \nonumber \\ && +\, \lambda_{H\Phi}\, \left\vert H\right\vert^2\, \left\vert \Phi\right\vert^2\, +\, \lambda_{HP}\, \left\vert H\right\vert^2\, \left\vert P\right\vert^2\, +\lambda_{\Phi P}\,\left\vert P\right\vert^2\, \left\vert \Phi\right\vert^2.
\end{eqnarray}
\noindent
 \noindent Here $H$ is the SM Higgs $SU(2)_L$ doublet and $\mu^2$, $\mu_\Phi^2$ $<$ 0  and $\mu_P^2$ $>$ 0. After spontaneous symmetry breaking, CP even  scalar $\Phi$ picks up a VEV  and can be written as $\Phi=v_\Phi +\phi^0$, where $\left\langle\Phi\right\rangle \equiv v_\Phi$ = $\sqrt{\frac{-\mu^2_\Phi}{2\lambda_\Phi}}$ and $P\equiv i\, A^0$.

 The Lagrangian after the electroweak and $U(1)_d$ symmetry breaking  is re-written as 
\begin{equation}
\mathcal{L}=\mathcal{L_{SM}}+\mathcal{L}^{VLQ}+\mathcal{L}^{portal}+\mathcal{L_{DM}}^{scalar}+\mathcal{L_{DM}}^{vector}+\mathcal{L_{DM}}^{fermion}
\end{equation} 
where
\begin{eqnarray}
\mathcal{L_\psi}^{VLQ} &=& \bar{\psi}\, (i\, \gamma^\mu\, \slash \! \! \! \! D_\mu\, -\, m_\psi)\, \psi\, +\, y_{\phi^0}\, \bar{\psi}\, \psi\, \phi^0\, +\, y_{A^0}\, \bar{\psi}\, \gamma_5\, \psi\, A^0 \label{LDMVLQ}\\
 \mathcal{L}_{portal} &=& \frac{1}{2}\, \left\vert\partial_\mu\, \phi^0\right\vert^2\, -\, \frac{1}{2}\, m_{\phi^0}^2\, {\phi^0}^2\, +\, \frac{1}{2}\, \left\vert\partial_\mu\, A^0\right\vert^2\, -\, \frac{1}{2}\, m_{A^0}^2\, {A^0}^2 \label{LDMport}\\
 \mathcal{L_{DM}}^{scalar} &=& \frac{1}{2}\, \left\vert\partial_\mu\, \eta\right\vert^2\, -\, \frac{1}{2}\, m_\eta^2\, \eta^2\, +\, \frac{1}{2}\, v_{\Phi}\, \kappa_{\eta \phi^0}\, \eta^2\, \phi^0 \label{LDMscal}\\
 \mathcal{L_{DM}}^{vector} &=& -\frac{1}{4}\, {V^0}_{\mu\nu}\, {V^0}^{\mu\nu}\, -\frac{1}{2} m_{V^0}^2 \, {V^0}^\mu {V^0}_\mu +\, \frac{1}{2}\, v_{\Phi}\, \kappa_{V^0 \phi^0}\, {V^0}_\mu\, {V^0}^\mu\, \phi^0 \label{LDMvec}\\
 \mathcal{L_{DM}}^{fermion} &=& \bar{\chi}\, (i\, \gamma^\mu\, \partial_\mu\, -\, m_\chi)\, \chi\, +\, \kappa_{\chi \phi^0}\, \bar{\chi}\, \chi\, \phi^0\, +\, i\, \kappa_{\chi A^0}\, \bar{\chi}\, \gamma_5\, \chi\, A^0 \label{LDMferm}
\end{eqnarray}
where $D_\mu$ = $\partial_\mu$ - $g_s\, \lambda^a\, G_\mu^a$ - $g'\, y_f\, B_\mu$, $m^2_{A^0}$ = $\mu_P^2$ and $m_{\phi^0}^2$ = --\,4\, $\mu_\Phi^2$. The Quantum number  assignments of new  scalars, pseudo-scalar, vector-Boson and fermions of the dark $U(1)_d$ sector under the gauge symmetry group $SU(3)_C\times SU(2)_L\times U(1)_Y\times U(1)_d$ is given in Table \ref{tab_no:Q_NO}.
\par
\begin{wrapfigure}{l}{0.5\textwidth}
  \centering
  %\vskip -2.5 cm
  \includegraphics[width=0.75\textwidth]{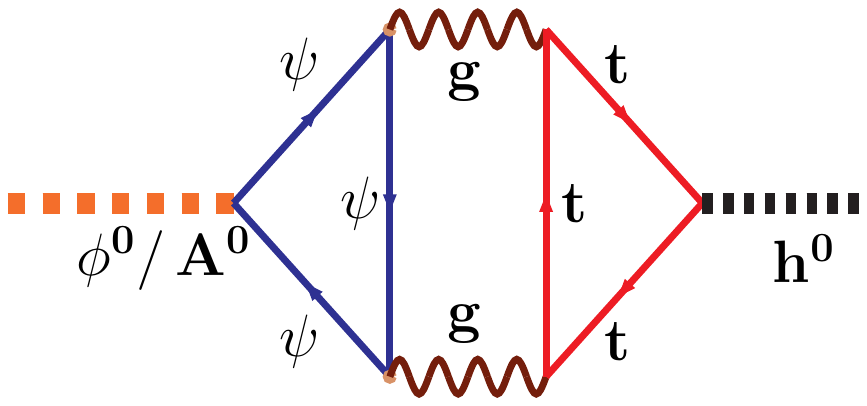}
  \vskip -13cm
  \caption{\small \em{ Radiatively generated $h^0-\phi^0$ ($h^0-A^0$) mixing diagram induced by the  VLQ and top quark  at the highly suppressed three loop level. }}
\label{fig:multiloop}
\end{wrapfigure}

 In general $\Phi$ and $P$ will mix with SM Higgs and consequently the DM scalar $\eta$ will interact with SM particles through the Higgs portal unless $\lambda_{H\Phi}$ ($\lambda_{HP}$) is zero. Even in the absence of $\lambda_{H\Phi}$  ($\lambda_{H\Phi}$) term, a mixing term $h^0-\phi^0$ ($h^0-A^0$) will be generated radiatively through  multi-loop diagrams as shown in Figure \ref{fig:multiloop} and will be highly suppressed.  Recently, the authors of reference \cite{Arcadi:2017kky} have reviewed the scalar induced DM models and  considered  such $\sim $ 10 \%  Higgs mixing with the scalar portal, to generate the required DM relic density in the universe. However, we will assume this term is absent and the radiatively induced portal $\phi^0/\, A^0$- Higgs mixing is negligible such that the DM interacts with SM particles albeit gauge-boson pairs only through the di-Boson portal. 
\par The decay of $(3,1,5/3)$ VLQ  is protected by $Z_2$ symmetry and  the lifetime of such singlet VLQs is naturally large, as there is no renormalizable, gauge invariant operator to mediate their decays into SM particles.  However,  VLQs form a bound state which can decay through  non-trivial mechanism. This subject has been  analysed in detail by the authors of reference \cite{Carena:2016bnq}, where they have also considered the possibility of introducing an  additional $Z_2$ odd VLQ doublet $\psi^\prime$ in the $(3,2,7/6)$ representation  facilitating the decay of the singlet VLQ {\it via} an off-shell $\psi^\prime$  with the constraint $m_{\Psi'} > m_\Psi$ such that the extra VLQ does not contribute appreciably to the di-photon spectrum.  On the other hand, the authors of reference \cite{Griest:1990kh} while exploring the various  co-annihilation scenarios where the colored VLQ's are slightly heavier than some new dark matter state so that they lie in the compressed spectra also found that VLQ's are likely to hadronize before they can decay. This bound state formation of VLQ  opens up the frontier to look for the new resonances at the ongoing and proposed particle accelerators. In LHC, one can expect to observe the second peak at $\sim 2\, m_\psi$  in the di-Boson invariant mass distribution in $pp\to \gamma\gamma,\,\gamma Z$, and $ZZ$ channels, following the primary peak  due to the scalar/ pseudo-scalar portal  at $m_{\phi^0/\,A^0}$ and the preliminary theoretical exercise has been performed  in reference  \cite{Carena:2016bnq}. 
\begin{table}[b]
\begin{center}
  \begin{tabular}{|l|r|r|r|r|r|r|}
    \hline
   Particle  & Spin &  $SU(3)_{c}$  &  $SU(2)_{L}$  &  $U(1)_{Y}$ & $U(1)_d$ & $Z_{2}$  \\
    \hline
   Di-Boson portal & & & & & & \\
   \, $\phi^0$ & 0 & 1 & 1 & 0 & 0 & + \\
   \, $A^0$ & 0 & 1 & 1 & 0 & 0 & +  \\
    \hline 
   Vector Like quark $\psi$ & $\frac{1}{2}$ & 3 & 1 & $\frac{5}{3}$ & a& - \\
    \hline
   Dark Matter particle & & & & & & \\ 
   \,  $\eta$ & 0 & 1 & 1 & 0 & 0& -  \\    
   \,  $\chi$ & $\frac{1}{2}$ & 1 & 1 & 0 & b & -  \\    
   \,  $V^0$ & 1 & 1 & 1 & 0 & 0 & -  \\
    \hline        
\end{tabular}
\caption{Quantum number  assignments of new  scalars, pseudo-scalar, vector-Boson and fermions of the dark $U(1)_d$ sector under the gauge symmetry group $SU(3)_C\times SU(2)_L\times U(1)_Y\times U(1)_d$.}
\label{tab_no:Q_NO}\end{center}
\end{table}
\begin{figure}[h!]
%\begin{wrapfigure}{l}{0.5\textwidth} 
 \centering
\begin{subfigure}{0.49\textwidth}
\centering
\includegraphics[width=\textwidth,height=5cm,clip]{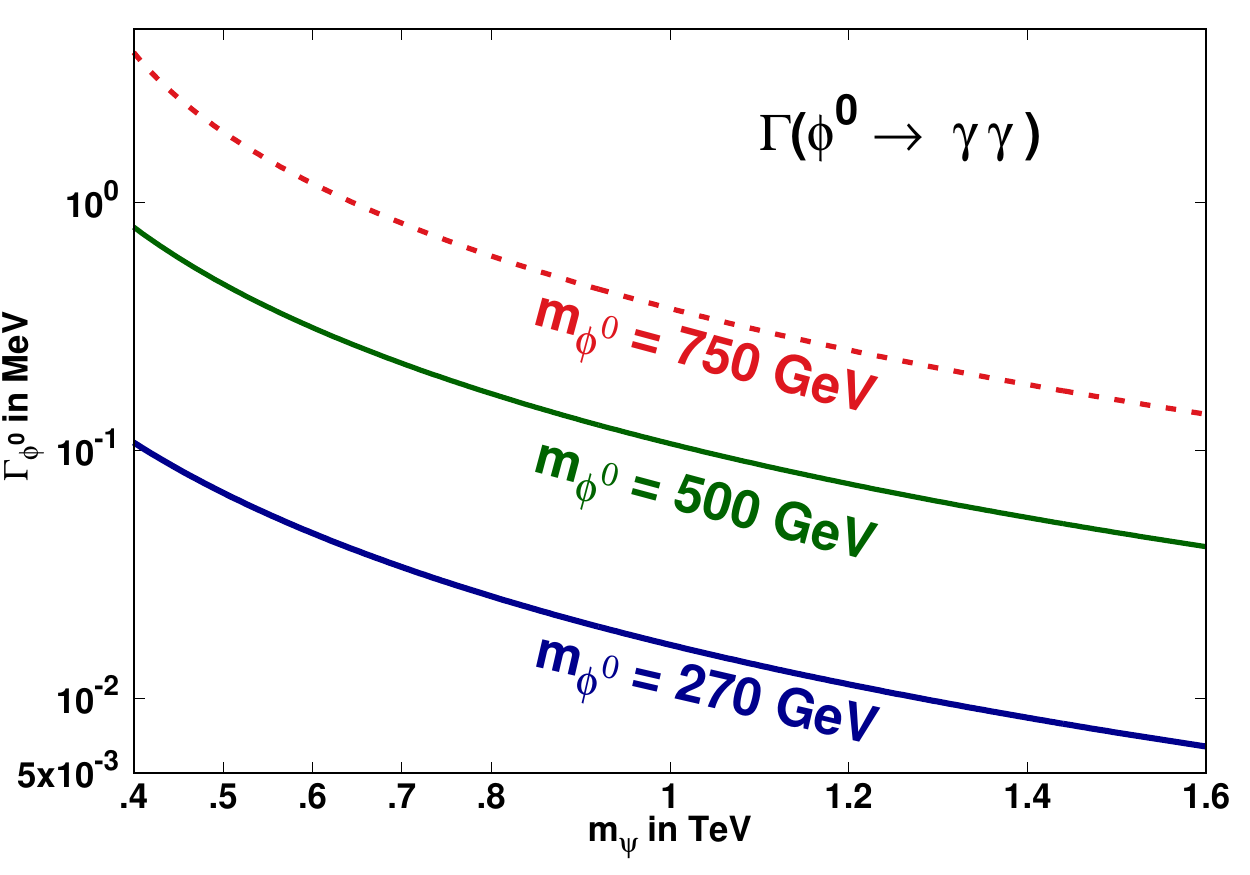}
 \caption{}\label{scaldecaytophotons}
\end{subfigure}%
\begin{subfigure}{0.5\textwidth}
\centering
\includegraphics[width=\textwidth,height=5cm,clip]{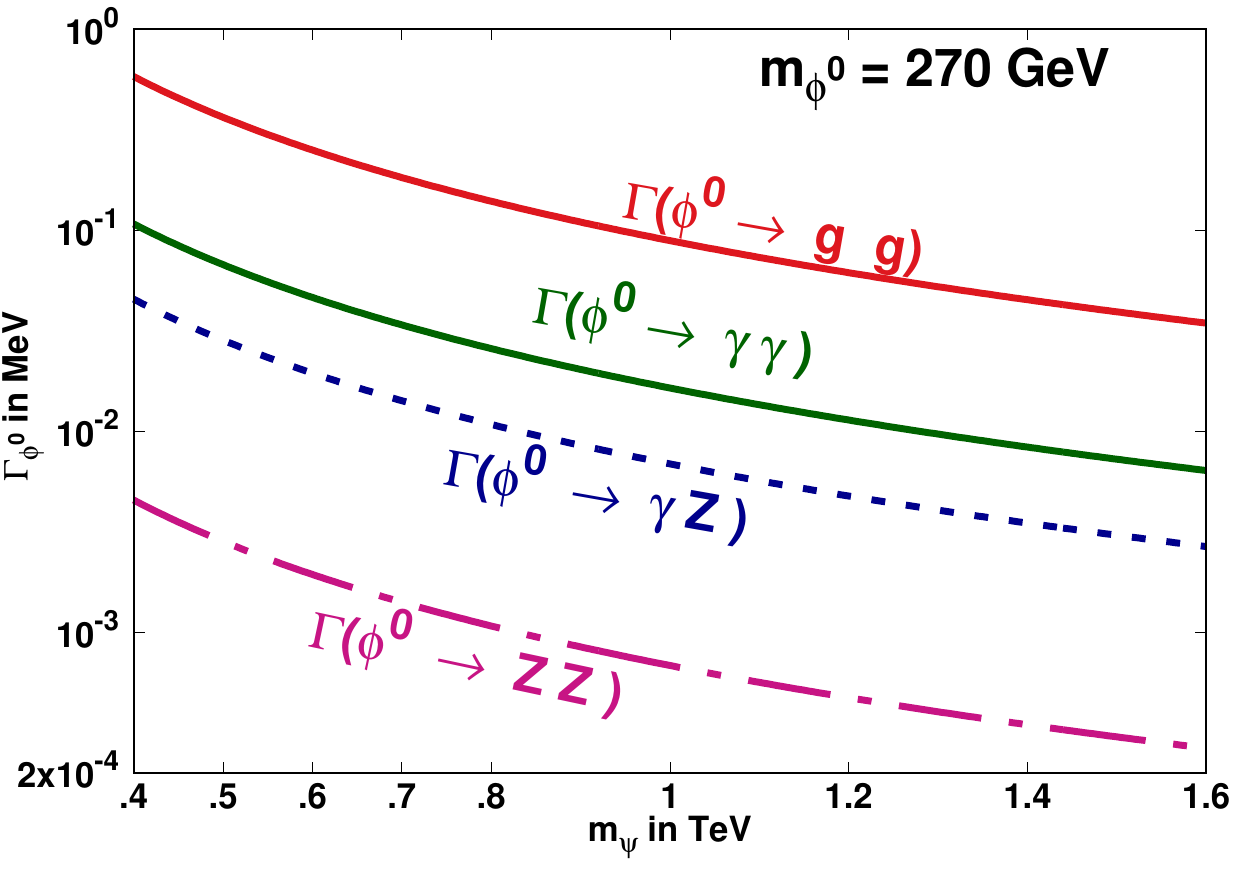}
\caption{}\label{scaldecaytobosons}
\end{subfigure}%
\caption{\small \em{   Figure \ref{scaldecaytophotons}, depicts the variation of the partial decay-widths of the scalar with  the VLQ mass $m_\psi$ for the three  values of the decaying  scalar masses 270, 500, 750 GeV respectively and Figure \ref{scaldecaytobosons}  shows   the partial decay-widths of the 270 GeV scalar to pair of $gg$, $\gamma\gamma$, $\gamma Z$ and $ZZ$ gauge Bosons respectively.}}
\label{fig:scaldkywdth}
\end{figure}
\begin{figure}[h!]
%\begin{wrapfigure}{l}{0.5\textwidth} 
 \centering
 \begin{subfigure}{0.49\textwidth}
      \centering
      \includegraphics[width=\textwidth,height=5cm,clip]{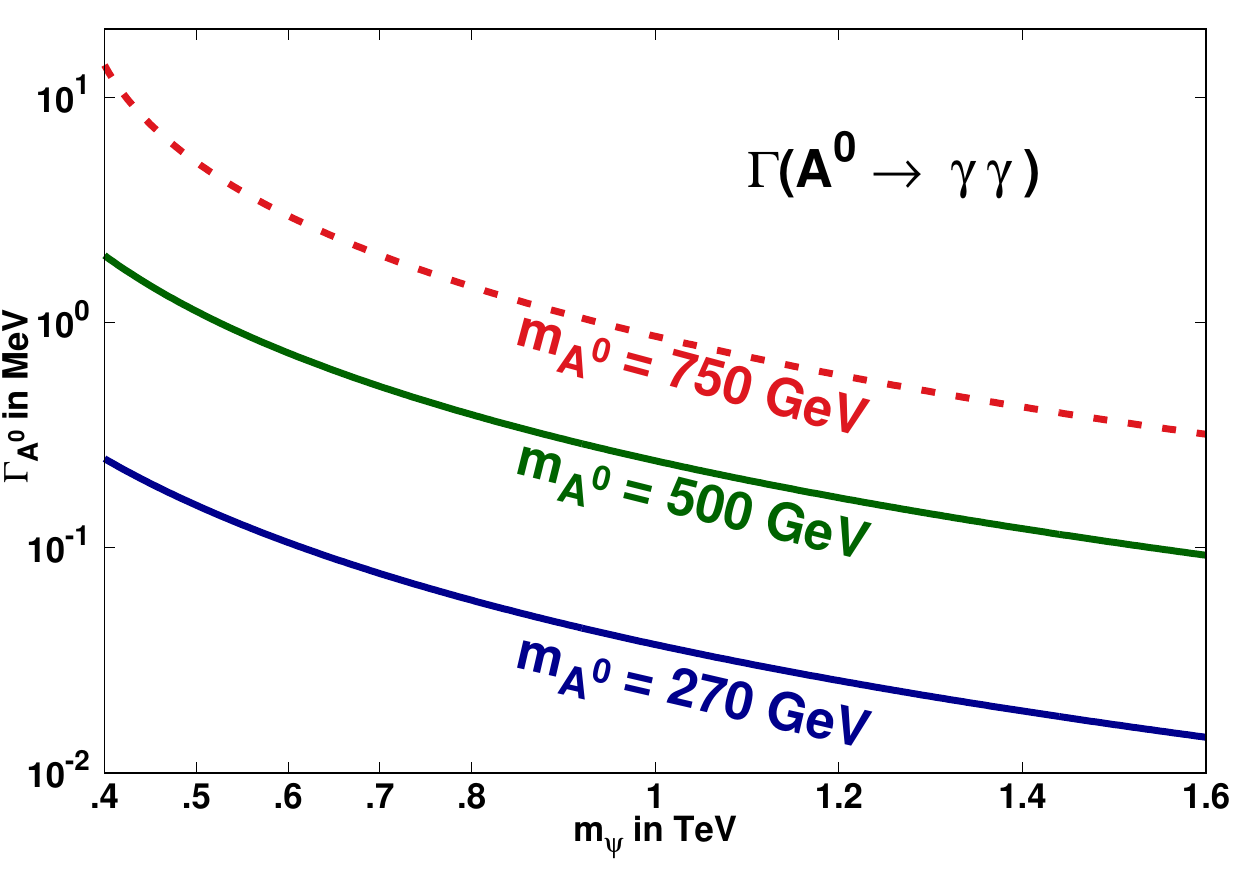}
      \caption{}\label{pscaldecaytophotons}
    \end{subfigure}%
  \begin{subfigure}{0.49\textwidth}
      \centering
      \includegraphics[width=\textwidth,height=5cm,clip]{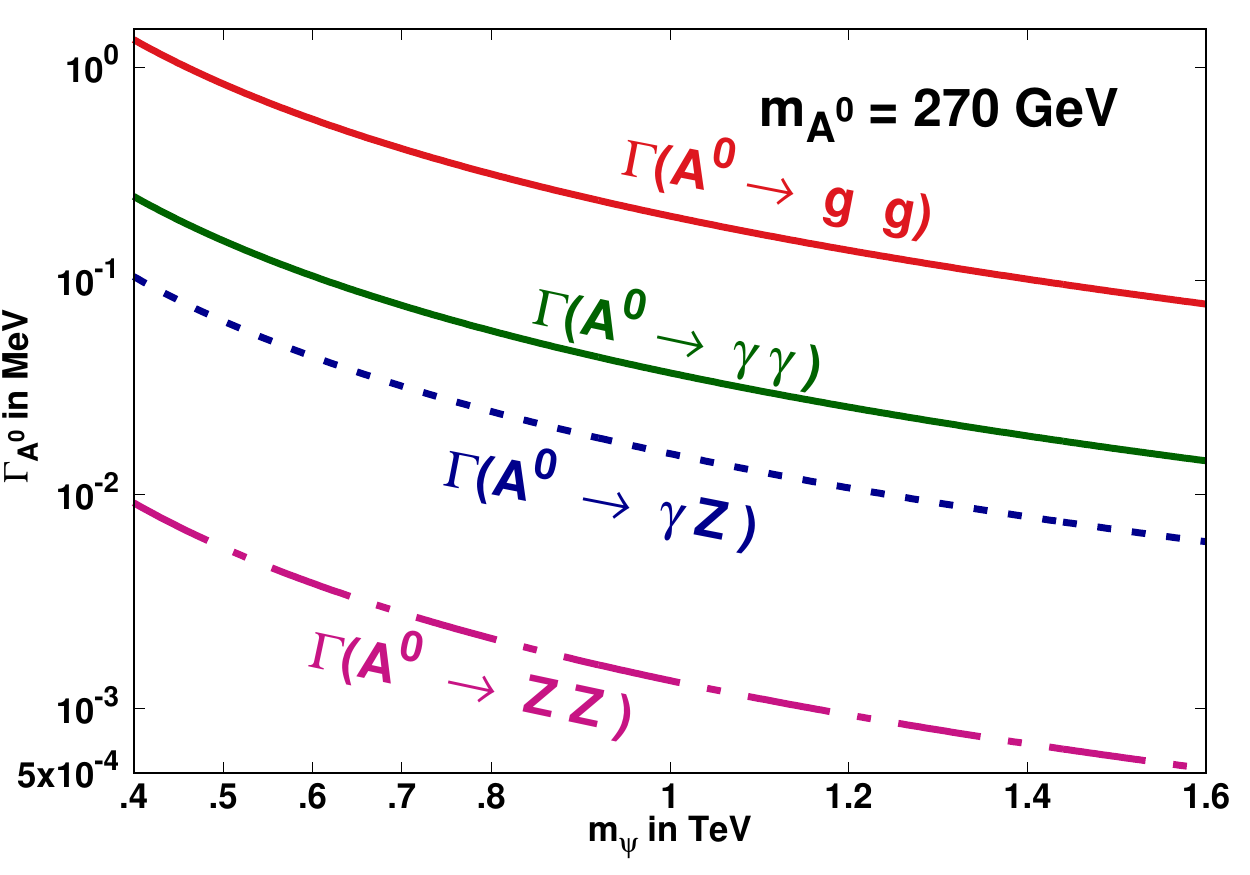}
      \caption{}\label{pscaldecaytobosons}
    \end{subfigure}%
\caption{\small \em{  Figure \ref{pscaldecaytophotons}, depicts the variation of the partial decay-widths of the pseudo-scalar  with  the VLQ mass $m_\psi$ for the three  values of the decaying  pseudo-scalar masses 270, 500, 750 GeV respectively and Figure \ref{pscaldecaytobosons}  shows   the partial decay-widths of the 270 GeV pseudo-scalar to pair of $gg$, $\gamma\gamma$, $\gamma Z$ and $ZZ$ gauge Bosons respectively. }}
\label{fig:pscaldkywdth}
%\end{wrapfigure}
\end{figure}

\subsection{Partial decay-widths to Gauge Bosons}
\label{partialdkywidths}
Interactions  described in equations \eqref{LDMVLQ}-\eqref{LDMferm} indicate that there are no tree level couplings of the singlet scalar and pseudo-scalar with the SM neutral gauge Bosons. However, the non-vanishing couplings are generated at the level of  one loop  which   are induced by fermions. The dominant one loop contribution comes from the VLQ which is the heaviest fermion available in our model.   They  
are evaluated in the appendix \ref{Decaywidthcalculation}. These one loop amplitudes  can then be translated in the language  of the effective field theory as the effective couplings of the three interacting fields.  They become the coefficients of the  four distinct effective  interacting three point vertices, each for scalar and  pseudo-scalar.  The resulting effective   Lagrangians can then be written   as 
 \begin{subequations}
 \begin{eqnarray}
   {\cal L}^{\phi^0}_{\rm eff} &=& \kappa_{gg} \,\phi^0\, G^{a}_{\mu\nu} \,G_{a}^{\mu\nu} + \kappa_{\gamma\gamma} \,\phi^0\, F_{\mu\nu}\, F^{\mu\nu} + \kappa_{Z\gamma}\, \phi^0\, F_{\mu\nu}\, Z^{\mu\nu} + \kappa_{ZZ}\, \phi^0\, Z_{\mu\nu}\, Z^{\mu\nu} \label{lag:effscal}\\
   {\cal L}^{A^0}_{\rm eff} &=& \tilde\kappa_{gg}\, A^0 \,G^{a}_{\mu\nu} \,\tilde{G_{a}}^{\mu\nu} + \tilde\kappa_{\gamma\gamma}\, A^0\, F_{\mu\nu} \,\tilde{F}^{\mu\nu} + \tilde\kappa_{Z\gamma}\, A^0\, F_{\mu\nu}\, \tilde{Z}^{\mu\nu} + \tilde\kappa_{ZZ}\, A^0\, Z_{\mu\nu}\, Z^{\mu\nu} \label{lag:effpscal}
 \end{eqnarray}
 \end{subequations}
 
\par Using the effective Lagrangians \eqref{lag:effscal} and \eqref{lag:effpscal}  the 
partial decay-widths of a CP even and odd scalars are calculated in terms of the loop integrals  in Appendix A. We studied the variation of the partial decay-widths $\Gamma_{\phi^0\to \gamma\, \gamma}$  with   the VLQ mass $m_\psi$ varying between 400 GeV - 1.6 TeV for the  three  choice of the scalar portal masses namely $m_{\phi^0}$ = 270, 500 and 750 GeV  in Figure \ref{scaldecaytophotons}. We also compare the partial decaywidth of the scalar to the di-photon channel with the  dominant channel $\phi^0\to g\,g$  and the suppressed channels of $\phi^0\to\gamma Z$ and $\phi^0\to Z\, Z$ in Figure \ref{scaldecaytobosons}.  
\par Similarly the pseudo-scalar partial decay-width variations are shown in Figures \ref{pscaldecaytophotons} and \ref{pscaldecaytobosons}.

\subsection{Constraints on the model from LHC}
\label{PortalSection}
We study the   production cross-section of these exotic scalar and/ or pseudo-scalar at the LHC ($pp\to \phi^0/\,\,A^0\to V_1\,V_2$). The di-photon production cross-section $\sigma(p\, p\to \phi^0/\,A^0 \to \gamma\,\gamma)$ is mainly through gluon fusion.  As shown in the previous section that   the partial decay-widths of scalar / pseudo-scalar being negligibly small in comparison to their masses, we can calculate the production of SM di-Bosons $V_1$ and $V_2$ $\left(V_{1,\,2}\equiv \gamma,\,Z\right)$ in the narrow width approximation of the scalar (or pseudo-scalar).  The cross-section is given as 
 \begin{equation}
  \sigma_{V_1\,V_2} = \frac{\pi^{2}}{8 m_{\phi^0/ A^0}}\, \Gamma\left(\phi^0/ A^0\to g\,g\right) \,\frac{1}{s} \,\,{\rm BR}\left(\phi^0/A^0\to V_1\,V_2\right)\,\, \int \frac{dx}{x}\, g\left(x\right) \,g\left(\frac{m_{\phi^0/A^0}^2}{s \,\, x}\right).\label{crosssectionfromdecaywidth}
 \end{equation}
\begin{figure}[htb]
%\begin{wrapfigure}{l}{0.5\textwidth} 
 \centering
  \begin{subfigure}{0.49\textwidth}
      \centering
      \includegraphics[width=\textwidth,clip]{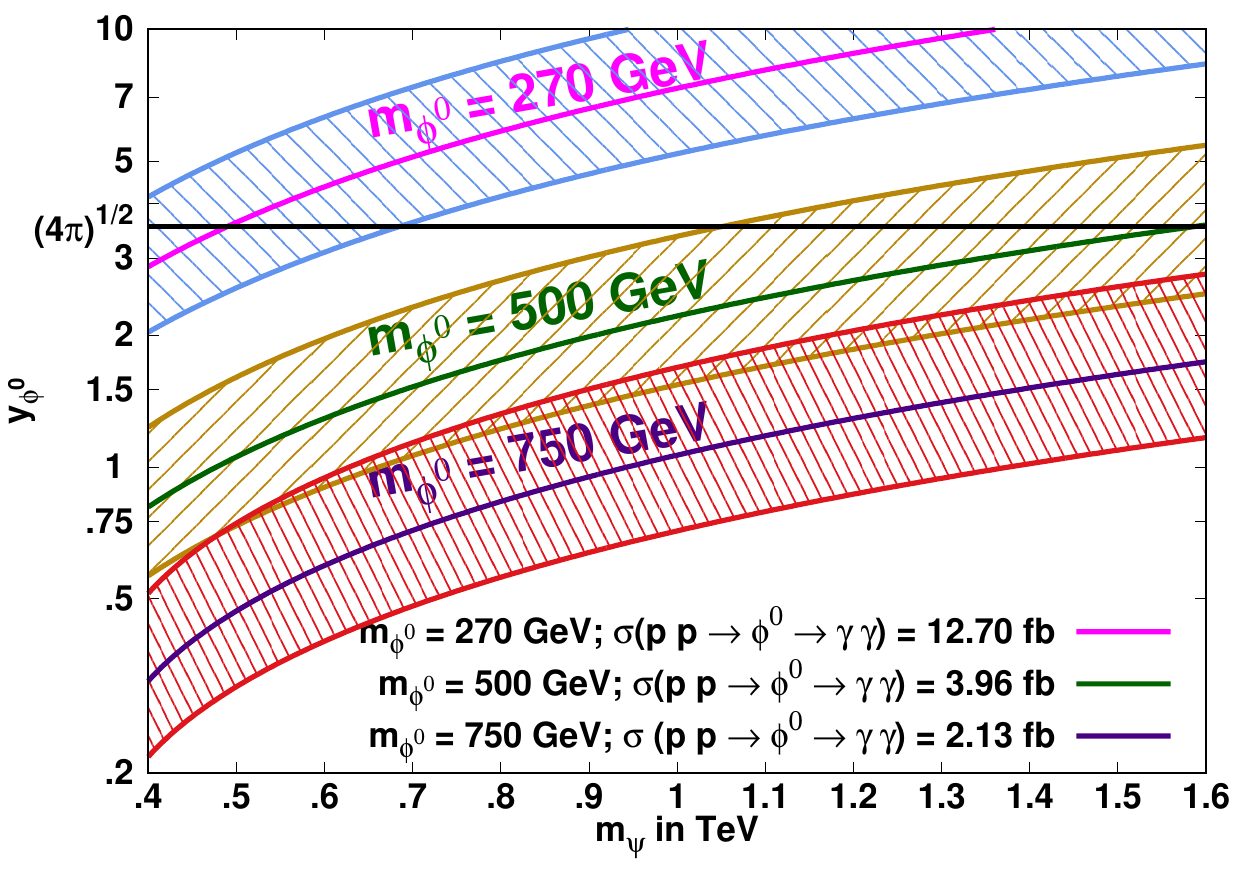}
      \caption{}\label{pptoStobosons}
    \end{subfigure}%
  \begin{subfigure}{0.49\textwidth}
      \centering
      \includegraphics[width=\textwidth,clip]{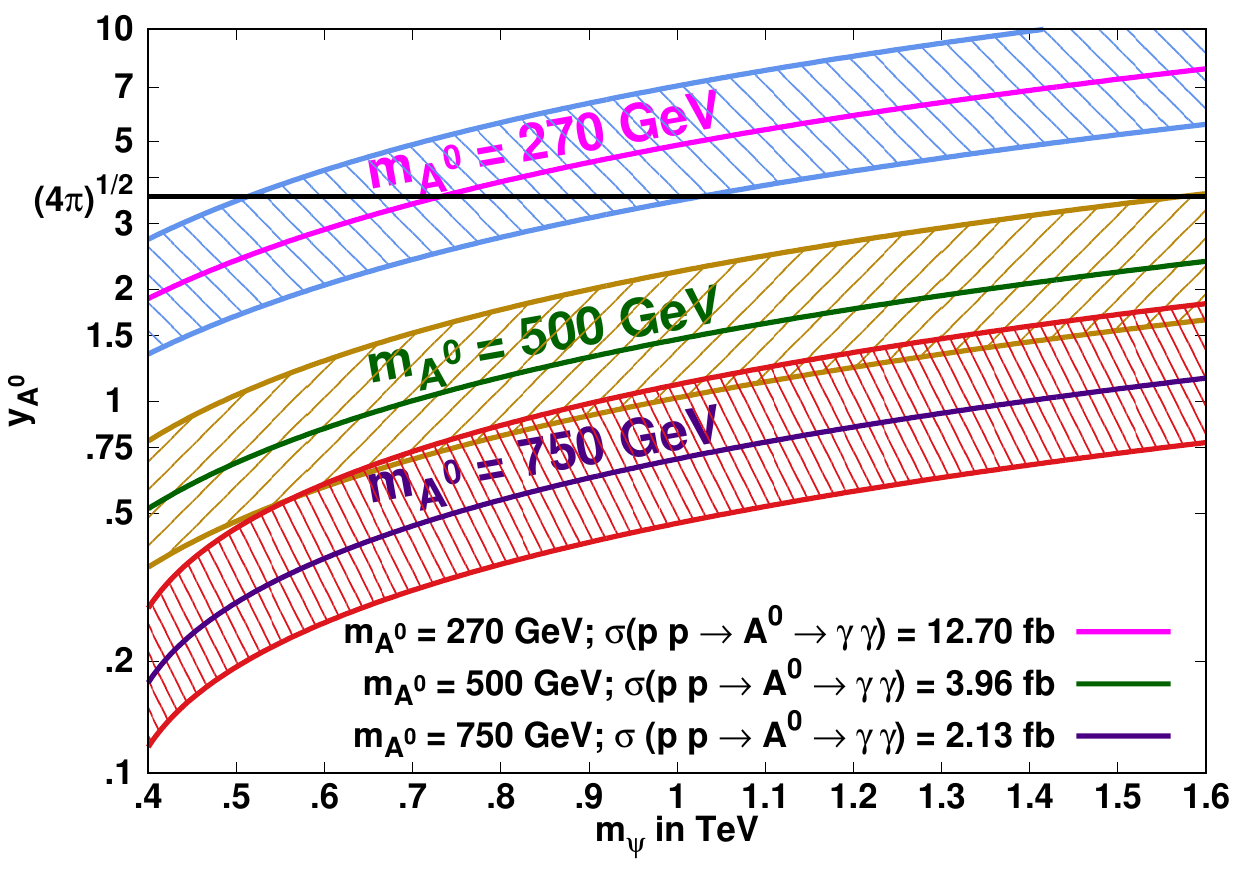}
      \caption{}\label{pptoPStobosons}
    \end{subfigure}%
\caption{\small \em{ Three band of contours  showing the  2 $\sigma$ limits on the coupling   in the  $m_\psi-y_{\phi^0/\,A^0}$ plane, {\it w.r.t.} the central value  of the  di-photon production  cross-section as observed  at $\sqrt{s}$ = 13 TeV from the CMS and ATLAS collaborations in Run 2 \cite{Aaboud:2016tru,Aad:2015zqe,Aaboud:2016trl} corresponding to the three  scalar/ pseudo-scalar  masses 270, 500 and 750 GeV respectively. The left and right panels are for the scalar and pseudo-scalar panels respectively. }}
\label{fig:scalgaugebosonprod}
%\end{wrapfigure}
\end{figure}
 
\begin{figure}[h!]
%\begin{wrapfigure}{l}{0.5\textwidth} 
 \centering
 \begin{subfigure}{0.49\textwidth}
      \centering
      \includegraphics[width=\textwidth,height=7cm,clip]{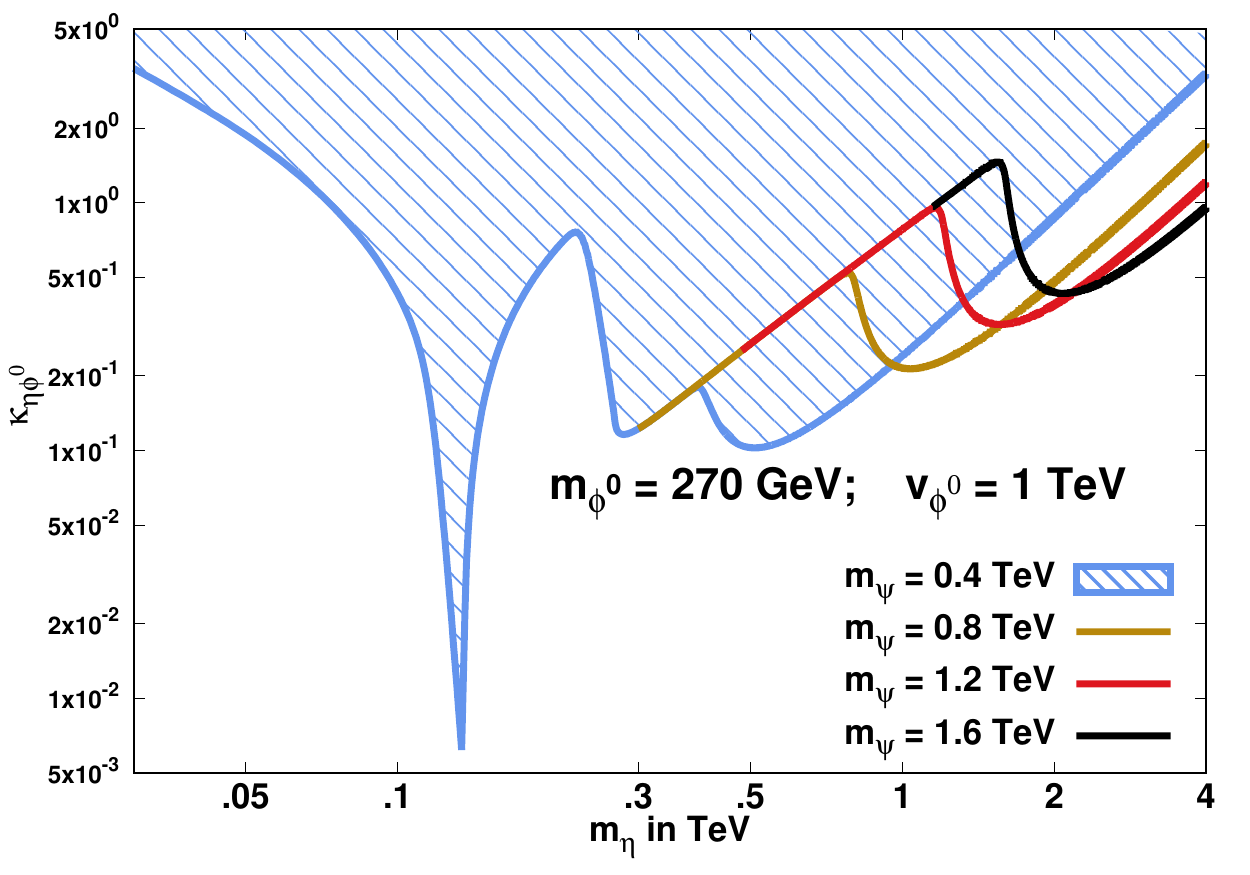}
      \caption{}\label{scalportrelicscaldm270}
    \end{subfigure}%
  \begin{subfigure}{0.49\textwidth}
      \centering
      \includegraphics[width=\textwidth,height=7cm,clip]{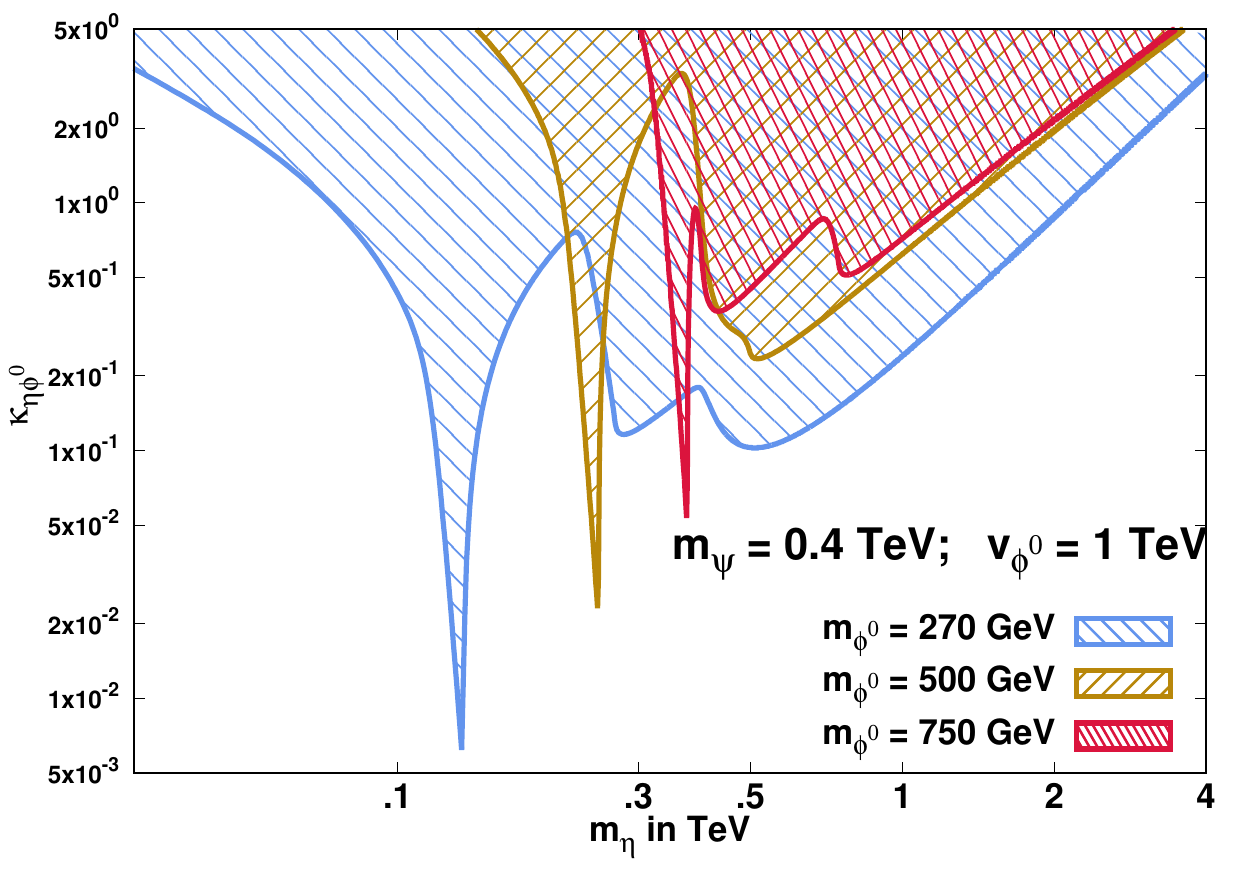}
      \caption{}\label{scalportrelicscaldm}
    \end{subfigure}%

 \begin{subfigure}{0.49\textwidth}
      \centering
      \includegraphics[width=\textwidth,height=7cm,clip]{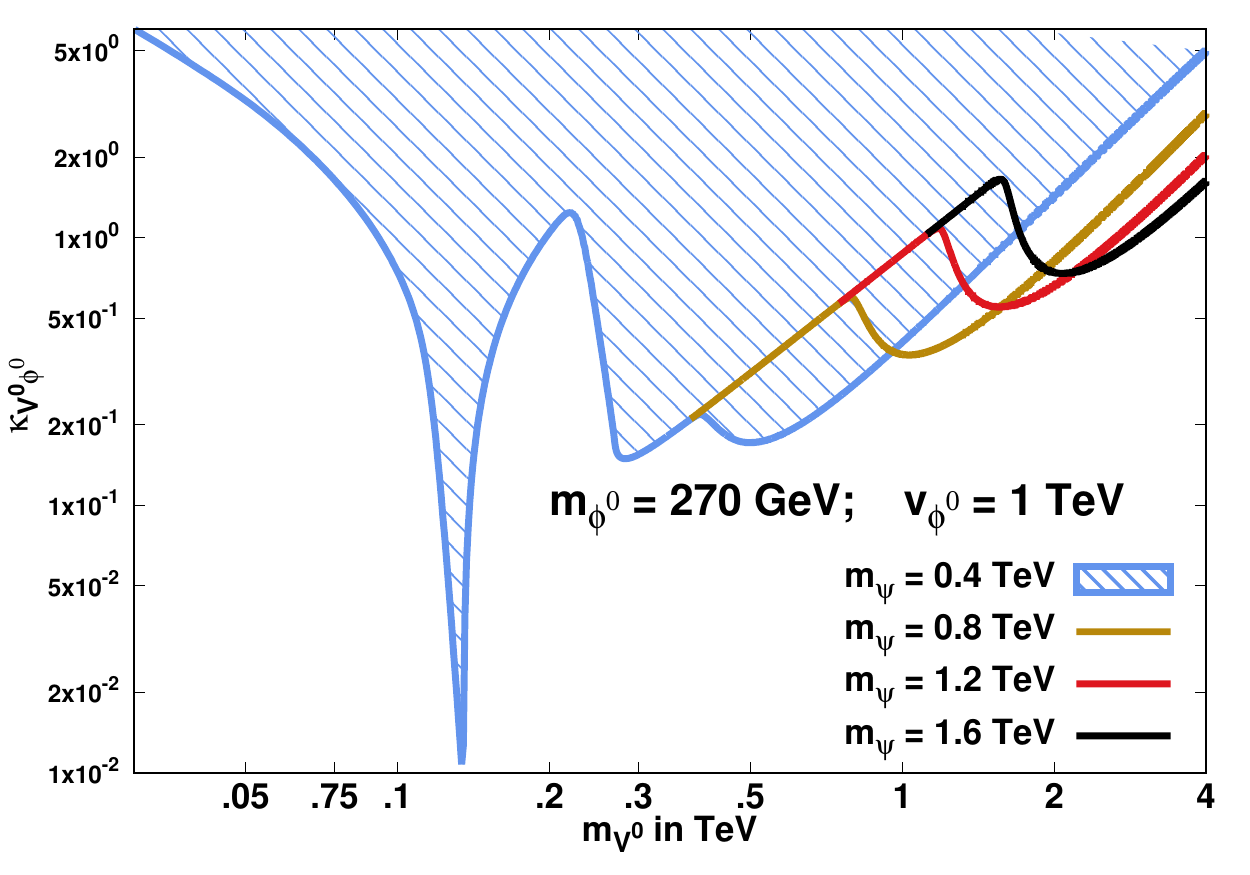}
      \caption{}\label{scalportrelicvecdm270}
    \end{subfigure}%
  \begin{subfigure}{0.49\textwidth}
      \centering
      \includegraphics[width=\textwidth,height=7cm,clip]{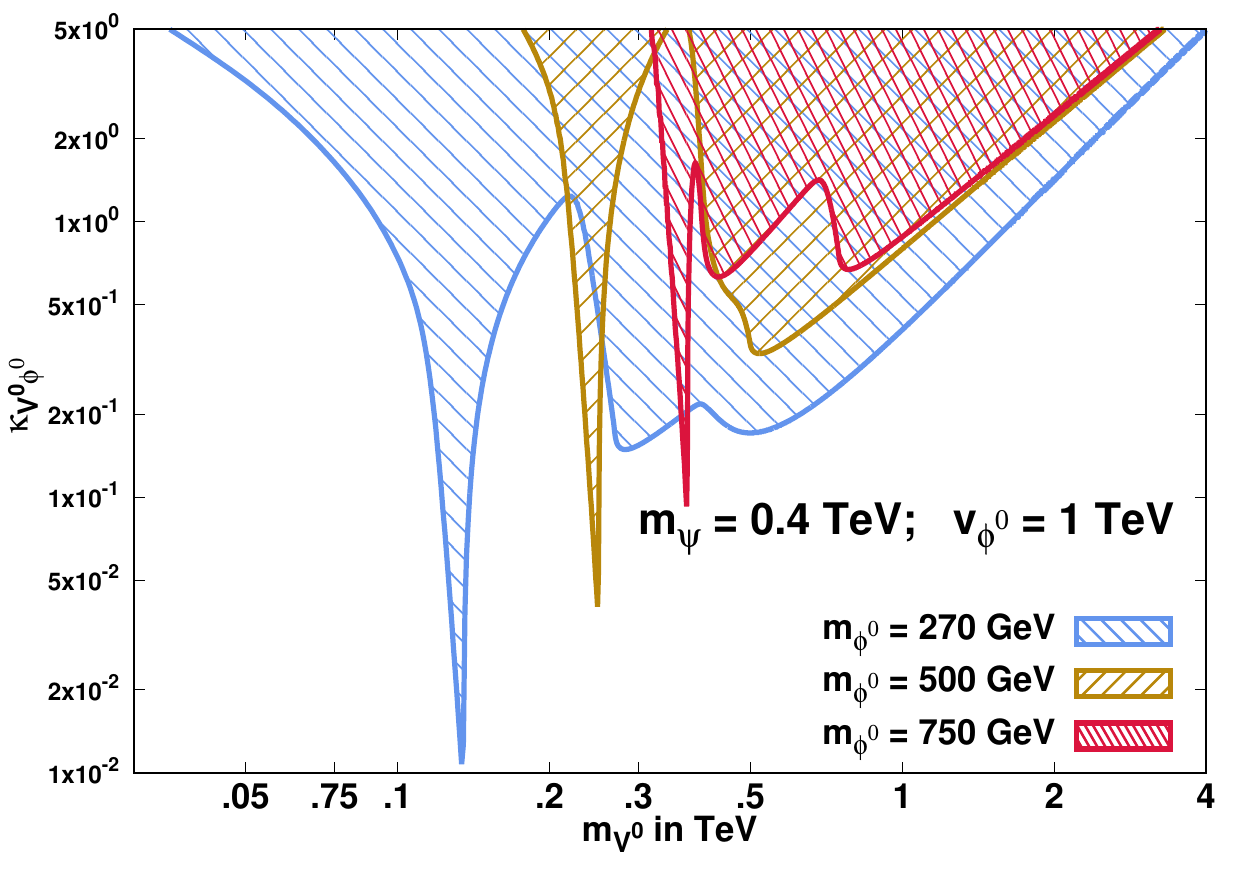}
      \caption{}\label{scalportrelicvecdm}
    \end{subfigure}%
\caption{\small \em{ In the left panel we depict relic density contours satisfying $\Omega_ch^2$ = $0.1138\pm .0045$ in the plane defined by scalar (vector)   DM mass $m_\eta$ ($m_{V^0}$) and scalar (vector) DM - mediator coupling $\kappa_{\eta\phi^0}$ ($\kappa_{V^0\phi^0}$) for a fixed scalar mediator mass of 270 GeV corresponding to  the four  choice of VLQ masses 400, 800, 1200 and 1600 GeV respectively. In the right panel we have exhibited   the constant relic density contours for a fixed VLQ mass of 400 GeV  corresponding to the three different choices of the portal scalar mass 270, 500 and 750 GeV respectively. Shaded regions appearing in blue, golden yellow  and red correspond to the relic density allowed regions  for the  portal mass of 270, 500 and 750 GeV respectively.}}
\label{scalvecDMcont}
\end{figure}
\begin{figure}
\begin{subfigure}{0.5\textwidth}
      \centering
      \includegraphics[width=\textwidth,clip]{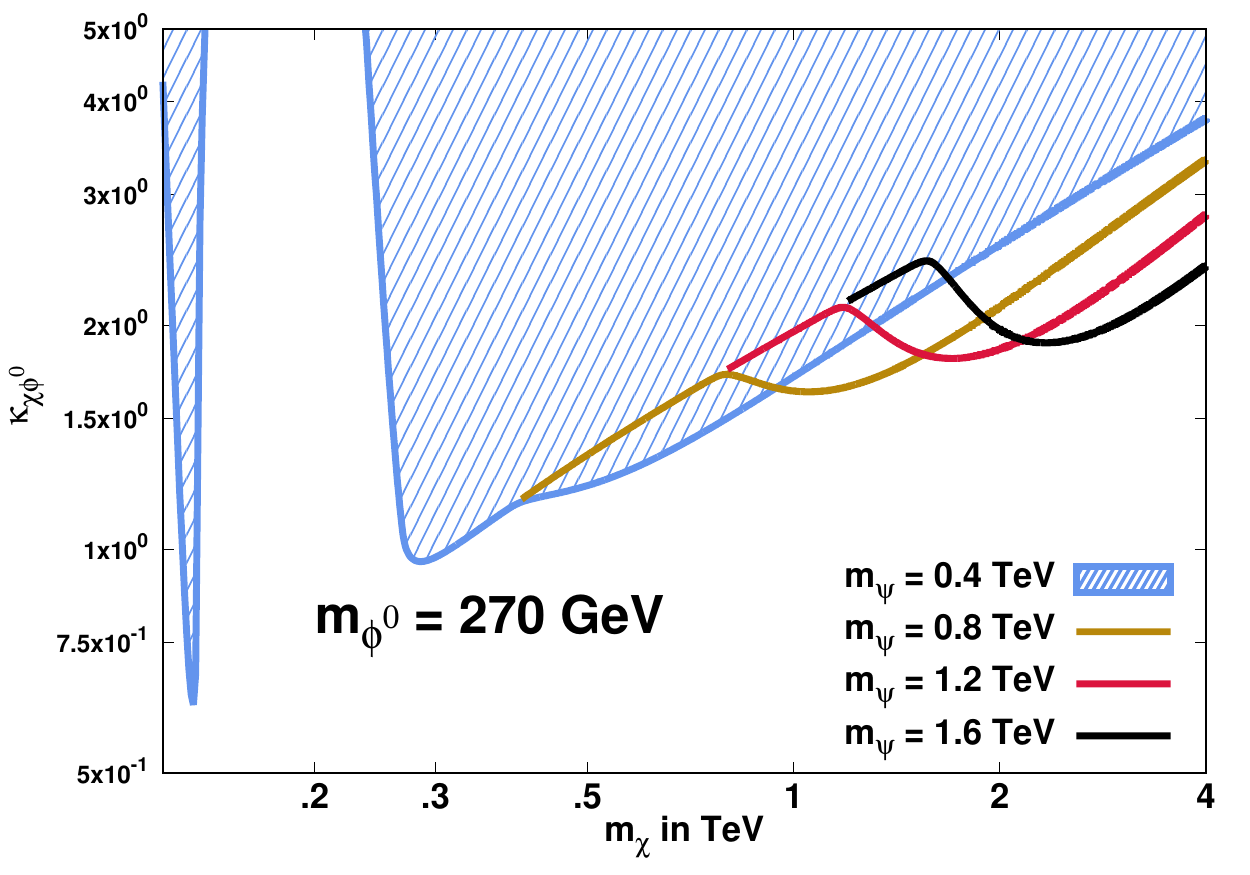}
      \caption{}\label{scalportrelicfermdm270}
    \end{subfigure}%
  \begin{subfigure}{0.5\textwidth}
      \centering
      \includegraphics[width=\textwidth,clip]{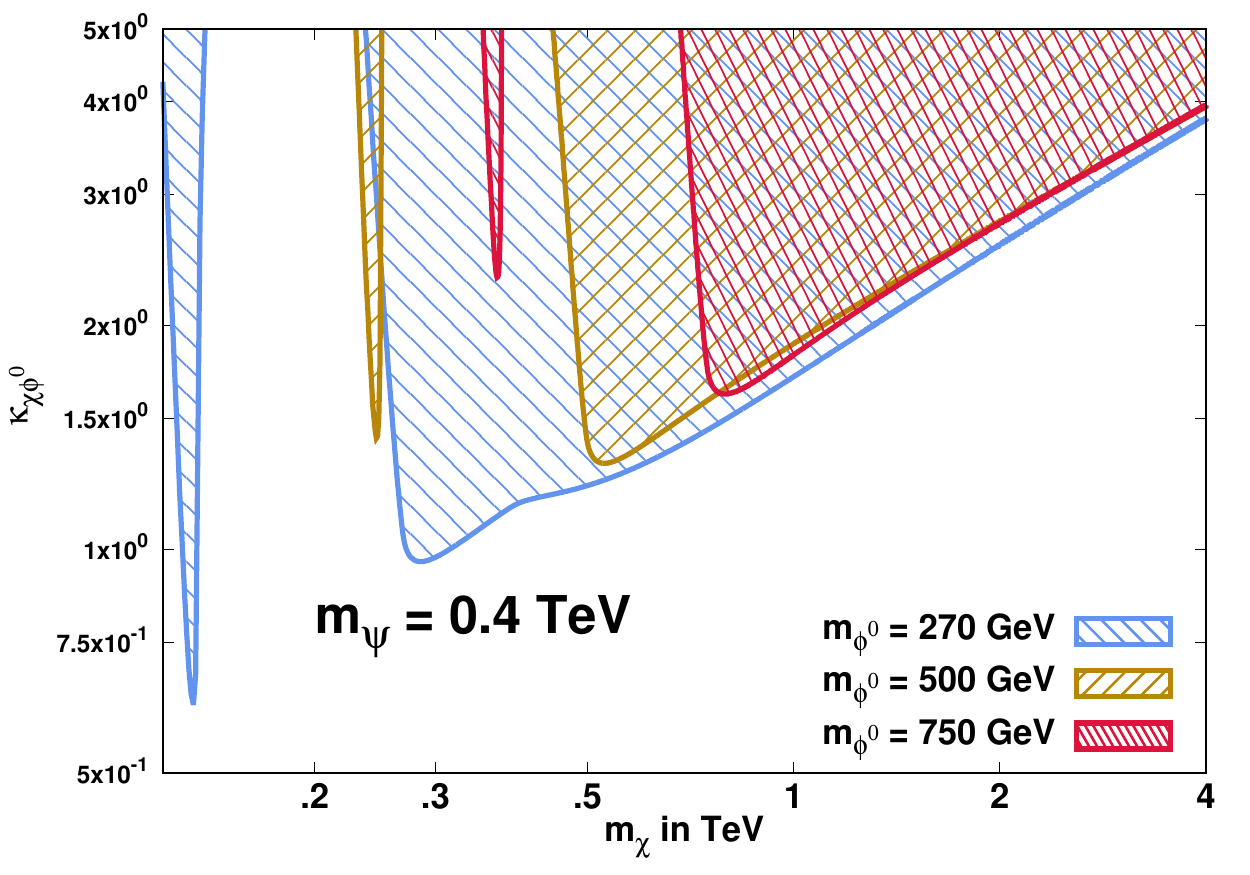}
      \caption{}\label{scalportrelicfermdm}
    \end{subfigure}%

    \begin{subfigure}{0.5\textwidth}
      \centering
      \includegraphics[width=\textwidth,clip]{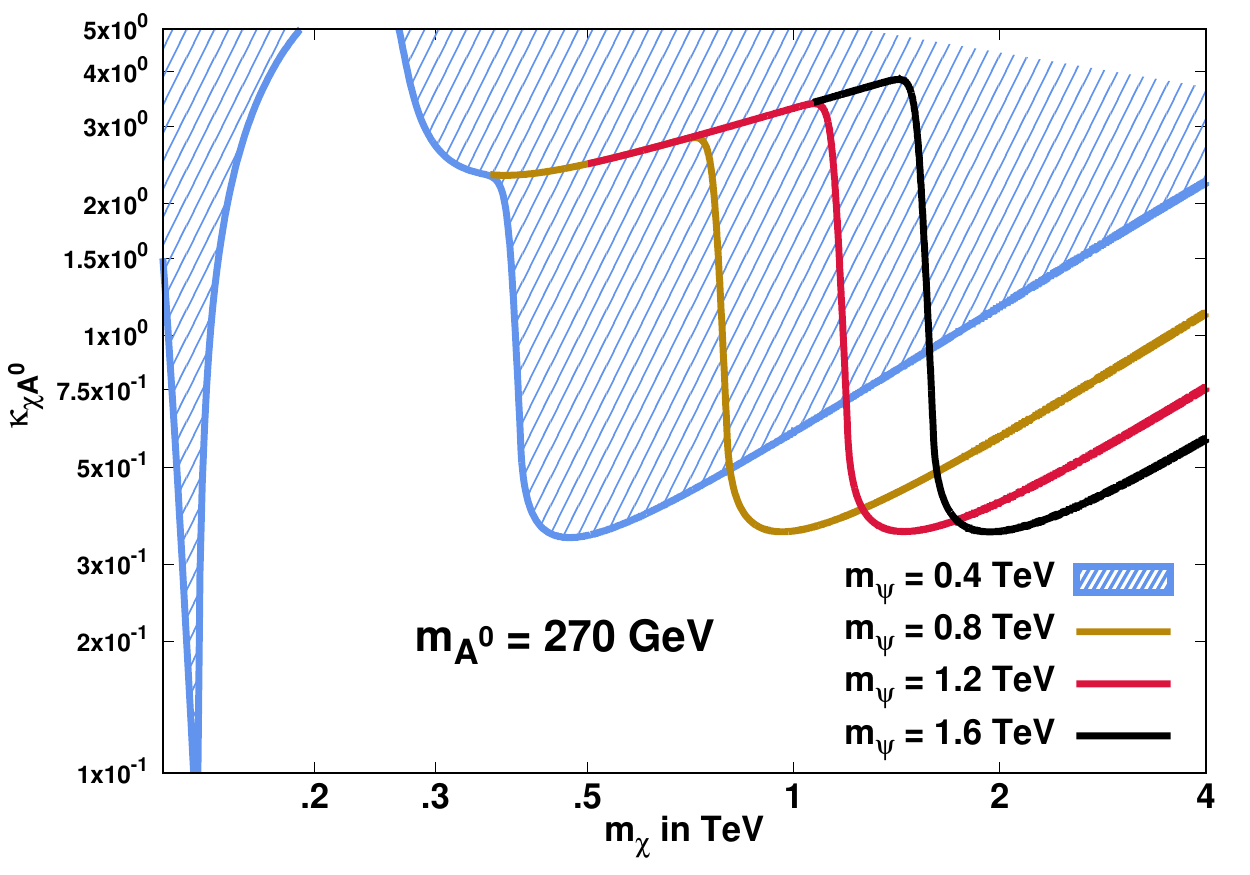}
      \caption{}\label{pscalportrelicfermdm270}
    \end{subfigure}%
  \begin{subfigure}{0.5\textwidth}
      \centering
      \includegraphics[width=\textwidth,clip]{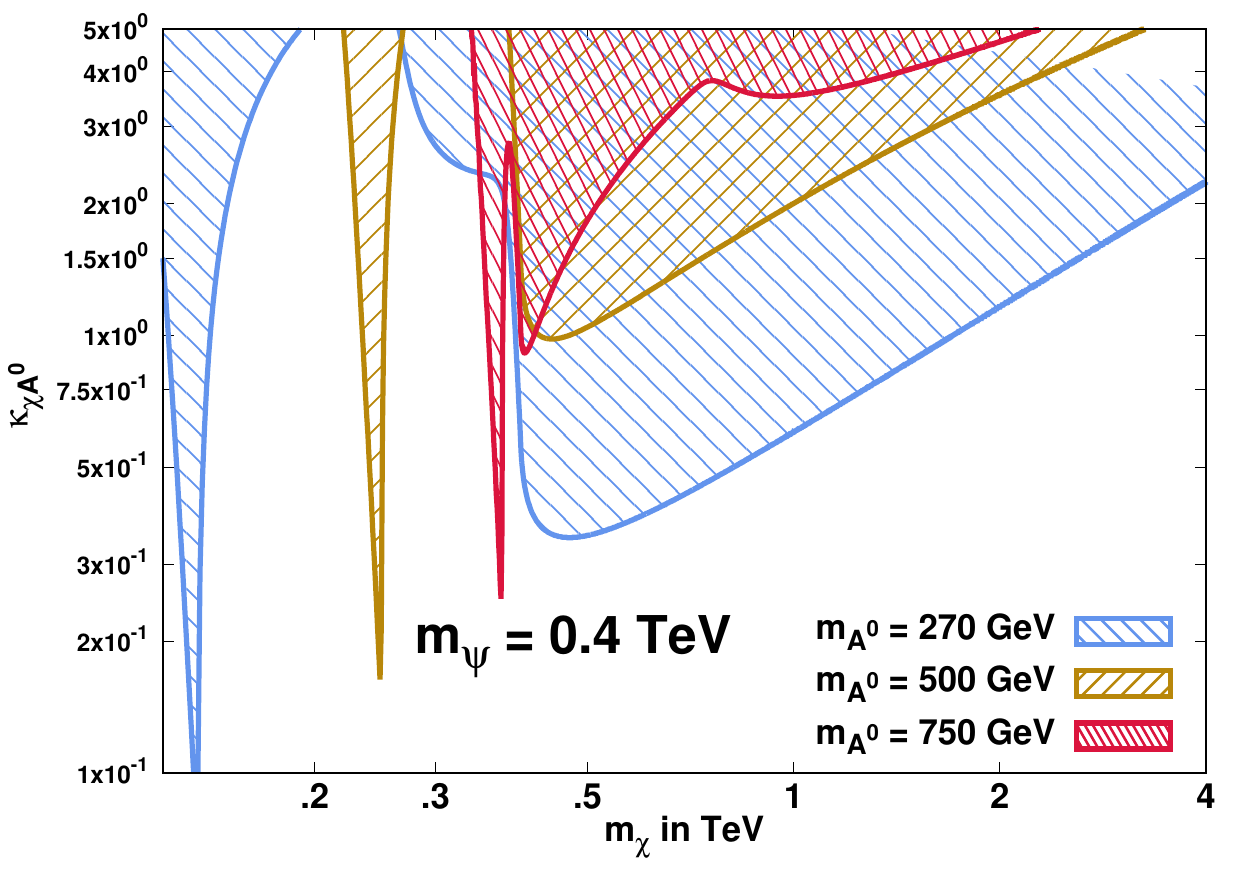}
      \caption{}\label{pscalportrelicfermdm}
      \end{subfigure}
\caption{\small \em{ In  Figure \ref{scalportrelicfermdm270} we depict the relic density contours satisfying $\Omega_ch^2$ = $0.1138\pm .0045$ in the plane defined by fermionic DM mass $m_\chi$ and  scalar DM mediator coupling $\kappa_{\chi\phi^0}$  for a fixed scalar mediator mass of 270 GeV corresponding to  the four  choice of VLQ masses 400, 800, 1200 and 1600 GeV respectively. In  Figure \ref{scalportrelicfermdm} we have exhibited   the constant relic density contours for a fixed VLQ mass of 400 GeV  corresponding to the three different choices of the portal scalar mass 270, 500 and 750 GeV respectively. Similarly, in Figures  \ref{pscalportrelicfermdm270} and  \ref{pscalportrelicfermdm}
 the constant relic density contours are drawn for the fixed pseudo-scalar mass of 270 GeV corresponding to four VLQ masses and for the fixed VLQ mass of 400 GeV corresponding to the three pseudo-scalar masses respectively. Shaded regions appearing in blue, golden yellow and red   correspond to the relic density allowed regions  for the  portal mass of 270, 500 and 750 GeV respectively.}}
\label{FermiDMcont}
%\end{wrapfigure}
\end{figure}

 \par Using equation \eqref{crosssectionfromdecaywidth}, we evaluate the cross-sections for the processes $pp\to \gamma\gamma,\, gg,$ $\, \gamma Z,\, ZZ $ and compare with the observations for a spin $0^\pm$ resonance mass in the narrow width approximation from the CMS and ATLAS run II collaboration  at  $\sqrt{s}$ = 13 TeV   and  an integrated luminosity of 3.2 fb$^{-1}$\cite{Aaboud:2016tru,Aad:2015zqe,Aaboud:2016trl}.  We have tabulated these results only for the  $\gamma\gamma$ and $\gamma Z$ channels for the benchmark resonance masses $m_{\phi^0}/\, m_{A^0}$ = 270 GeV, 500 GeV and 750 GeV in  Table \ref{table:AtlasCrossSection}.  
\begin{table}[h!]
\centering
	\begin{tabular}{c|c|c}
		\hline
		\hline
$m_{\phi^0}/\, m_{A^0}$  & \hspace*{1.7cm} $\sigma_{pp\rightarrow\gamma\gamma}$ (fb)  \hspace*{1.7cm} & \hspace*{1.7cm} $\sigma_{pp\rightarrow\gamma Z}$ (fb) \hspace*{1.7cm}  \\ & &\\
		\hline
		\hline
		270 GeV & 12.70 & 32.49 \\
	
		500 GeV & 3.96 &  10.58   \\
		
		750 GeV & 2.13 & 6.41 \\
		\hline
		\hline
	\end{tabular}
	\caption{\small \em{
	Upper limits on the cross-section for  spin $0^\pm$ resonances in the narrow width approximation  from  run II ATLAS collaboration  at $\sqrt{s}$ = 13 TeV and  an integrated luminosity of 3.2 fb$^{-1}$ \cite{Aaboud:2016tru,Aad:2015zqe,Aaboud:2016trl}.}}\label{table:AtlasCrossSection}\end{table}
 
Figures \ref {pptoStobosons} and \ref{pptoPStobosons}  show  the    2-sigma limits  on  the couplings allowed by the LHC data at  $\sqrt{s}$ = 13 TeV (as given in Table \ref{table:AtlasCrossSection}) in the plane defined by the  VLQ mass $m_\psi$  and   its coupling with the $y_{\phi^0}/\, y_{A^0}$ for the scalar and pseudo-scalar portal respectively. Each plot depicts the three different bands of allowed region corresponding to the three choices of scalar or pseudo-scalar masses  270, 500 and 750 GeV respectively.

\section{Portal induced Dark matter Scenarios }
\label{DMDescription}
\par  DM which are popularly known as  weakly interacting
massive particle (WIMP) do  not have  either electromagnetic  or strong interaction.  
One of the most challenging tasks today is to identify
the nature of the DM particle \cite{Beringer:1900zz}. 
\par Since the investigation of the nature of the DM particles needs an understanding of the underlying
physics of the model and vice-versa, we would like to begin our analysis  by considering the spin of DM particle to be either   0, 1 and/ or 1/2. Before, proceeding with the analysis, we consider the existing cosmological constraints on such a DM candidate from the  the WMAP  \cite{WMAP1,WMAP2}
and Planck data \cite{Ade:2015xua}.
\subsection{DM pair-Annihilation and Relic Density}  
\label{RelicDensity}
 \par In the model described above in section \ref{TheModel}, the proposed scalar  $\eta$, vector $V^0$ and fermion $\chi$ DM candidates can  interact with the SM gauge-bosons through CP even $\phi^0$ and odd $A^0$  scalars respectively. In the early universe  SM particles remained in thermal equilibrium as long as their reaction rate was faster than expansion rate of the universe. As the universe cooled, the reaction rate fell below the expansion rate and DM particles de-coupled from the thermal bath and contributed to the relic density observed today. The equilibrium in the early universe was maintained via the leading DM pair annihilation processes {\it viz} into pair of SM particles. The vector-like quark - antiquark pair and a pair of portal scalar/ pseudo-scalar can also be produced as a result of the annihilation of DM particles provided mass of the DM particle is higher than those of $\psi$ and/ or $\phi^0/A^0$.

\par Therefore as a  next logical step we  compute the thermal averaged DM pair-annihilation cross section. The DM pair annihilation is facilitated through   the portal mediated $s$ channel processes assuming the momentum transfer in the scattering to be much less than the portal mass. These annihilation processes lead to the following visible final states: $g\,g$, $\gamma\gamma$, $\gamma Z$ and  $ZZ$. Therefore, the $s$ channel processes  at such low energy appears to be an effective point interaction among the SM vector bosons and pair of DM candidates, suppressed by the portal mass squared. 
As to the couplings of Dark Matter with the portal, we consider two different cases, namely
 \begin{enumerate}
\item[(a)]  Scalar portal couplings to the scalar, vector and Fermion DM pair which are  defined by $\kappa_{\eta\phi^0}$, $\kappa_{V^0\phi^0}$, $\kappa_{\chi\phi^0}$ respectively.   
\item[(b)]  Pseudo-scalar portal couplings to the fermions which is defined as $\kappa_{\chi A^0}$.
\end{enumerate}

In the Appendix \ref{ThermalAvCalculationscal}, we calculate  the thermal averaged cross-section for the scalar DM pair annihilation {\it via} scalar portal to the above visible states and are given in equations \eqref{thav:etaetatogg}-\eqref{thav:etaetatozz}. In addition, to these the pair annihilation also lead to  a production of the vector like quark- antiquark pair {\it via } the scalar portal and its thermal averaged cross-section is given in equation \eqref{thav:etaetatopsipsi}, which is kinematically possible, only for DM mass greater than VLQ mass. 

\par The corresponding thermal averaged cross-section for the annihilation of vector DM to SM gauge Bosons can be  obtained directly by substituting  $\kappa_{\eta\phi^0}$ by $\kappa_{V^0\phi^0}/3$, $v_{\Phi}$ by $v_{\Phi}$ and $m_\eta$ by $m_{V^0}$ in equations \eqref{thav:etaetatogg} - \eqref{thav:etaetatozz}.  For the VLQ pair production  we multiply a factor 1/6 to the contribution given by scalar DM in  \eqref{thav:etaetatopsipsi}.

 \par The  thermalized fermionic DM pair annihilation cross-section for the  corresponding  final states are given in equations \eqref{thav:chichitogg}-\eqref{thav:chichitopsipsi}. These cross-sections for every annihilation channel are $p$-wave suppressed unlike the scalar DM case and thus result in low thermally averaged cross-sections. Therefore, the sensitivity of the fermionic DM - scalar mediator coupling is an order of magnitude less sensitive than those of the corresponding couplings of the scalar and vector DM candidates with the portal.

\par In addition, the  $t$-channel annihilation diagrams  also contribute to the relic density, where  DM pair annihilation to a pair of   portal scalars  can become kinematically feasible   for DM mass   $ \ge m_{\phi^0}$ . The thermal averaged cross-sections for these processes  are given in equations \eqref{thav:etaetatophiphi} and \eqref{thav:chichitophiphi} for pair annihilation of scalar and fermionic DM respectively. The contribution of the vector DM pair annihilation to the  pair production of such scalars  {\it via} $t$  channel is found to be identical to that of the scalar as given in \eqref{thav:etaetatophiphi}.

\par Unlike the scalar portal case, pseudo-scalar cannot decay to two scalar  or vector DM pairs and leaving no choice but to consider relevant spin 1/2 DM candidate. The thermal averaged $s$    channel annihilation  cross-sections are computed in Appendix \ref{ThermalAvCalculationpscal} and are given in \eqref{pthav:chichitogg}-\eqref{pthav:chichitopsipsi}.  The $t$ channel thermalized annihilation cross-section to the pair of pseudo-scalars is given in equation \eqref{pthav:chichitoAA}.

 \par  We are now well equipped to calculate the present day relic abundance of DM  by solving the Boltzmann equation:  
\begin{equation}
	\frac{dn_{DM}}{dt} + 3 H n_{DM} = - \langle \sigma \vert v \vert \rangle \Big((n_{DM})^2 - (n_{DM}^{EQ})^2\Big)
\end{equation}
where $H = \frac{\dot{a}}{a}$ = $\sqrt{\frac{8\pi \rho}{3M\rho_l}}$, $\langle \sigma \vert v \vert \rangle$ is the thermally averaged cross-section and $n_{DM}^{EQ}$ = g $\bigg(\frac{m_{DM} T}{2\pi}\bigg)^{3/2}$ $\exp(\frac{-m_{DM}}{T})$ where the number of degrees of freedom $g$ are 1, 2 and 3 for scalar, fermionic and vector DM respectively. The freeze-out occurred when DM is non-relativistic with  $v\ll c$  and then $\langle \sigma \vert v \vert \rangle$ can be written as 
$\langle \sigma \vert v \vert \rangle = a + b \, v^2 + \mathcal{O}(v^4)$.
\par The Boltzman equation is solved numerically following the reference \cite{Kolb:1990vq}  to give the thermal relic density
\begin{eqnarray}
\Omega_{DM}h^2\simeq \frac{1.07\times 10^9\,x_F}{M_{Pl}\,\sqrt{{g^\star}(x_F)}\,\left(a+\frac{6\,b}{x_F}\right)}\label{omegadmhsq}
\end{eqnarray}
\noindent where ${g^\star}(x_F)$ is the total  number of effective degrees of freedom  at the freeze-out temperature $T_F$ and $x_F = m_{DM}/T_F$ is obtained by solving 
\begin{eqnarray}
x_F=\ln \left[ C\left(C+2\right)\,\sqrt{\frac{45}{g}} \, \frac{g\, M_{Pl}\, m_{DM}\, \left(a+\frac{6\,b}{x_F}\right)}{2\,\pi^3\,\sqrt{{g^\star}(x_F)}\,\sqrt{x_F}}\right]\label{analyticxf}
\end{eqnarray}
\noindent where $C$ is of order 1. For the Dirac fermionic DM, the additional contribution from the anti-particle will make the  $\Omega_{DM}h^2$ twice of that is given in equation \eqref{omegadmhsq}. 
\par We compute the relic density numerically using MadDM \cite{Backovic:2013dpa, Backovic:2015tpt}, which require the pair annihilation cross-sections to be calculated by the event generator MadEvent \cite{Alwall:2007st,Maltoni:2002qb}. We have generated the input model files containing all the Feynman rules from the Lagrangian given in equations \eqref{LDMVLQ}-\eqref{LDMferm}, \eqref{lag:effscal} and \eqref{lag:effpscal} for the MadEvent using FeynRules \cite{Alloul:2013bka,Christensen:2008py}.  
\par To analyse and study the model we consider a single DM candidate with a specific intrinsic spin quantum  number associated with a given portal at a time. In other words, all DM-portal couplings   bar the one under discussion shall be switched off to zero. To keep our calculation in compliance with the LHC data, we use the central values of the portal-VLQ coupling $y_{\phi^0}$ ($y_{A^0}$) for a given  $m_\psi$ and the portal mass $m_{\phi^0}$ ($m_{A^0}$) as computed from the experimental cross-section curves for the gauge-boson production at ATLAS \cite{Aaboud:2016tru,Aad:2015zqe,Aaboud:2016trl} and given in the Figures \ref{pptoStobosons} and \ref{pptoPStobosons} respectively. 

\par We have verified  analytically the relic DM abundance by taking $g^\star(x_F)$ = 92 and $C$ = 1/2 in equations \eqref{omegadmhsq} and \eqref{analyticxf} and found them to be in agreement with the numerical calculations done by MadDM. 

In Figure \ref{scalportrelicscaldm270}, we depict the contours (for the scalar portal mass of 270 GeV and VLQ masses 0.4, 0.8, 1.2 and 1.6 TeV)  in the plane defined by the varying scalar DM mass between  0.02 - 4.0 TeV  and the  scalar portal - scalar DM coupling $\kappa_{\eta\phi^0}$ which generate  the correct amount of present day energy density for the scalar DM. In Figure \ref{scalportrelicscaldm}, we show the variation of relic density $\sim 0.11$ curves corresponding to the three   scalar portal masses 270, 500 and 750 GeV  for a fixed VLQ mass of 400 GeV. In Figures \ref{scalportrelicvecdm270} and \ref{scalportrelicvecdm} we plot the relic density $\sim 0.11$  contours in the plane defined by the varying vector DM mass between  0.02 - 4.0 TeV and the  scalar portal - vector DM coupling $\kappa_{V^0\phi^0}$ for the  values of portal and VLQ masses. These contours are evaluated corresponding $v_{\Phi}$  = 1 TeV.  In  Figures \ref{scalportrelicfermdm270} and \ref{pscalportrelicfermdm270} we plot the constant relic density contours for the fermionic DM in the plane defined by the DM mass and DM-portal coupling {\it w.r.t.} corresponding to the  scalar and pseudo-scalar portals  respectively.   Three contours in  Figures \ref{scalportrelicscaldm},  \ref{scalportrelicvecdm}, \ref{scalportrelicfermdm} and \ref{pscalportrelicfermdm}  correspond to the   three  choices of the portal masses 270, 500 and 750 GeV respectively.
We observe that \begin{itemize}
\item the unshaded region below the curve  is disallowed as it would over-close the universe with the DM. The perturbativity requirement that  the  coupling should be less than  $\sqrt{4\,\pi}$  further shrinks   the allowed parameter region.
\item  the pair annihilation cross-section on varying with DM mass  maximizes at the DM mass $ m_{\phi^0}/2$ for which the  constant relic density contours drop sharply {w.r.t.} the coupling. The sharp fall in the constant relic density contours are  again observed at the two  different values of DM masses a) first at  DM mass $\approx m_{\psi}$ where the portal  mediated $s$ channel  pair annihilation process opens up for the pair production of VLQ's  and b) then at DM mass $\approx m_{\phi^0}\,\, (m_{A^0})$ where DM mediated $t$ channel annihilation process opens up for pair production of portal scalars  (pseudo-scalars) respectively.  
\item  the relic density curve corresponding to the lowest  VLQ mass spans  the minimal allowed region as depicted {\it via} blue shaded region in Figures \ref{scalvecDMcont} and  \ref{FermiDMcont}. Increasing the mass of VLQ requires its  coupling with the portal to be large so that it is consistent with  the gauge Boson pair production at LHC, which in turn pull down the DM - portal coupling to a much  lower value such that enough  relic density is generated. Therefore, the contribution of the  DM greater than 1 TeV to the relic density can be made favourable with the choice of high VLQ masses  > 400 GeV.
\item the absence of the portal VEV dependence in the interaction Lagrangian of the fermionic DM renders its coupling with the portal  to be more sensitive.  The portal coupling with the DM are found to be an order of magnitude higher than the scalar and vector DM for an appreciable range of DM mass to generate the same relic density. Consequently, the allowed parameter region for which the perturbativity is satisfied becomes highly constrained.
\end{itemize}
\par We have performed the rest of our analysis with the conservative choice for VLQ mass of 400 GeV, corresponding to the three  choices of portal masses 270, 500 and 750 GeV. 
\begin{figure}[h!]
\begin{subfigure}{0.5\textwidth}
      \centering
      \includegraphics[width=\textwidth,clip]{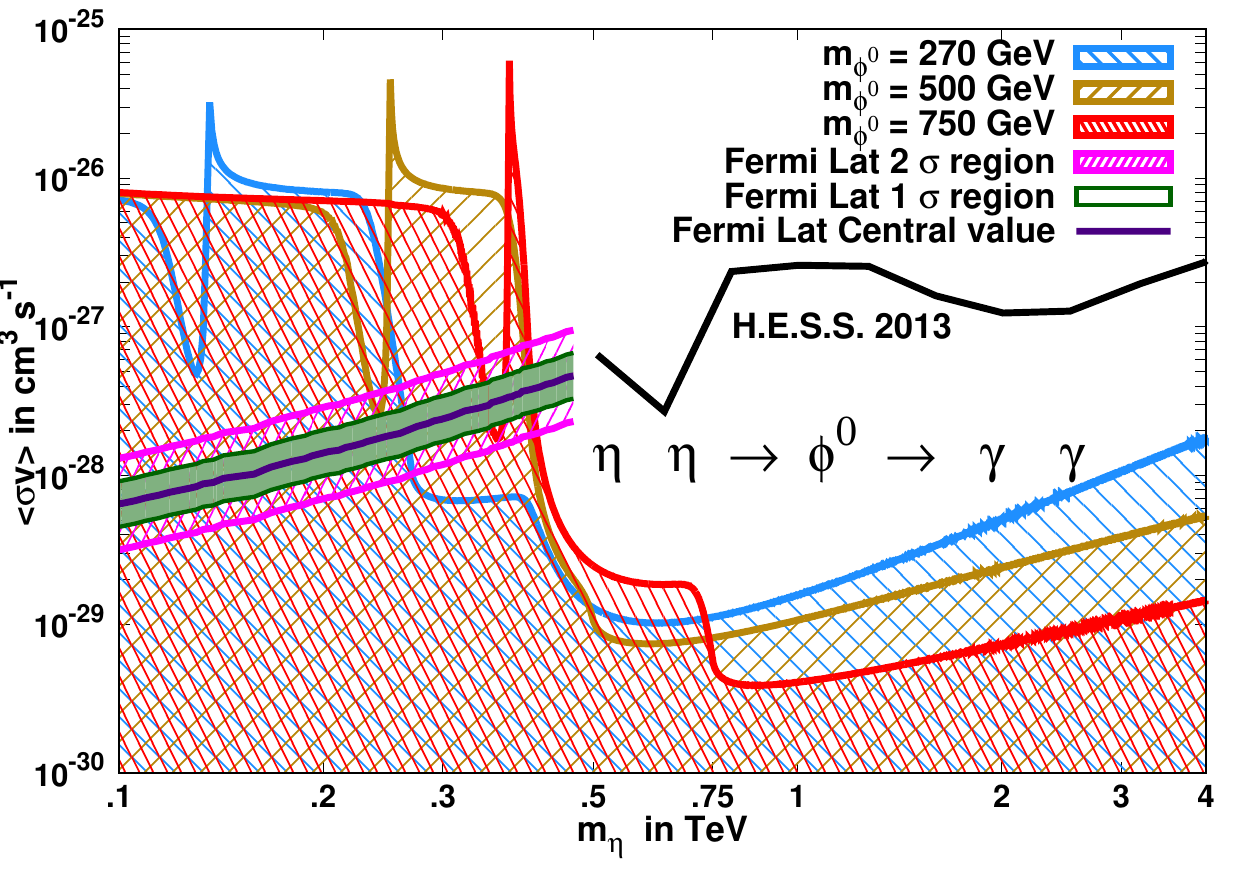}
      \caption{}\label{SSindirectpp}
    \end{subfigure}%
  \begin{subfigure}{0.5\textwidth}
      \centering
      \includegraphics[width=\textwidth,clip]{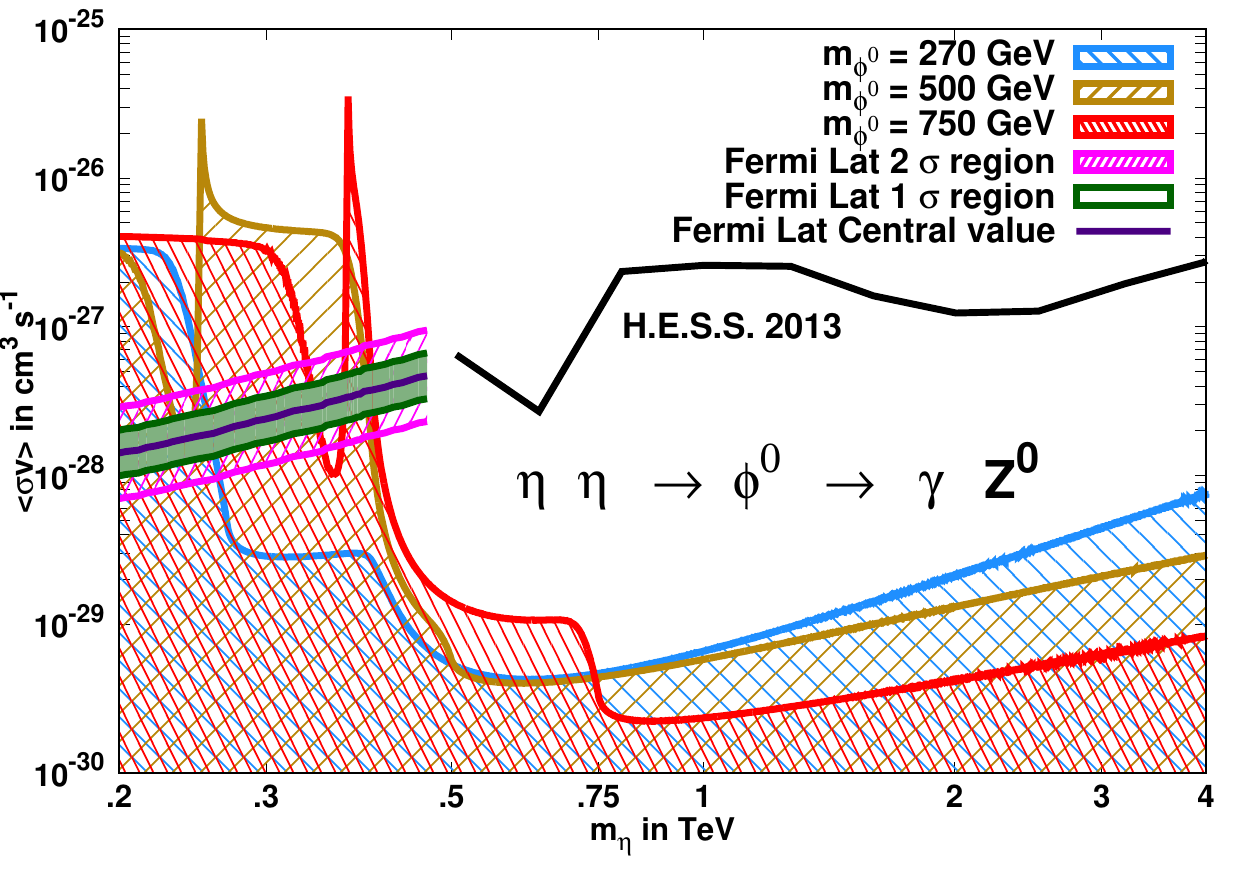}
      \caption{}\label{SSindirectpz}
    \end{subfigure}%

    \begin{subfigure}{0.5\textwidth}
      \centering
      \includegraphics[width=\textwidth,clip]{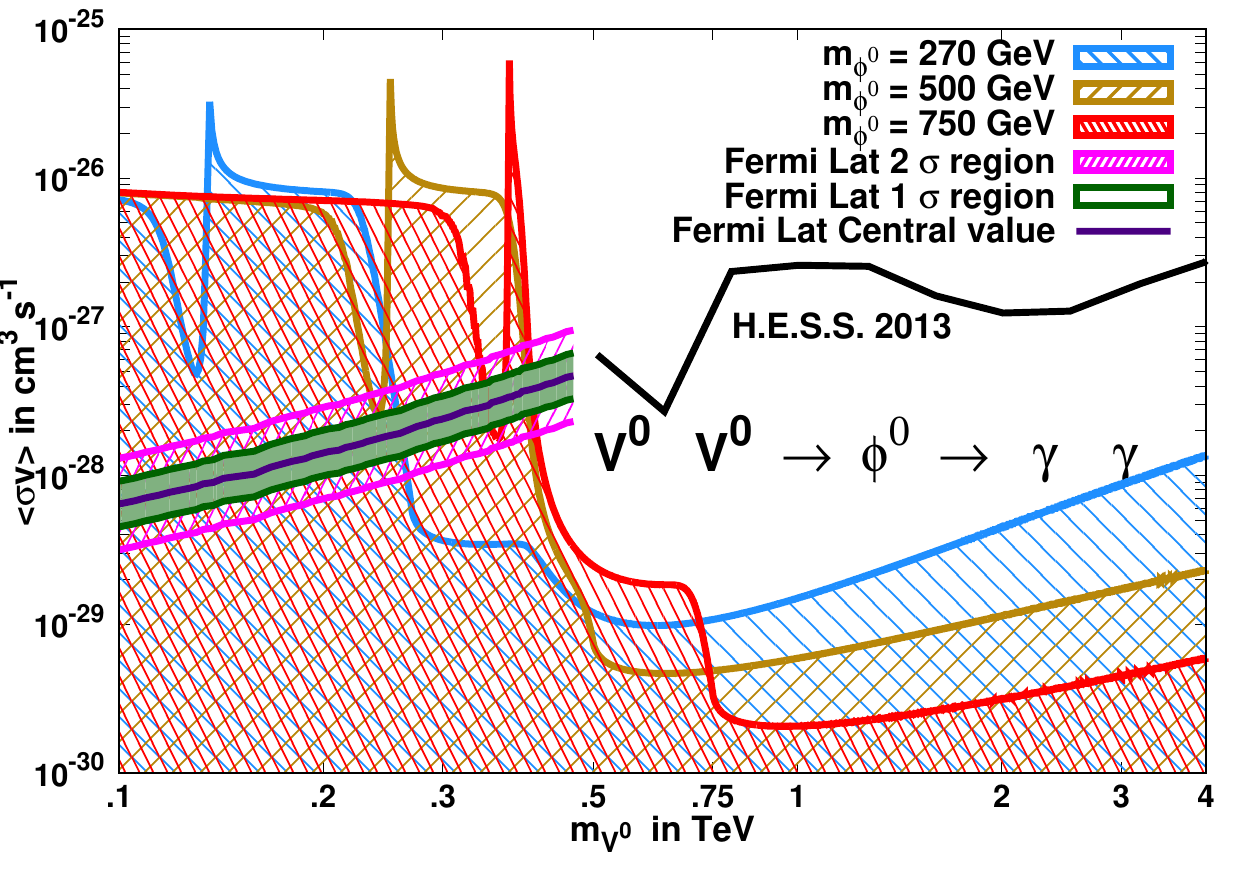}
      \caption{}\label{SVindirectpp}
    \end{subfigure}%
  \begin{subfigure}{0.5\textwidth}
      \centering
      \includegraphics[width=\textwidth,clip]{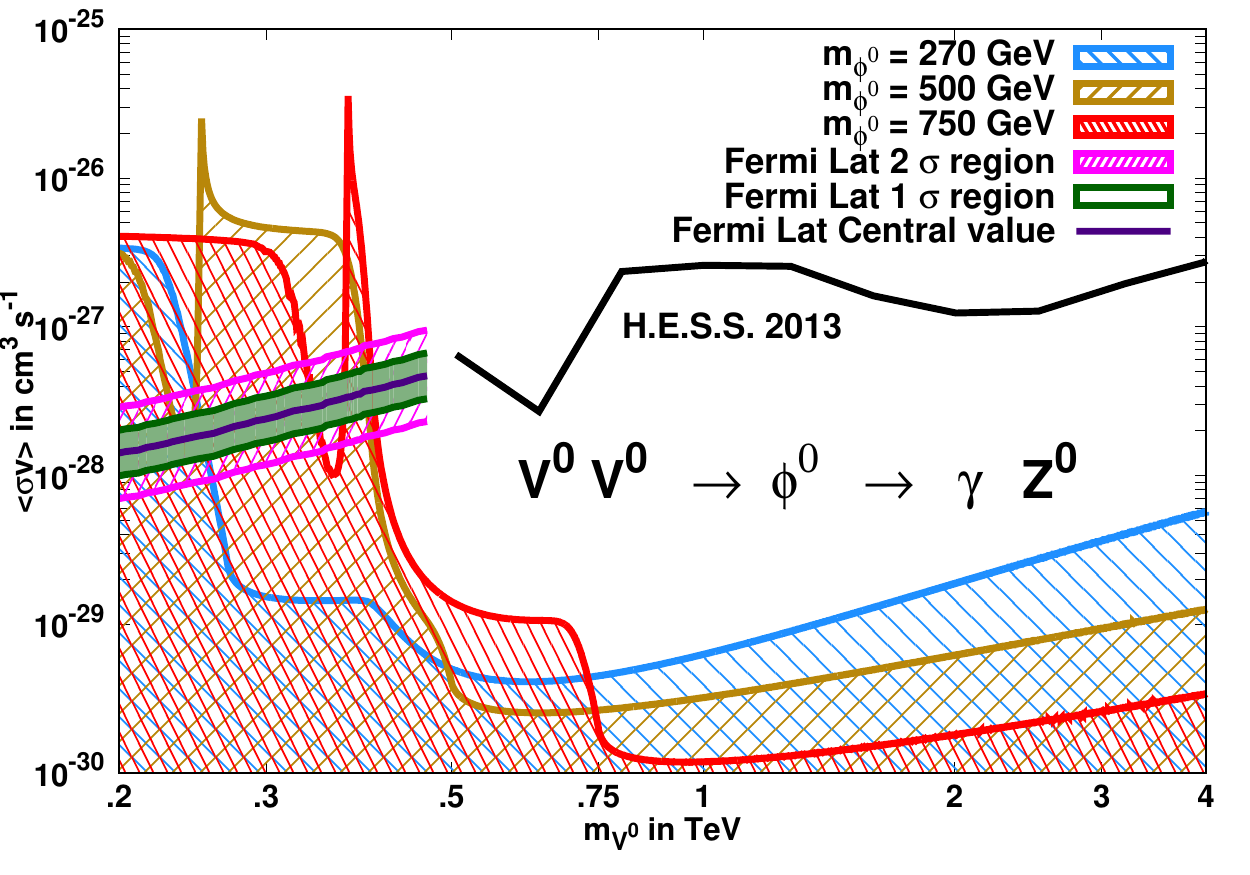}
      \caption{}\label{SVindirectpz}
      \end{subfigure}
\caption{\small \em{
Figures \ref{SSindirectpp} and \ref{SSindirectpz} depict the thermal averaged cross-sections for scalar DM pair annihilation $\eta\eta\to\phi^0\to\gamma\gamma$ and $\eta\eta\to\phi^0\to\gamma Z$  processes  {\it via} scalar portal respectively. Figures \ref{SVindirectpp} and \ref{SVindirectpz} show the thermal averaged cross-sections for vector DM pair annihilation $V^0V^0\to \phi^0\to\gamma\gamma$ and $V^0V^0\to \phi^0\to\gamma Z$  processes  {\it via} scalar portal respectively. All the cross-sections are drawn for the upper limit on DM couplings allowed by the relic density constraints for a given DM mass.
We have also exhibited the Fermi-LAT 1$\sigma$ and 2$\sigma$  limits for DM mass range < 500 GeV \cite{Ackermann:2015lka} and H.E.S.S. 2013 upper limit on the  thermal averaged cross-section for DM mass range > 500 GeV \cite{Abramowski:2013ax} corresponding to $\gamma\gamma$ and $\gamma Z$ channels. Shaded regions appearing in blue, golden yellow and red   correspond to the relic density forbidden regions  for the  portal mass of 270, 500 and 750 GeV respectively.}}
\label{fig:DMInDirectDetectionSV}
%\end{wrapfigure}
\end{figure}
\begin{figure}[h!]
%\begin{wrapfigure}{l}{0.5\textwidth} 
\begin{subfigure}{0.5\textwidth}
      \centering
      \includegraphics[width=\textwidth,clip]{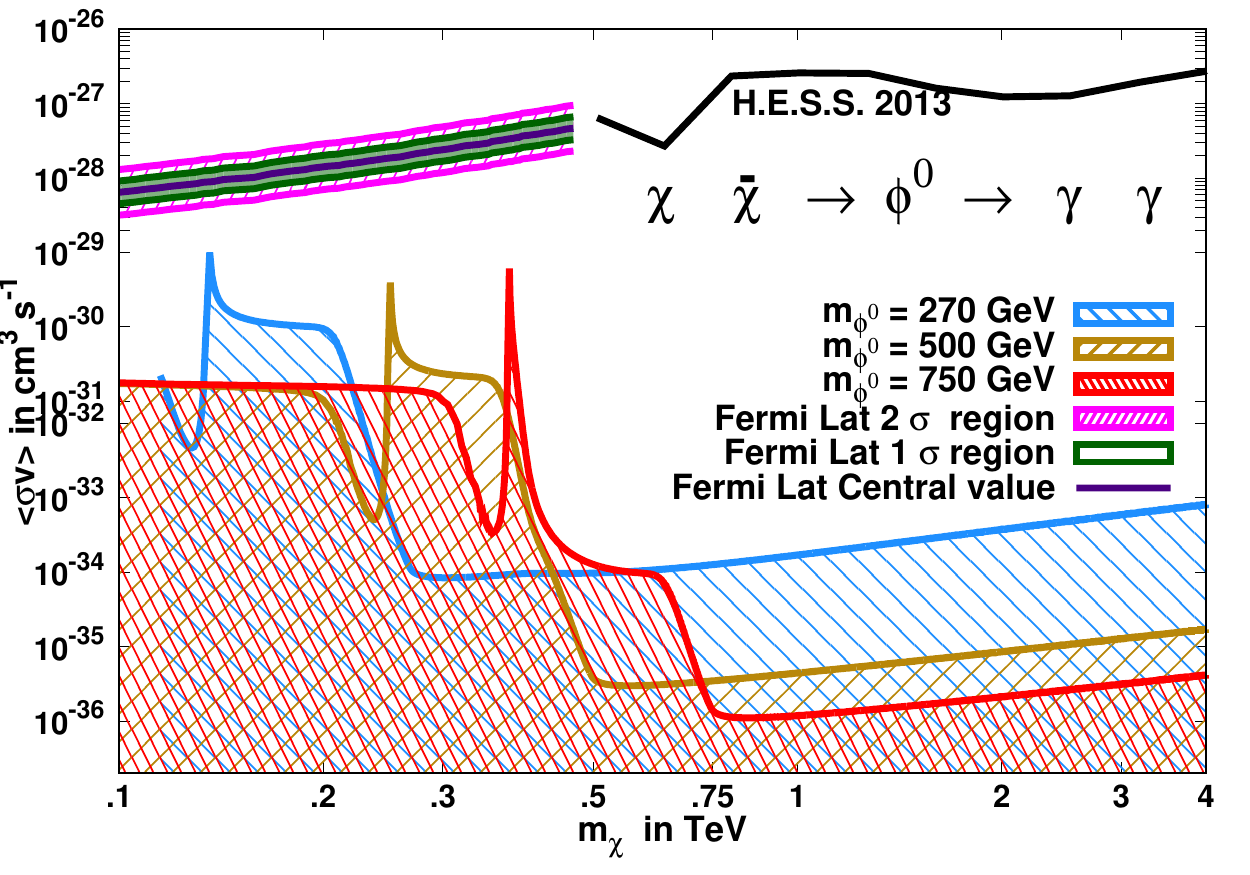}
      \caption{}\label{SFindirectpp}
    \end{subfigure}%
  \begin{subfigure}{0.5\textwidth}
      \centering
      \includegraphics[width=\textwidth,clip]{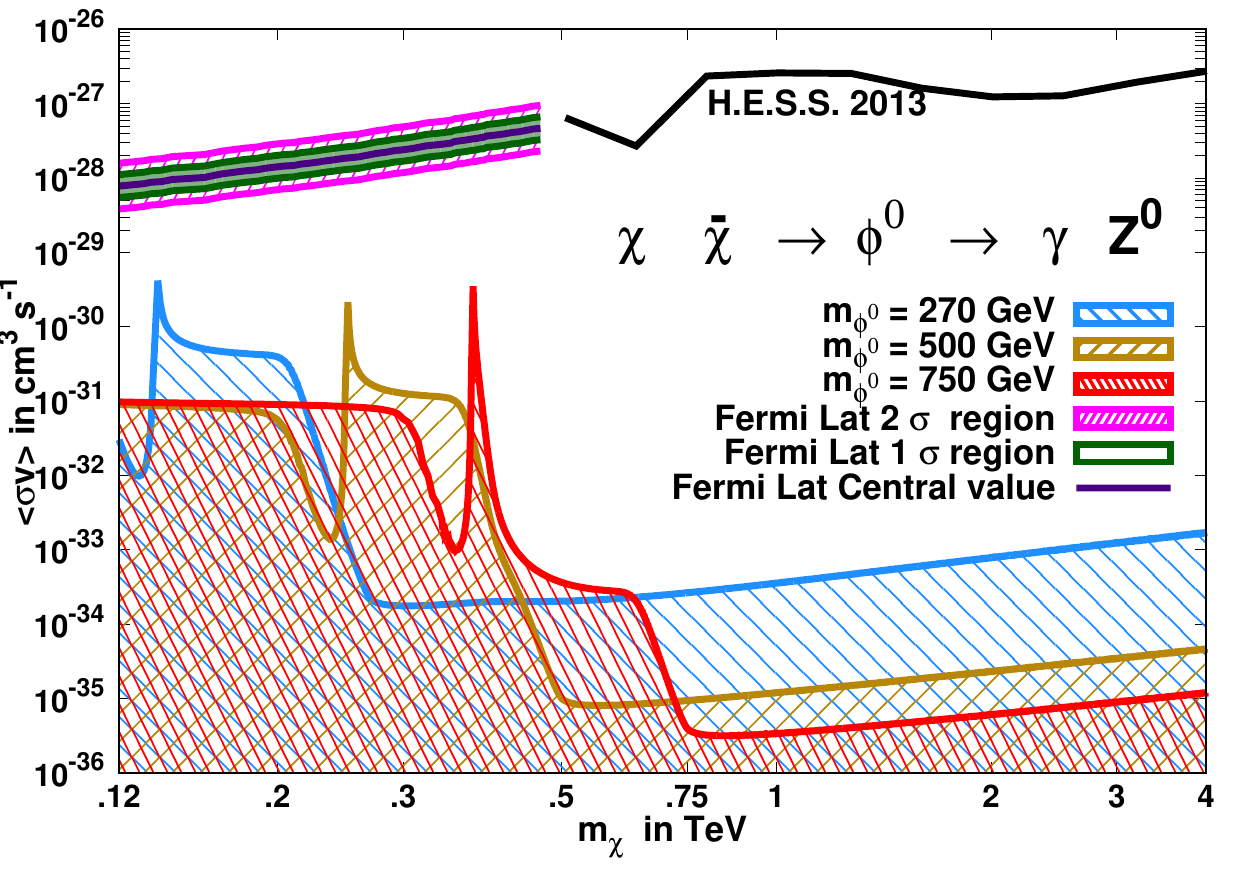}
      \caption{}\label{SFindirectpz}
    \end{subfigure}%

    \begin{subfigure}{0.5\textwidth}
      \centering
      \includegraphics[width=\textwidth,clip]{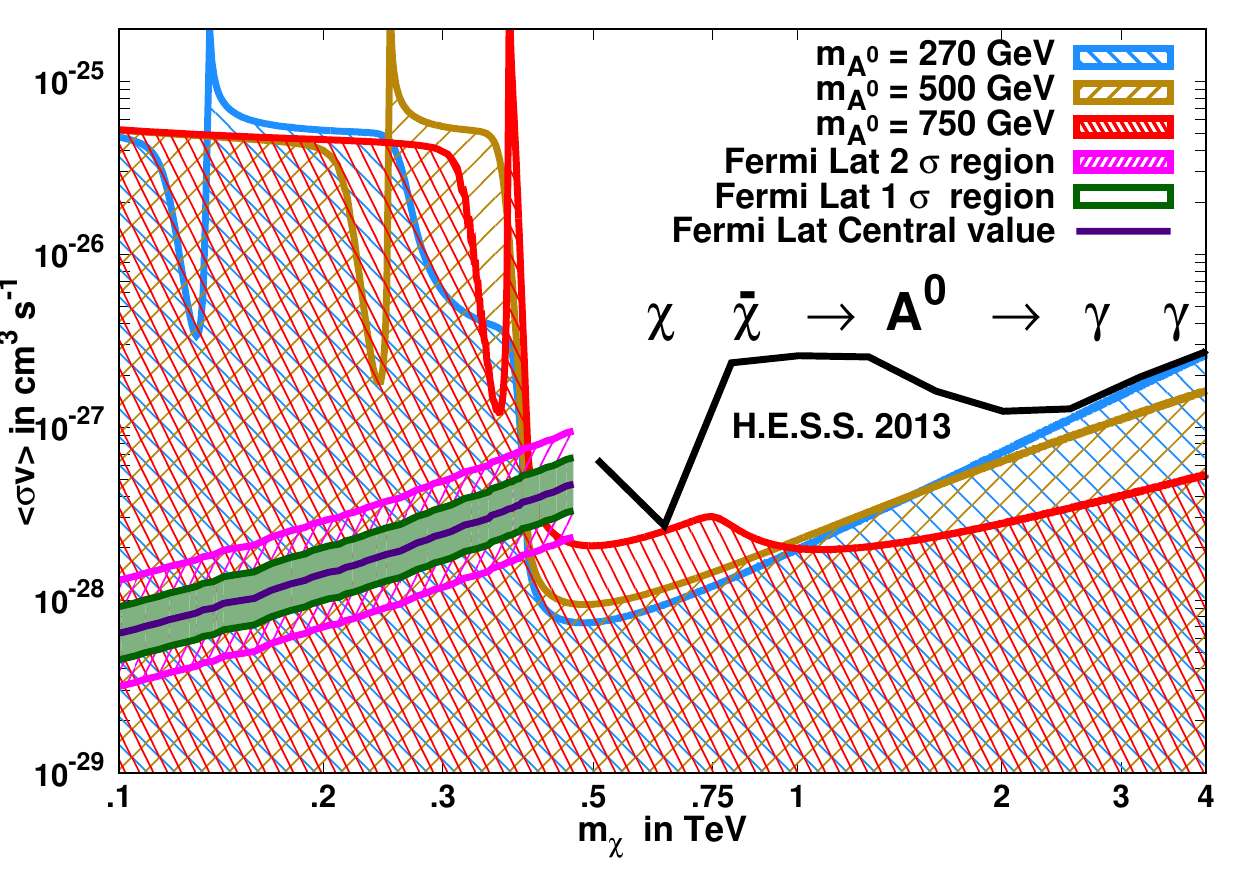}
      \caption{}\label{PSFindirectpp}
    \end{subfigure}%
  \begin{subfigure}{0.5\textwidth}
      \centering
      \includegraphics[width=\textwidth,clip]{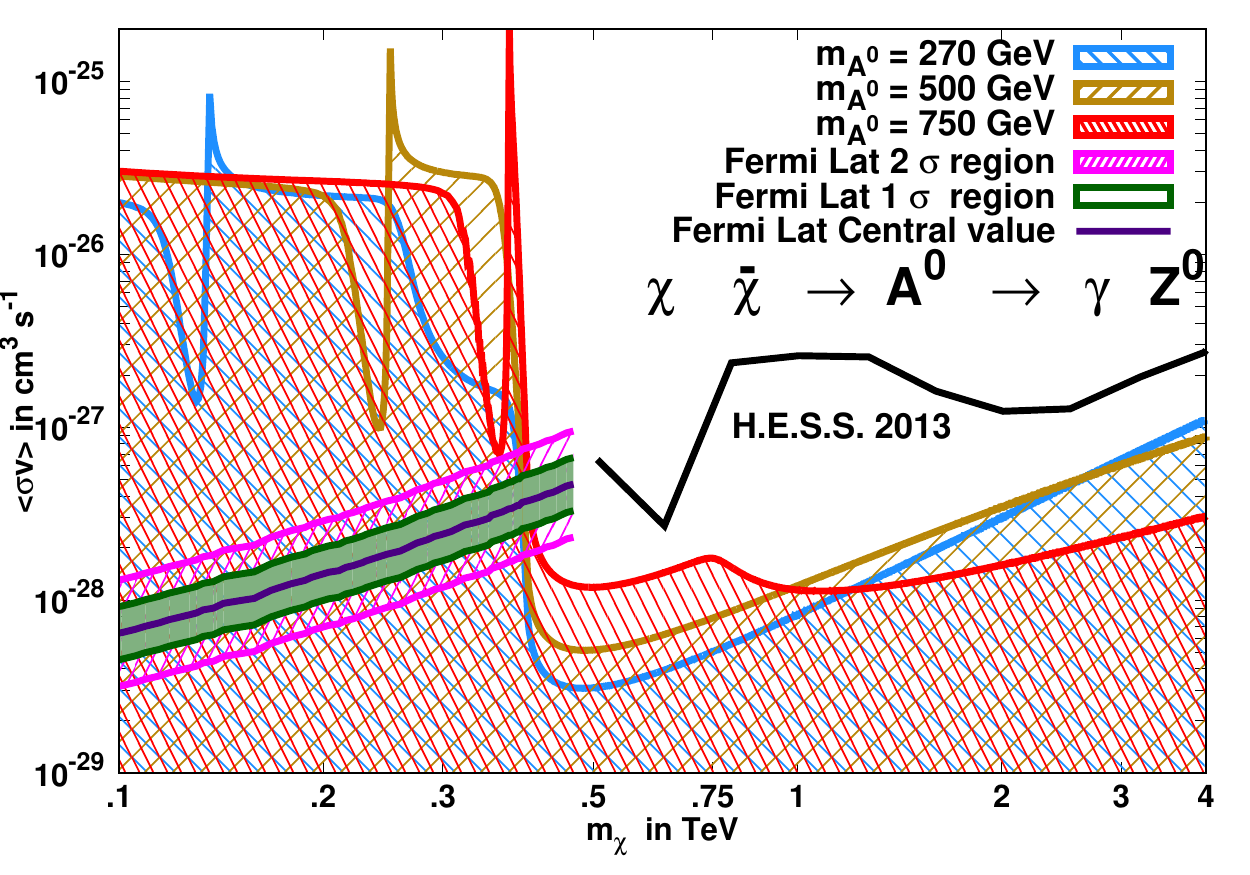}
      \caption{}\label{PSFindirectpz}
      \end{subfigure}
\caption{\small \em{Figures \ref{SFindirectpp} and \ref{SFindirectpz} depict the thermal averaged cross-sections for fermionic DM pair annihilation $\chi\bar\chi\to\phi^0\to\gamma\gamma$ and $\chi\bar\chi\to\phi^0\to\gamma Z$  processes  {\it via} scalar portal respectively. Figures \ref{PSFindirectpp} and \ref{PSFindirectpz} show the thermal averaged cross-sections for fermionic DM pair annihilation $\chi\bar\chi\to A^0\to\gamma\gamma$ and $\chi\bar\chi\to A^0\to\gamma Z$  processes  {\it via} pseudo-scalar portal respectively. All the cross-sections are drawn for the upper limit on DM couplings allowed by the relic density constraints for a given DM mass. We have also exhibited the Fermi-LAT 1$\sigma$ and 2$\sigma$  limits for DM mass range < 500 GeV \cite{Ackermann:2015lka} and H.E.S.S. 2013 upper limit on the  thermal averaged cross-section for DM mass range > 500 GeV \cite{Abramowski:2013ax} corresponding to $\gamma\gamma$ and $\gamma Z$ channels. Shaded regions appearing in blue, golden yellow and red   correspond to the relic density forbidden regions for  the  portal masses of 270, 500 and 750 GeV respectively.}}
\label{fig:DMInDirectDetectionF}
%\end{wrapfigure}
\end{figure}
\subsection{Indirect Detection : Monochromatic Gamma Rays}
\label{Indirect}
DM annihilation to SM photons (high energy gamma rays) in galactic halos  can be generated from various astrophysical targets for example the Dwarf Spheroidal galaxies, Galactic centre and Galaxy  clusters \cite{Fermi-LAT:2016uux,Ackermann:2015lka,Conrad:2015bsa}. These gamma rays can travel galactic distances and their flux can be observed by the satellite based $\gamma$ ray observatory Fermi-LAT \cite{Fermi-LAT:2016uux,Ackermann:2015lka} and the ground-based Cherenkov telescope  H.E.S.S. \cite{Abramowski:2011hc,Abramowski:2013ax}. The annihilation rates are roughly velocity independent in the non-relativistic region. Here in particular we compare the bounds from the indirect detection experiments in the $\gamma\gamma$ and $\gamma Z$ channels.

\par In our model, the production of the monochromatic photons are realised in the DM pair annihilation to $\gamma\gamma$ and $\gamma Z$ two body final states. These processes are studied in the context of the  scalar portal induced interactions for  the scalar, vector and fermionic DM candidates while the pseudo-scalar induced interactions are only allowed for spin 1/2 DM candidates. We can directly use the thermal averaged cross sections which are expressed in terms of the local velocity of the DM particle  in the Appendix \ref{ThermalAvCalculation}. We analyse the variation of the thermal averaged DM annihilation
cross-sections to $\gamma\gamma$ and $\gamma Z$  {\it w.r.t.} the  DM mass in detail for all the cases and  are depicted in the Figures \ref{fig:DMInDirectDetectionSV} and \ref{fig:DMInDirectDetectionF}.  To calculate the thermal averaged cross-section for the indirect detection we shall use the conservative lower bound on the DM-portal effective coupling obtained from  the relic density criterion for a given DM mass.   Therefore, the region below the curves defined in the DM mass and DM-portal coupling plane become cosmologically disfavoured.  We plot and compare our results in the $\gamma\gamma$ mode with the limits obtained from  Fermi-LAT \cite{Ackermann:2015lka}, for the restricted DM mass range (< 500 GeV). 
\par We find that the model calculated thermal averaged  DM pair annihilation cross-section in the $\gamma \gamma $ channel lies below the limit obtained from the experimental results for an appreciable range of DM mass.    The region trapped between the experimental curve  and  our results from the top and below respectively depicts the  allowed region of the thermal averaged cross-section, which can further be translated in terms of the allowed model parameter space  {\it w.r.t.} relic density and the indirect detection. Therefore the indirect experimental results naturally provides the upper bound on the coupling, for a given DM mass.  We note that  the  $\left\langle\sigma
\left(\chi\bar\chi\to \phi^0 \to\gamma\gamma\right)\,\, v\right\rangle$ is $p$-wave suppressed and therefore lies  much below than that of the null result obtained from the FermiLAT. 

\subsection{Direct Detection}
\label{Direct}
Direct detection of the DM identifies the nature of the low-energy effective DM-nucleon scattering interaction. Therefore, the direct detection experiments aim to establish a first confirmed detection of DM particles.  Direct detection data analysis focuses on formulation and computation of the momentum (energy) and velocity dependent cross-section scenarios, where DM couples to target nuclei by spin-independent or spin-dependent interactions. Spin-independent cross-section scales coherently with the nucleon number, so nuclei of larger atomic mass are always more effective in the direct detection searches. The spin-dependent scattering cross-section on the other hand is most effectively probed by a nucleus with larger spin.
\par Direct detection experiments Dark-Side50 \cite{Agnes:2015ftt}, LUX \cite{Akerib:2016vxi, Savage:2015xta,Cui:2017nnn}, XENON1T \cite{Aprile:2017iyp} etc. are set to observe the recoil energy transferred to target nucleus in an elastic collision with the DM particles. The current experimental null results \cite{Aprile:2017iyp}, constrain the maximum value of the elastic
nucleon-DM cross-sections. The constraints on the upper limit of the elastic cross-sections  will be considerably lowered in the future projected sensitivities of the super CDMS experiments \cite{Witte:2017qsy}.

\par The elastic scattering of DM particles $\eta, V^0$ and $\chi$ from a heavy nucleus  can be illustrated in terms of the effective DM  scattering off the gluons, where the DM is attached to the triangle loop
of charged virtual VLQ {\it via} the portal which in turn interacts with the nucleons via two gluons exchange. Since, this occurs at very low energy and momentum $\lesssim $ O(1) GeV, all the propagators are approximated by their respective  masses.  The nucleus recoil which is of the order of few MeV  is then measured in the detector.
\begin{wrapfigure}{l}{.5\textwidth}
\vskip -4.0cm
  \includegraphics[width=0.8\textwidth]{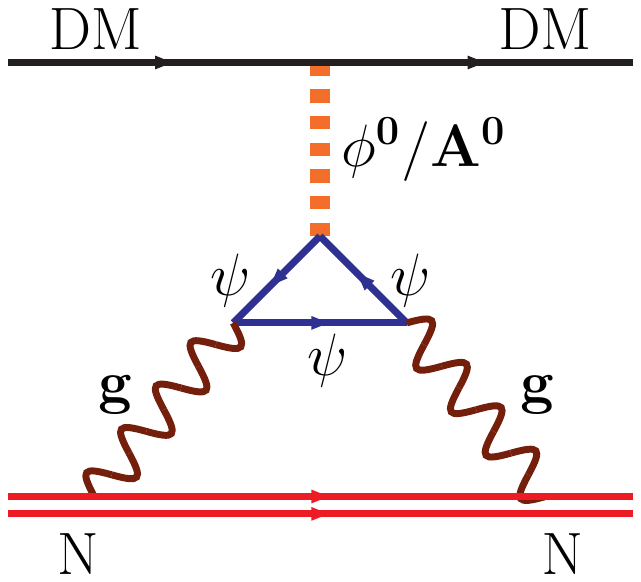}
  \vskip -9.75cm
  \caption{\small \em{ DM- Nucleon
scattering diagram induced by the portal - VLQ loop interaction.}}
\label{fig:DMgluonefffeyndia}
\vskip -0.5cm
\end{wrapfigure}
DM particles scatter-off nucleons through the $t$ channel exchange of portal scalar and the scattering of gluons {\it via} triangle VLQ loop, gives the dominant contribution to the DM-nucleon scattering cross-section. The scalar gluon-gluon coupling is described by the effective Lagrangian given in (2.9a) and (2.9b) as
\begin{eqnarray}
 {\cal L}_{\phi^0\,gg}^{\rm eff} &=& \frac{y_{\phi^0}}{m_{ \psi  }} \,\,\frac{\alpha_{s}(m_{\phi^0})}{12\,\pi} \,\, 3 I_{gg}\,\, G^{a}_{\mu\nu} \, G^{a}_{\mu\nu}\, \phi^0, \nonumber\\\label{direct3vertexscalleff}
 \end{eqnarray}
 \noindent where the loop integral $I_{gg}$ is given in Appendix A. The effective DM-gluon interaction can be described by the effective Lagrangian
 \begin{equation}
   {\cal L}_{\rm eff}^{\eta\eta gg} = \frac{\alpha_{s}(m_{\phi^0})\,v_{\Phi}\,\kappa_{\eta\phi^0}\, y_{\phi^0}}{24\,\pi\,m_{\phi^0}^{2} \,m_{ \psi  }} \,\,(3\,I_{gg}) \,\eta\,\eta\,\, G^{a}_{\mu\nu}\, G_{a}^{\mu\nu}
 \label{direct4vertexscalleff}
 \end{equation}
\noindent through the Feynman diagram given in Figure \ref{fig:DMgluonefffeyndia}. This interaction contributes to the spin-independent part of the DM- nucleon scattering cross-section to give
\begin{eqnarray}
 \sigma_{SI}(\eta N \rightarrow \eta N) = \frac{1}{729\,\pi} \frac{y_{\phi^0}^{2} \,v_{\Phi}^2\,\kappa_{\eta\phi^0}^{2} \,\mu_{N\eta}^{2}\, m_{N}^{2}}{2\,m_{\eta}^{2} \,m_{\phi^0}^{4}\, m_{ \psi  }^{2}} \left(\frac{\alpha_{s}(m_{\phi^0})}{\alpha_{s}(m_{H})}\right)^{2} \left(3\,I_{gg}\right)^2 \left(f_{TG}^N\right)^2 \end{eqnarray}
where $\mu_{N\eta} = \frac{m_{N} \, m_{\eta}}{(m_{\eta} + m_{N})}$ is the nucleon-DM reduced mass and 
$f_{TG}^N \equiv  -\,\frac{9\,\alpha_s(\mu)} {m_N\, 8\,\pi} \,\left\langle N\left\vert O_g\right\vert N\right\rangle $ is the gluon contribution to  the zero-momentum hadronic matrix element.  $f_{TG}^N $ can be extracted   in terms of  light quark contribution   to the zero-momentum hadronic matrix element as 
\begin{eqnarray}
f_{TG}^N= 1-\sum_{q = u,\,d,\,s} f_{Tq}^N \equiv 1-\sum_{q = u,\,d,\,s} \frac{1}{m_N}\left\langle N\left\vert O_q\right\vert N\right\rangle.
\end{eqnarray}
\noindent Here the hadronic matrix element  refers to a definite spin state of the nucleon. 
The $f_{TG}^N$  is  calculated to be 0.92  using the values for $f_{Tq}^N $ quoted in the literature \cite{DelNobile:2013sia,Belanger:2013oya,Cheng:2012qr,Beringer:1900zz}. 
 \par
\begin{figure}[H]
%\begin{wrapfigure}{l}{0.5\textwidth}
 \centering
 \begin{subfigure}{0.49\textwidth}
      \centering
      \includegraphics[width=\textwidth,clip]{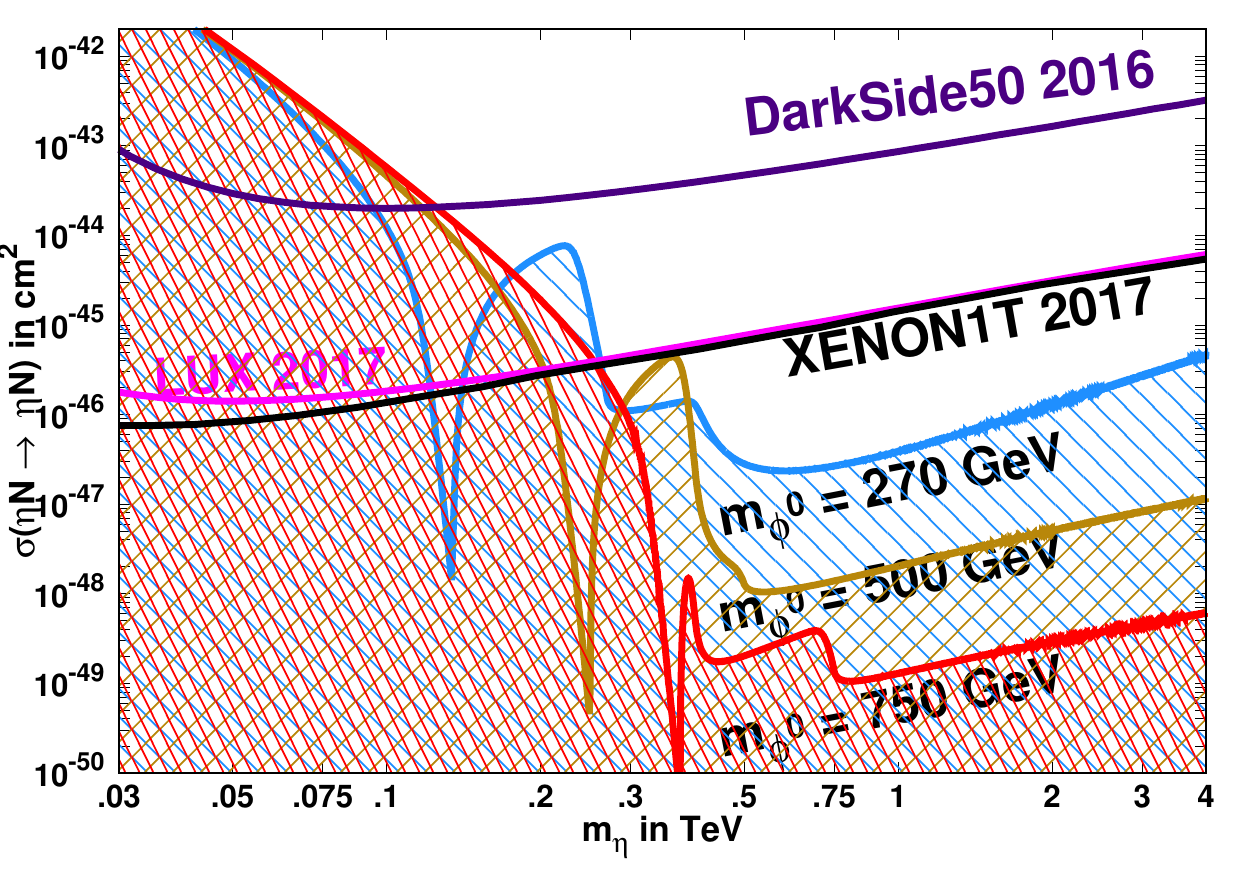}
      \caption{}\label{ssdirect}
    \end{subfigure}%
  \begin{subfigure}{0.49\textwidth}
      \centering
      \includegraphics[width=\textwidth,clip]{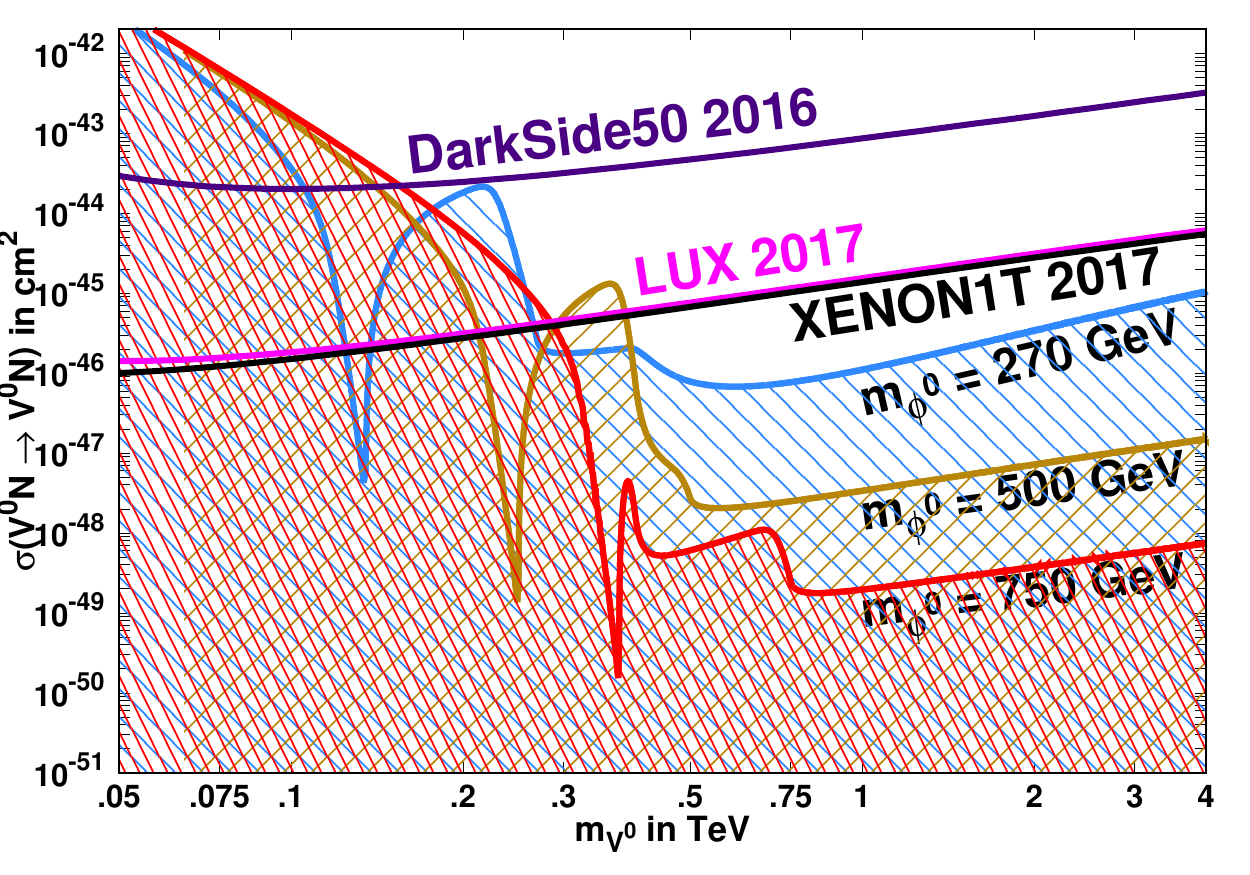}
      \caption{}\label{svdirect}
    \end{subfigure}%
    \vskip 0.01 cm
      \begin{subfigure}{0.49\textwidth}
      \centering
      \includegraphics[width=\textwidth,clip]{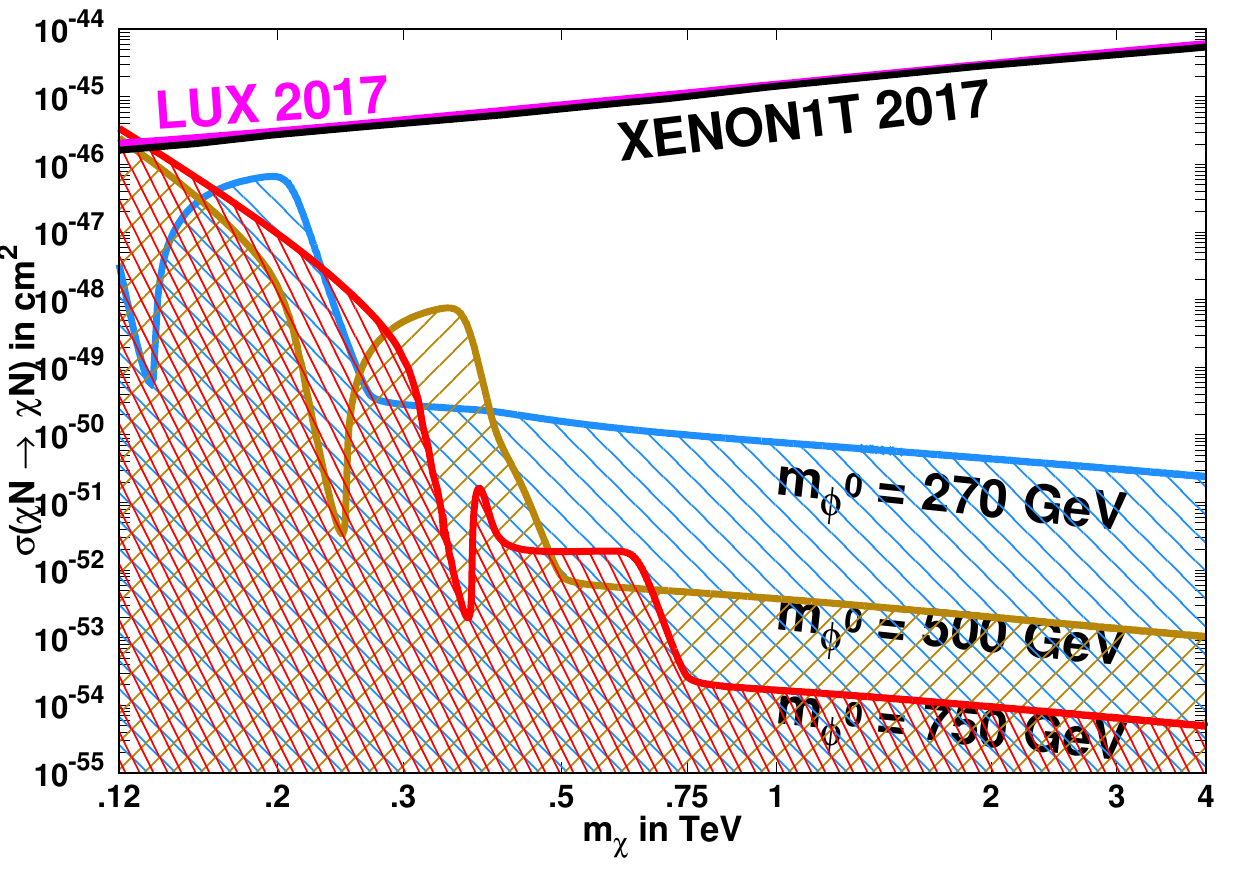}
      \caption{}\label{sfdirect}
    \end{subfigure}%
\caption{\small \em{The spin-independent DM-nucleon elastic cross-sections are depicted as a function of DM mass for the scalar, vector and fermionic dark matter in Figures \ref{ssdirect}, \ref{svdirect} and \ref{sfdirect} respectively. In each of the figure three graphs are exhibited for the fixed VLQ mass 400 GeV and scalar portal masses 270, 500, 750 GeV respectively.
All the cross-sections are drawn for the upper limit on the respective DM couplings allowed by the relic density constraints for a given DM mass. We have also exhibited the experimental cross-section from LUX (2017) \cite{Akerib:2016vxi,Savage:2015xta}, XENON (2017) \cite{Aprile:2017iyp} and Dark-Side50 (2016) \cite{Agnes:2015ftt}. Shaded regions appearing in blue, golden yellow and red   correspond to the relic density forbidden regions  for the  portal masses of 270, 500 and 750 GeV respectively.}}
\label{fig:DMDirectDetection}
%\end{wrapfigure}
\end{figure}

  We estimate the direct detection cross-section using the central value of $y_{\phi^0}$ obtained from the LHC for a given portal scalar mass  and  $m_\psi$ = 400 GeV  along with the  cosmologically allowed lower bound on the DM - scalar portal coupling for a given DM mass. The variation of the scalar DM - Nucleon scattering cross-section {\it via} scalar portal is depicted in the Figure \ref{ssdirect} corresponding to the three   scalar portal masses 270, 500 and 750 GeV respectively.

\par On the similar note, the  spin-independent cross-sections for the vector and fermionic dark matter interaction {\it via} the scalar mediator are given by
\begin{eqnarray}
\sigma_{SI}(\chi N \rightarrow \chi N) &=& \frac{2\, m_N^2\, \mu_{N\chi}^2}{\pi}\ \frac{\kappa_{\chi\phi^0}^2\, y_{\phi^0}^2}{729\ m_{\phi^0}^4\ m_\psi^2} \left(\frac{\alpha_{s}(m_{\phi^0})}{\alpha_{s}(m_{H})}\right)^{2} \,\,\left(3\,I_{gg}\right)^2 \,\,(f_{TG}^N)^2 \\
\sigma_{SI}(V^0N \rightarrow V^0N) &=& \frac{\mu_{NV^0}^2\, m_N^2}{2 \pi}\ \frac{\kappa_{V^0\phi^0}^2\ v_{\Phi}^2\  y_{\phi^0}^2}{729\, m_{V^0}^2\, m_{\phi^0}^4\ m_\psi^2} \left(\frac{\alpha_{s}(m_{\phi^0})}{\alpha_{s}(m_{H})}\right)^{2} \,\,\left(3\,I_{gg}\right)^2 \,\,(f_{TG}^N)^2
\end{eqnarray}
\noindent respectively where $\mu_{N\chi}=\frac{m_N\,m_\chi}{m_N+m_\chi}$ and $\mu_{NV^0}=\frac{m_N\,m_{V^0}}{m_N+m_{V^0}}$.
The variation of the vector and fermionic  DM - Nucleon scattering cross-sections {\it via} scalar portal are depicted in the Figures \ref{svdirect} and \ref{sfdirect} respectively.
Each figure depicts  the scattering cross-section for three  scalar portal masses 270, 500 and 750 GeV respectively.

\par On comparison we find that for the most of the DM mass range our direct detection cross-section curves  in Figures \ref{ssdirect}, \ref{svdirect} and \ref{sfdirect} lie much below the null result of the experiment \cite{Aprile:2017iyp} which in turn gives an upper bound on the DM -portal coupling for a given DM mass and thus constrains the portal induced DM parameter space. We note that the scalar and vector DM scattering cross-section is an  order of magnitude higher than that of the fermionic DM contribution. Thus the DM scattering experimental constraints do not shrink the allowed parameter space for the fermionic DM below the mass region 300 GeV which is in contrary to the scalar and vector DM contributions.
\par For the direct detection cross-section induced by  the CP odd pseudo-scalar mediator the corresponding $A^0\, g\,g$ coupling and the effective DM-gluon  interaction is given by
\begin{eqnarray}
 {\cal L}_{A^0\,gg}^{\rm eff} &=& \frac{y_{A^0}}{m_{\psi}} \frac{\alpha_{s}(m_{A^0})}{8\,\pi} 2\, \tilde I_{gg} G^{a}_{\mu\nu} \tilde{G}^{a}_{\mu\nu}\, A^0 \label{direct3vertexpscalleff} \\
  {\cal L}_{\rm eff}^{\chi\,\chi \,g\,g} &=& i \,\kappa_{\chi A^0}\, y_{A^0} \,\frac{1}{m_{A^0}^{2} \,m_{ \psi  }} \frac{\alpha_{s}}{ \pi} \,\tilde I_{gg} \,\overline{\chi} \,\gamma_{5} \,\chi G^{a}_{\mu\nu} \,\tilde{G}_{a}^{\mu\nu}. \label{direct4vertexpscalleff}
 \end{eqnarray}

\par The effective Lagrangian \eqref{direct4vertexpscalleff} generates a spin-dependent  DM-nucleon cross-section which is however  suppressed by the square of the momentum  exchanged and is therefore sub-dominant.  The available experimental results on the spin-dependent cross-sections are comparatively less tightly constrained \cite{Savage:2015xta,Marcos:2015dza,Cui:2017nnn}, hence not shown  graphically.
\begin{figure}[tbh]
%\begin{wrapfigure}{l}{0.5\textwidth} 
 \centering
 \begin{subfigure}{0.5\textwidth}
      \centering
      \includegraphics[width=\textwidth,clip]{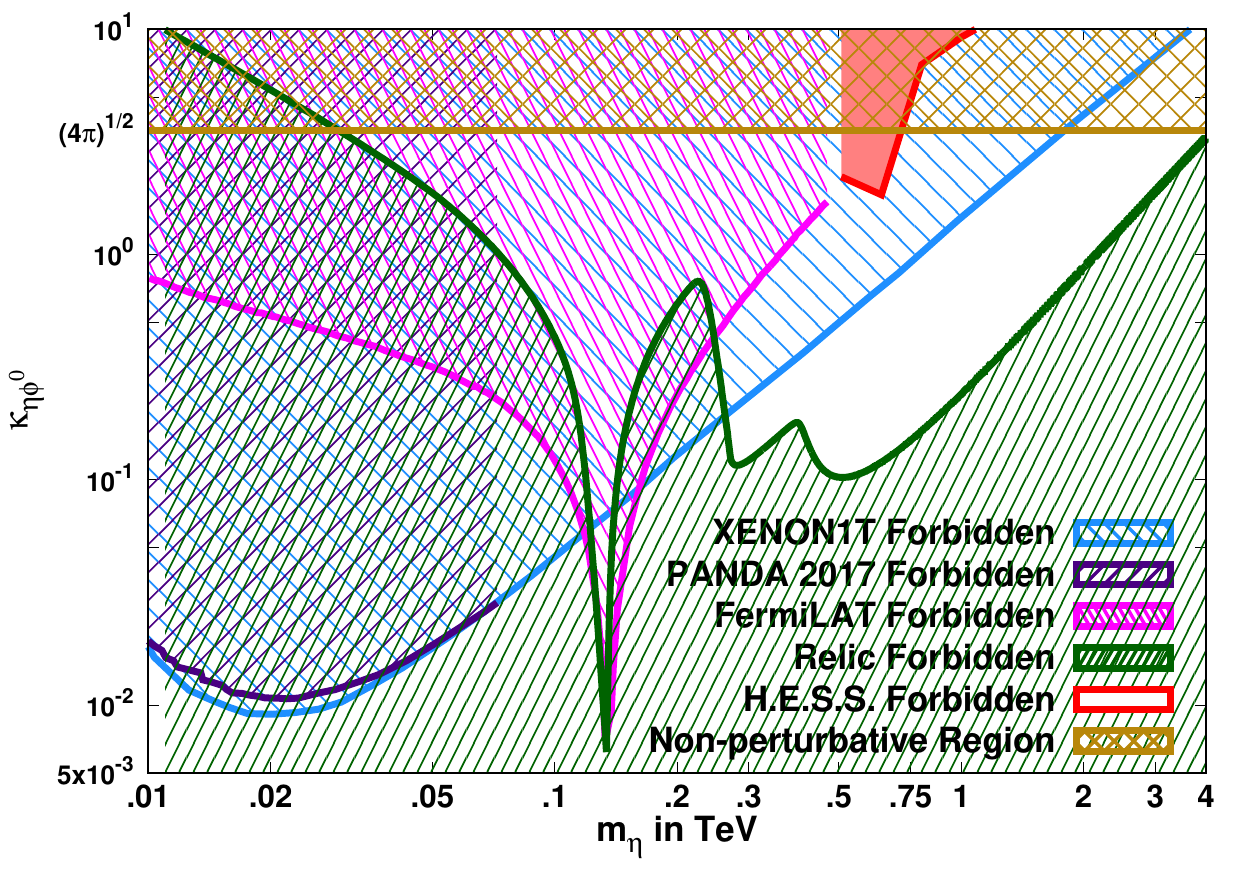}
      \caption{\small{\em Scalar DM with Scalar Portal}}\label{sscons270}
    \end{subfigure}%
    \begin{subfigure}{0.5\textwidth}
     \centering
      \includegraphics[width=\textwidth,clip]{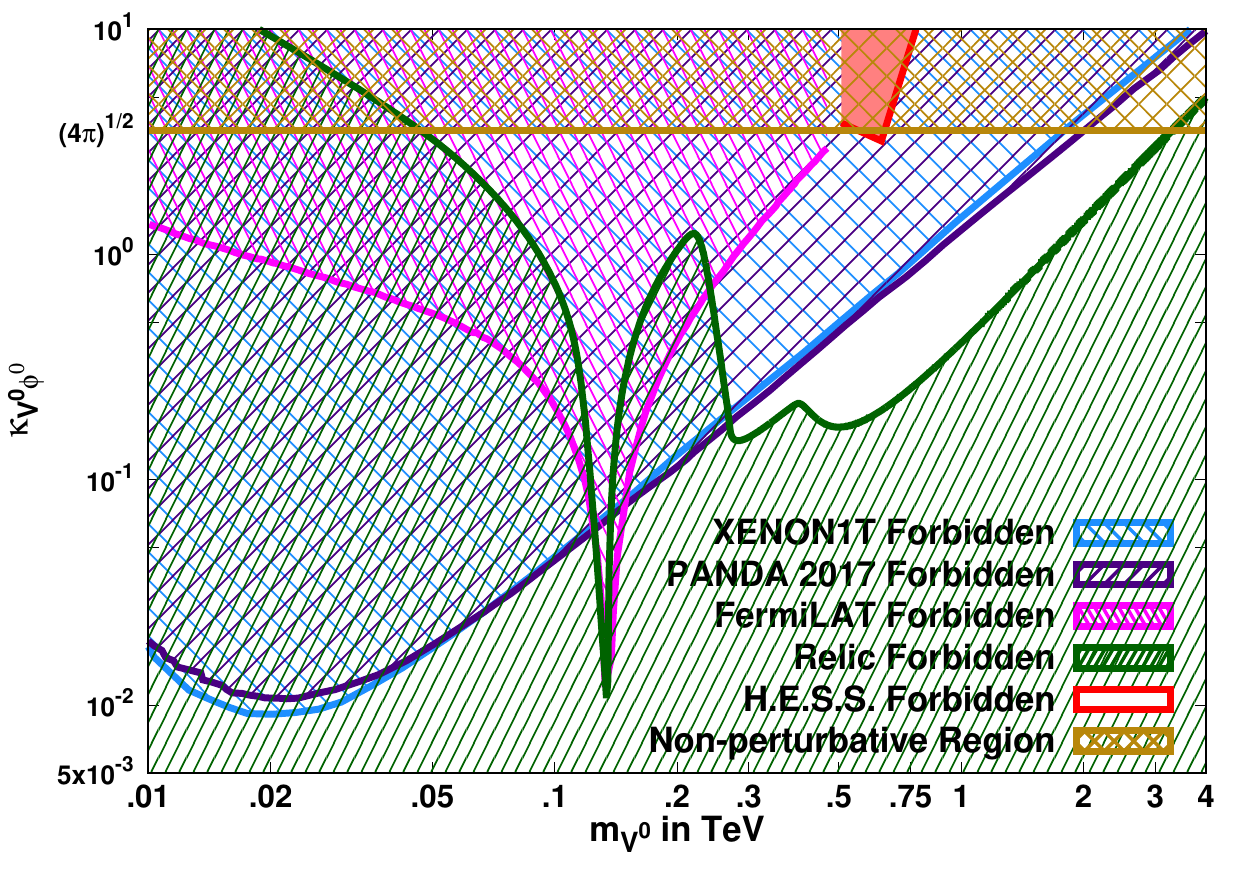}
      \caption{\small{\em Vector DM with Scalar Portal}}\label{svcons270}
    \end{subfigure}%

  \begin{subfigure}{0.5\textwidth}
      \centering
       \includegraphics[width=\textwidth,clip]{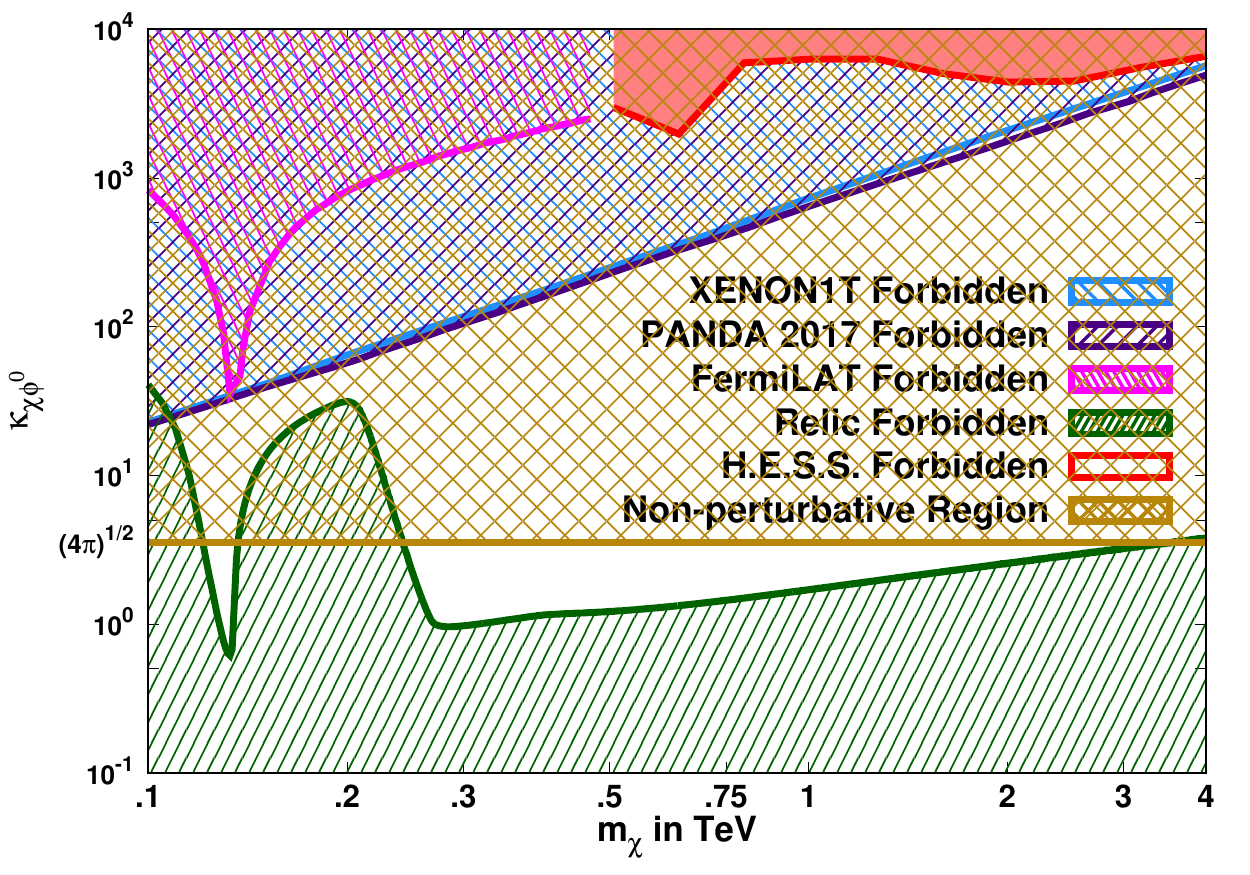}
      \caption{\small{\em Fermionic DM with Scalar Portal}}\label{sfcons270}
    \end{subfigure}%
      \begin{subfigure}{0.5\textwidth}
      \centering
      \includegraphics[width=\textwidth,clip]{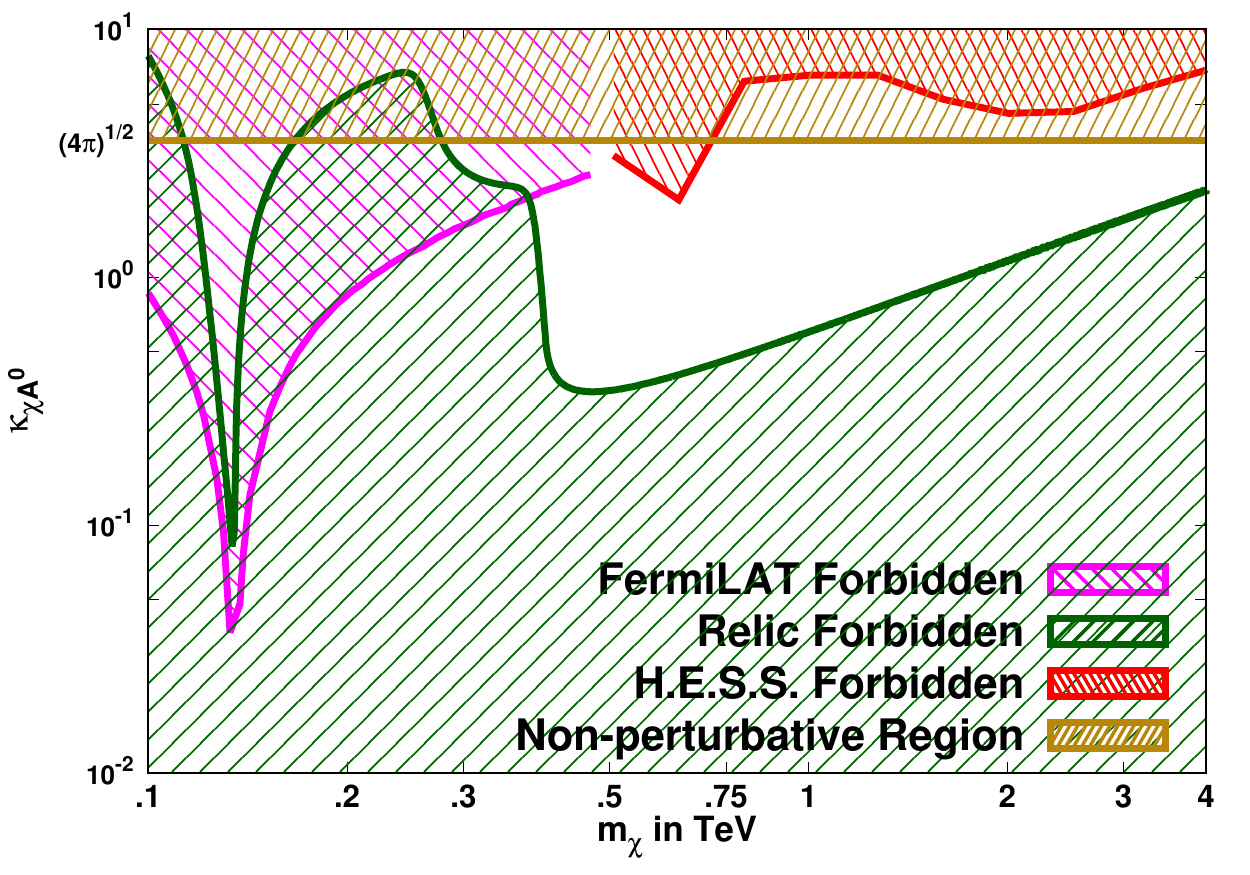}
      \caption{\small{\em Fermionic DM with Pseudo-scalar Portal}}\label{pfcons270}
    \end{subfigure}%
      \caption{\small{\em All model constraints are drawn for the portal mass of 270 GeV and VLQ mass 400 GeV. The shaded region correspond to  the forbidden  regions by relic density \cite{WMAP1,Ade:2015xua} (green), XENON1T 2017 \cite{Aprile:2017iyp} (blue), PANDA 2017 \cite{Cui:2017nnn} (indigo), Fermi-LAT \cite{Ackermann:2015lka} (pink), H.E.S.S. \cite{Abramowski:2013ax}  (red) and the perturbativity condition (golden yellow) respectively in the plane defined by  $m_\eta - \kappa_{\eta\phi^0}$, $m_{V^0} - \kappa_{V^0\phi^0}$, $m_\chi - \kappa_{\chi\phi^0}$ and  $m_\chi - \kappa_{\chi A^0}$ in Figures \ref{sscons270}, \ref{svcons270}, \ref{sfcons270}, and \ref{pfcons270}  respectively.  Constraint from the  direct detection experiments  for the pseudo-scalar portal are not shown (see text).}}\label{DMcons270}
    \end{figure}%
    \begin{figure}[tbh]
%\begin{wrapfigure}{l}{0.5\textwidth} 
 \centering
 \begin{subfigure}{0.5\textwidth}
      \centering
      \includegraphics[width=\textwidth,clip]{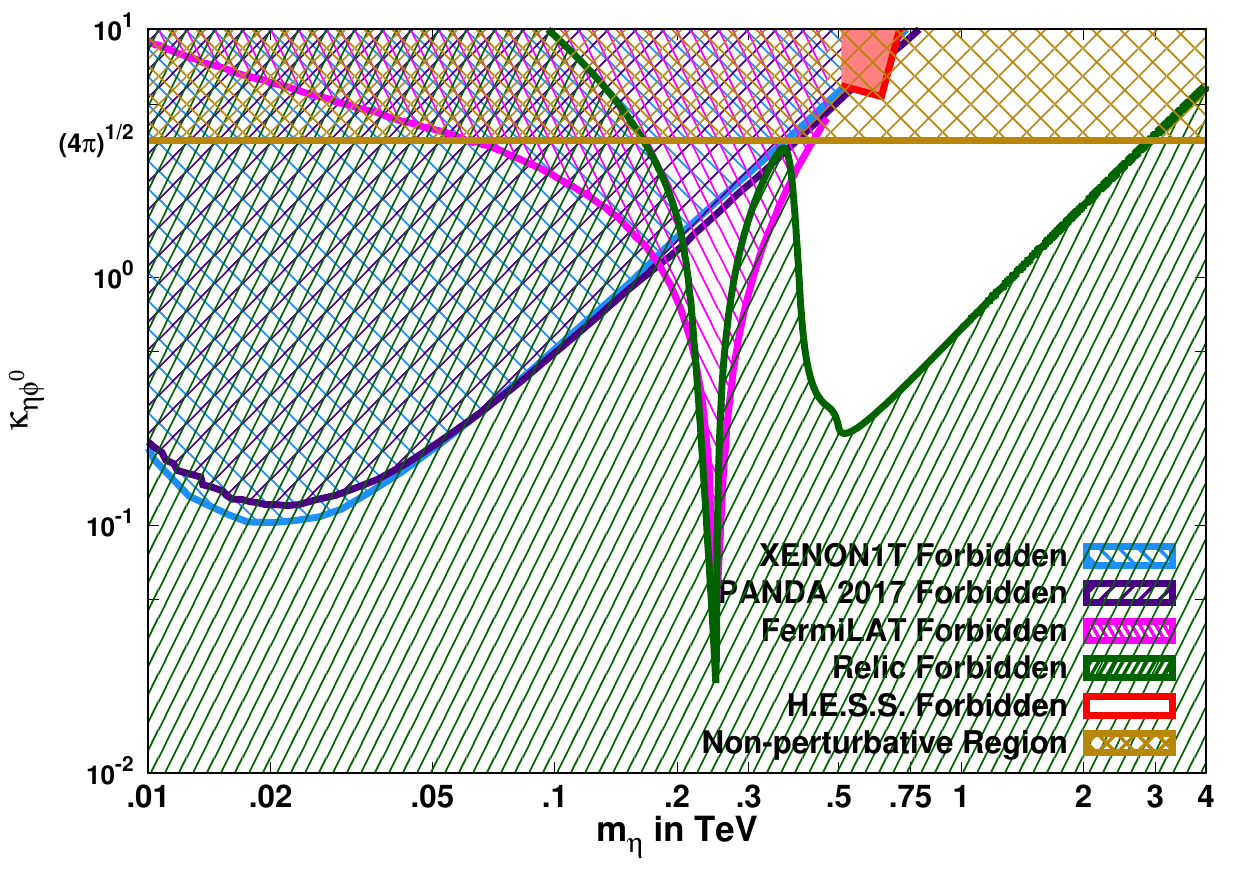}
      \caption{\small{\em Scalar DM with Scalar Portal}}\label{sscons500}
    \end{subfigure}%
    \begin{subfigure}{0.5\textwidth}
      \centering
         \includegraphics[width=\textwidth,clip]{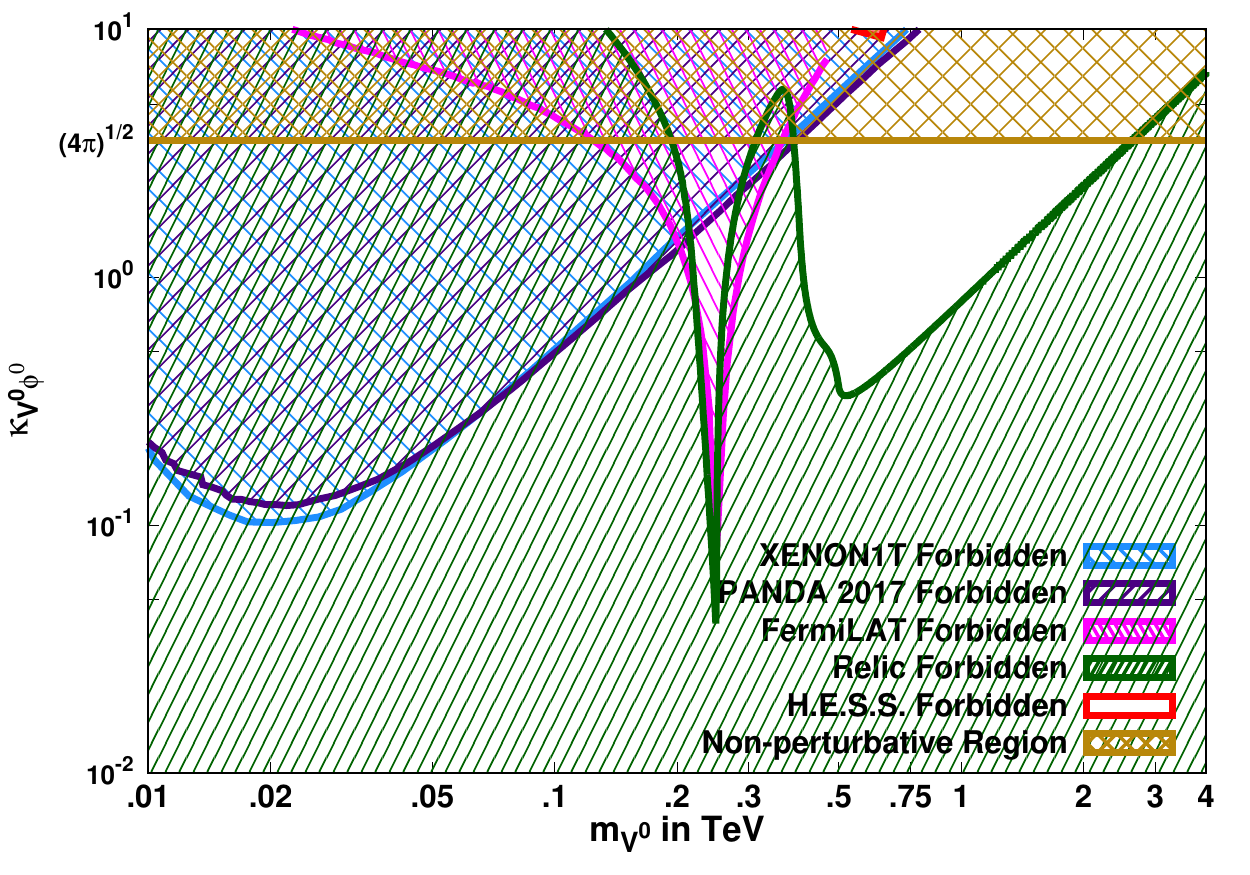}
      \caption{\small{\em Vector DM with Scalar Portal}}\label{svcons500}
    \end{subfigure}%
    
  \begin{subfigure}{0.5\textwidth}
      \centering
      \includegraphics[width=\textwidth,clip]{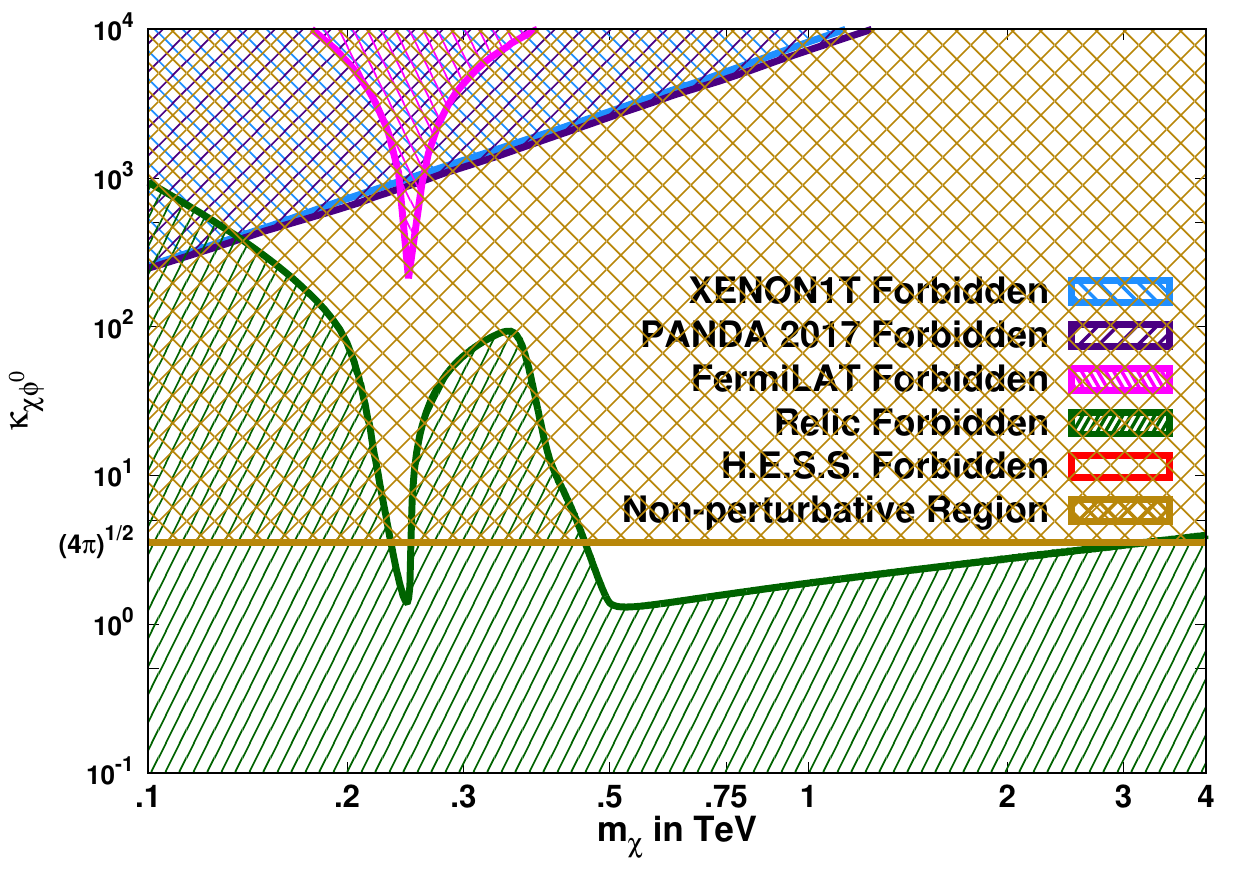}
      \caption{\small{\em Fermionic DM with Scalar Portal}}\label{sfcons500}
    \end{subfigure}%    
      \begin{subfigure}{0.5\textwidth}
      \centering
      \includegraphics[width=\textwidth,clip]{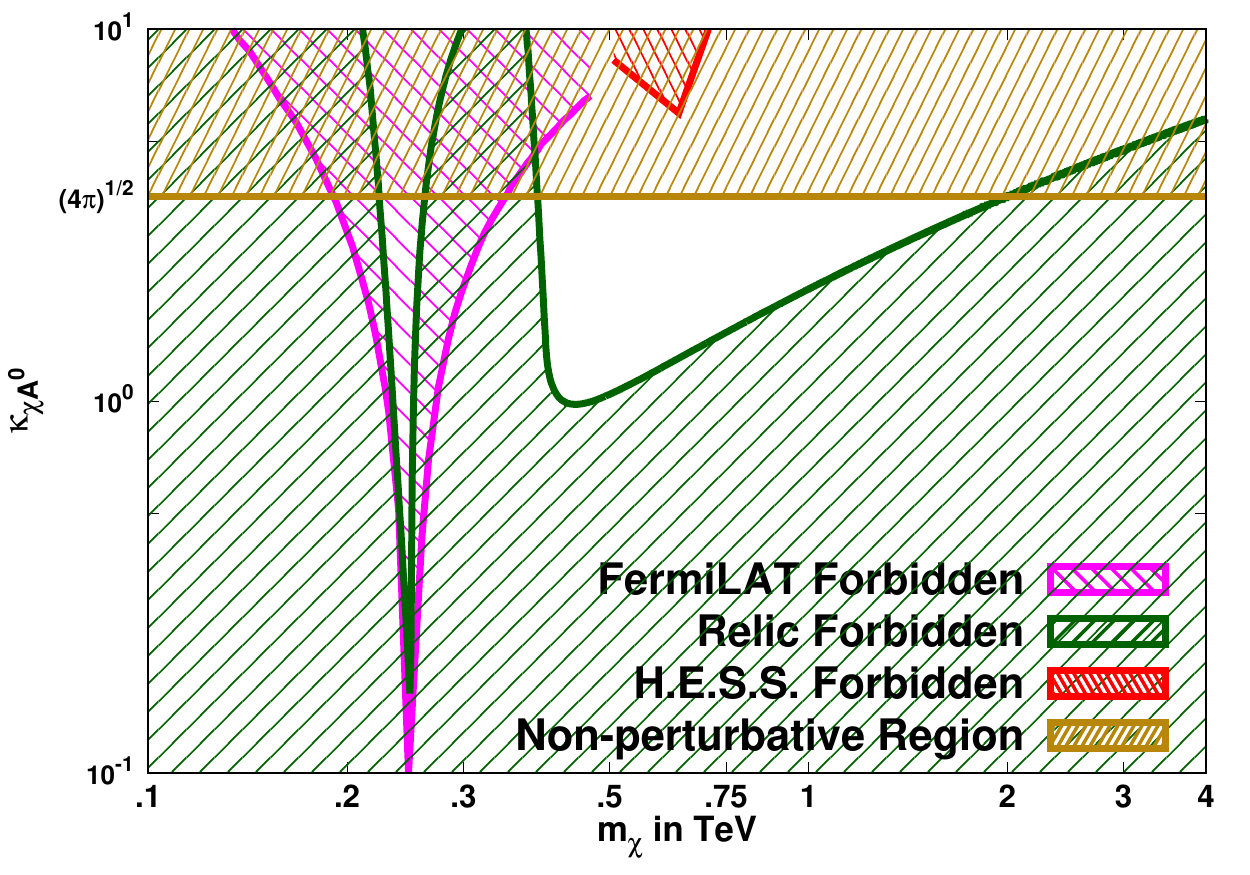}
      \caption{\small{\em Fermionic DM with Pseudo-scalar Portal}}\label{pfcons500}
    \end{subfigure}%
    \caption{\small{\em All model constraints are drawn for the portal mass of 500 GeV and VLQ mass 400 GeV. The shaded region correspond to  the forbidden  regions by relic density \cite{WMAP1,Ade:2015xua} (green), XENON1T 2017 \cite{Aprile:2017iyp} (blue), PANDA 2017 \cite{Cui:2017nnn} (indigo), Fermi-LAT \cite{Ackermann:2015lka} (pink), H.E.S.S. \cite{Abramowski:2013ax}  (red) and the perturbativity condition (golden yellow) respectively in the plane defined by  $m_\eta - \kappa_{\eta\phi^0}$, $m_{V^0} - \kappa_{V^0\phi^0}$, $m_\chi - \kappa_{\chi\phi^0}$ and  $m_\chi - \kappa_{\chi A^0}$ in Figures \ref{sscons500}, \ref{svcons500}, \ref{sfcons500}, and \ref{pfcons500}  respectively.  Constraints from the  direct detection experiments  for the pseudo-scalar portal are not shown (see text).}}\label{DMcons500}
    \end{figure}%
      \begin{figure}[tbh]
%\begin{wrapfigure}{l}{0.5\textwidth} 
 \centering
 \begin{subfigure}{0.5\textwidth}
      \centering
      \includegraphics[width=\textwidth,clip]{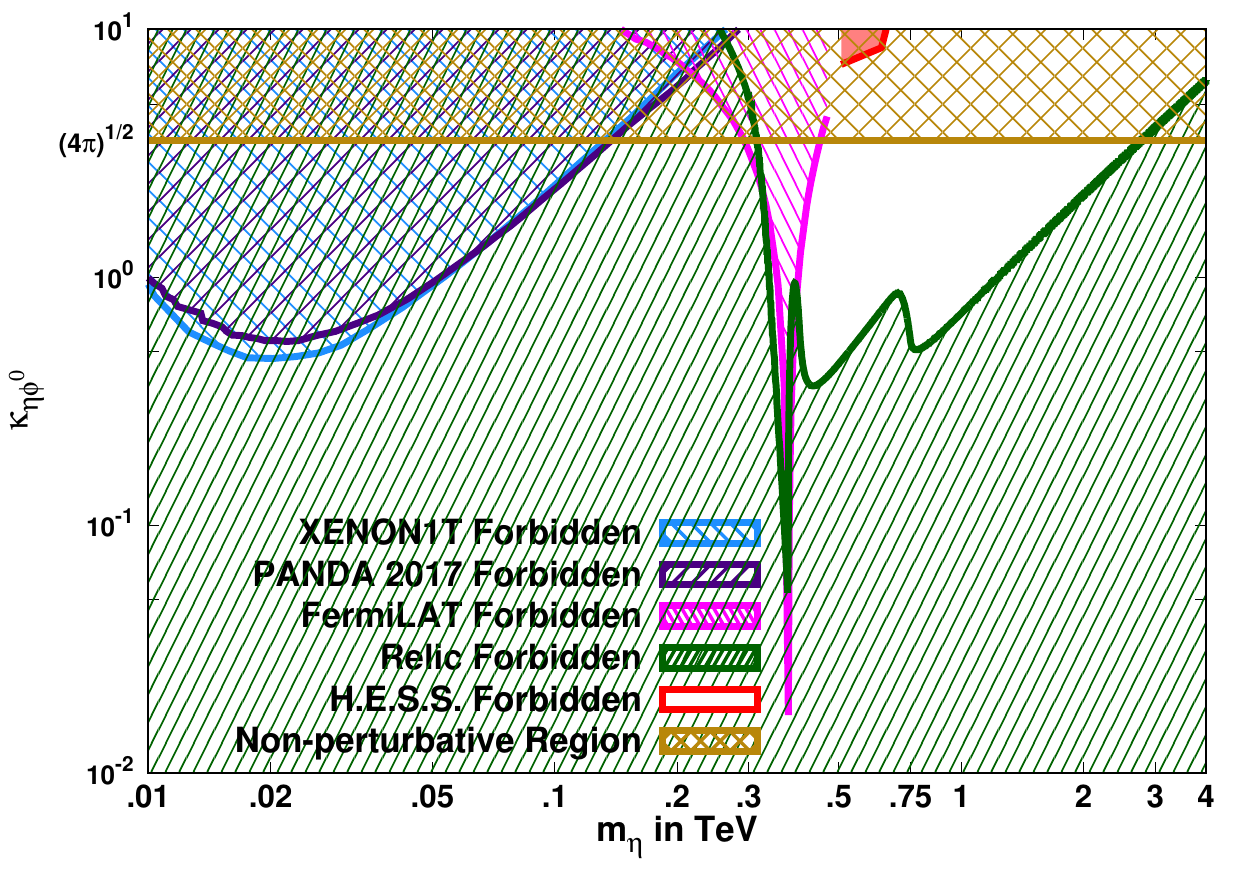}
      \caption{\small{\em Scalar DM with Scalar Portal}}\label{sscons750}
    \end{subfigure}%
     \begin{subfigure}{0.5\textwidth}
      \centering
      \includegraphics[width=\textwidth,clip]{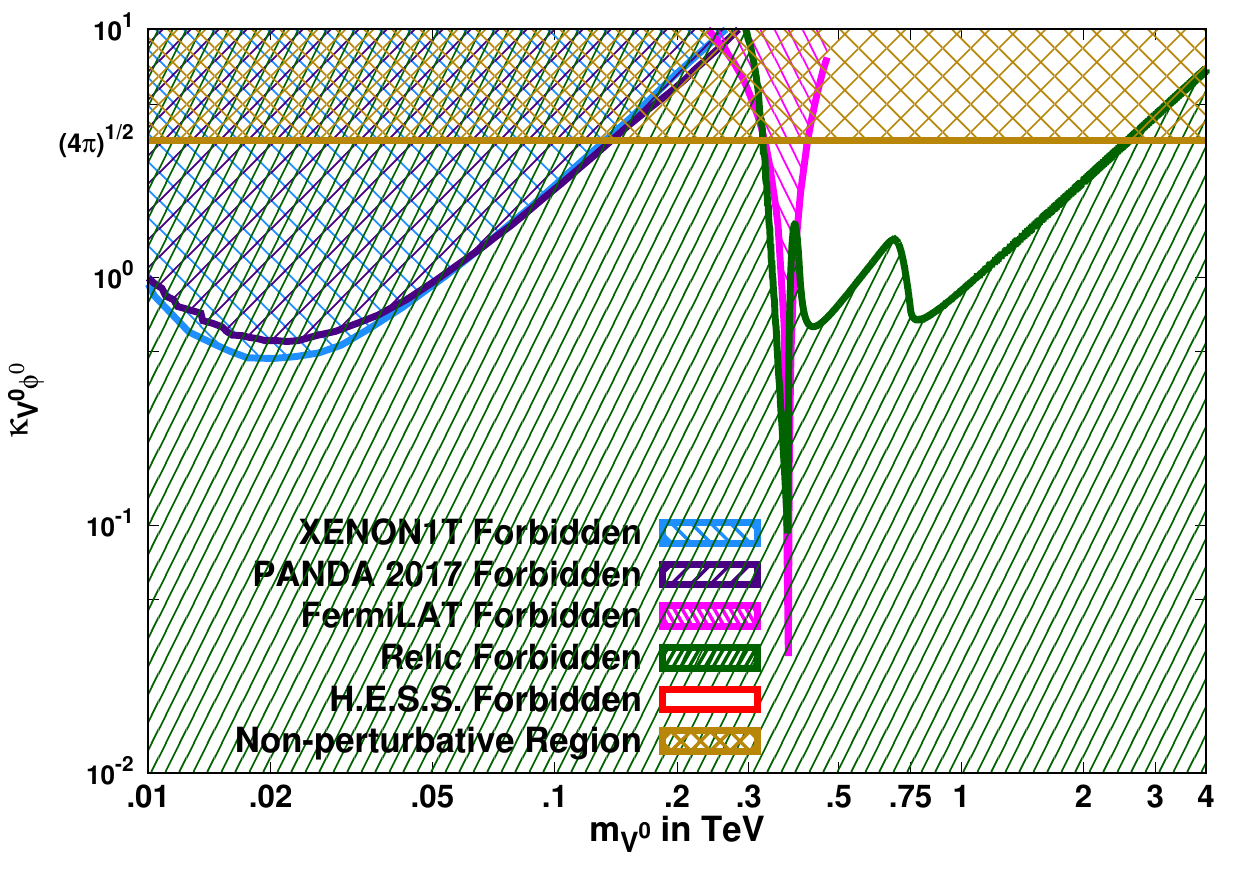}
      \caption{\small{\em Vector DM with Scalar Portal}}\label{svcons750}
    \end{subfigure}%

  \begin{subfigure}{0.5\textwidth}
      \centering
            \includegraphics[width=\textwidth,clip]{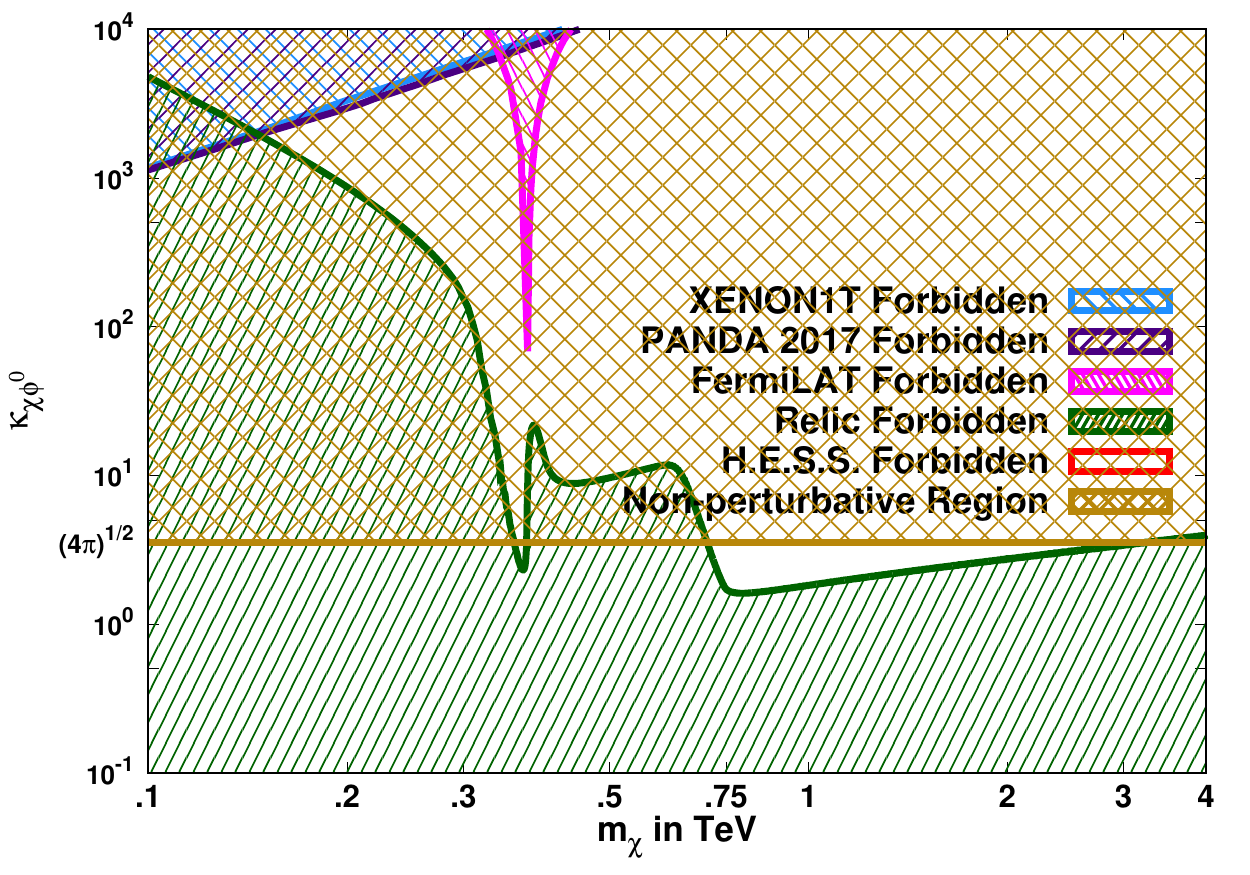}
      \caption{\small{\em Fermionic DM with Scalar Portal}}\label{sfcons750}
    \end{subfigure}%
      \begin{subfigure}{0.5\textwidth}
      \centering
      \includegraphics[width=\textwidth,clip]{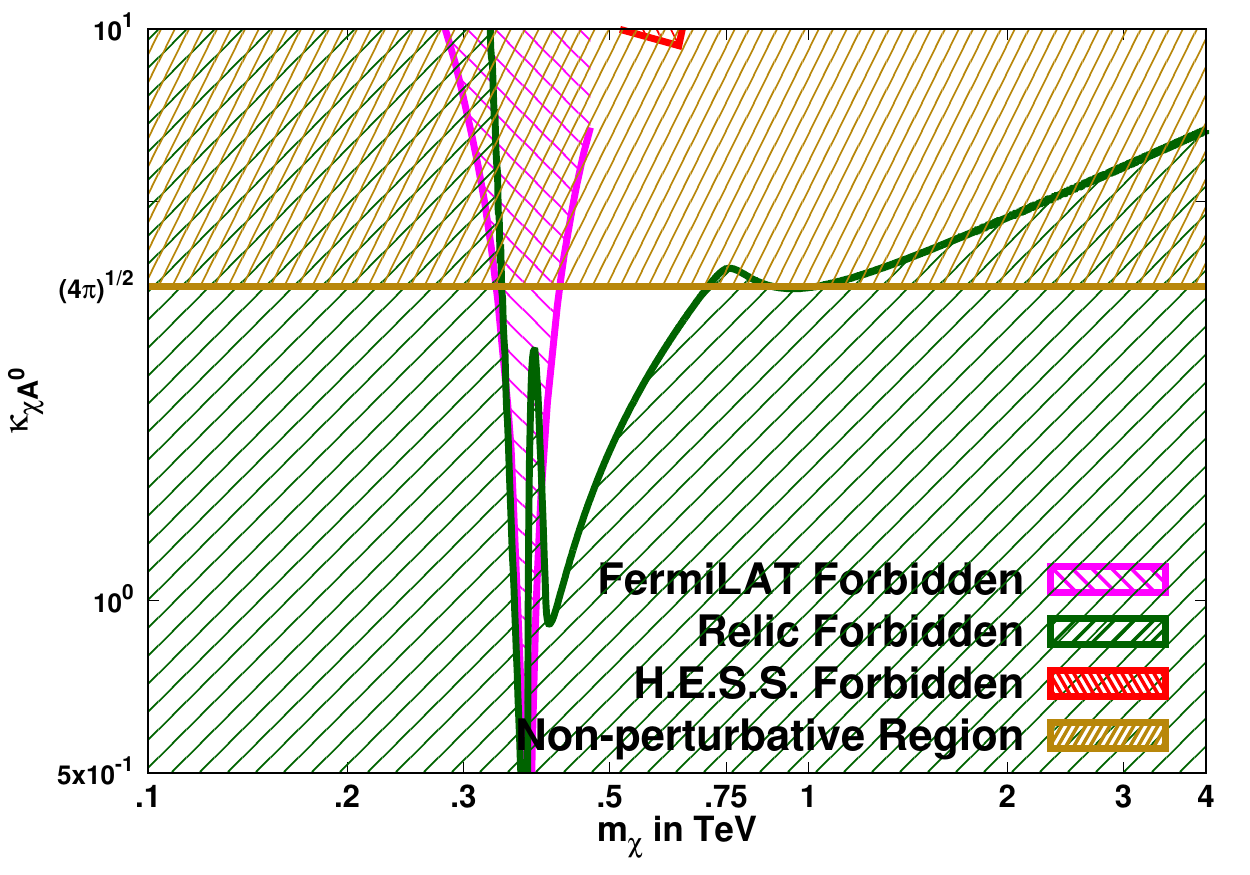}
      \caption{\small{\em Fermionic DM with Pseudo-scalar Portal}}\label{pfcons750}
    \end{subfigure}%
    \caption{\small{\em All model constraints are drawn for the portal mass of 750 GeV and VLQ mass 400 GeV. The shaded region correspond to  the forbidden  regions by relic density \cite{WMAP1,Ade:2015xua} (green), XENON1T 2017 \cite{Aprile:2017iyp} (blue), PANDA 2017 \cite{Cui:2017nnn} (indigo), Fermi-LAT \cite{Ackermann:2015lka} (pink), H.E.S.S. \cite{Abramowski:2013ax}  (red) and the perturbativity condition (golden yellow) respectively in the plane defined by  $m_\eta - \kappa_{\eta\phi^0}$, $m_{V^0} - \kappa_{V^0\phi^0}$, $m_\chi - \kappa_{\chi\phi^0}$ and  $m_\chi - \kappa_{\chi A^0}$ in Figures \ref{sscons750}, \ref{svcons750}, \ref{sfcons750}, and \ref{pfcons750}  respectively.  Constraints from the  direct detection experiments  for the pseudo-scalar portal are not shown (see text).}}\label{DMcons750}
    \end{figure}%
\section{Summary and Conclusions}
\label{conclusion}
Our analysis provide a conservative complementary scenario to that appeared in the review \cite{Arcadi:2017kky}, where we have neglected the portal mixing with the SM Higgs and instead we have generated the required relic density of DM by considering an effective loop induced interaction
of the VLQ with the portal. 
\par In this paper we considered a $U(1)_d$ extension of the standard model with a dark sector and a singlet scalar or a pseudo-scalar di-Boson resonance which interacts with the SM gauge Bosons through a vector-like SM colour triplet fermion of exotic charge $Q$ = 5/3. The dark matter particle considered here is a neutral SM singlet real scalar, a real vector or a  spin-1/2 fermion interacting with the standard model gauge Bosons through a scalar/pseudo-scalar di-Boson resonance. As a first step we obtained the constraints on the coupling of di-Boson resonance with the vector-like fermions from the ATLAS and CMS experimental searches through spin-zero di-photon production cross-section $\sigma$ (pp $\rightarrow \phi^0/A^0 \rightarrow$ $\gamma\gamma$). The constrained parameter region in ($y_{\phi^0/A^0} - m_\psi$) plane is depicted in Fig \ref{fig:scalgaugebosonprod} at the 2$\sigma$ level. With these constraints in place, we obtained relic density contours in the dark matter mass-coupling parameter space assuming that the dark matter particles considered here saturate the observed relic density $\Omega_c h^2 \simeq 0.1138\pm .0045$. We then study that  the prediction of   the LHC and relic density constrained model  for a) the  air-borne indirect detection of DM pair annihilation into  a pair of $\gamma\gamma/\, \gamma Z$ in the galactic halo   and b) the ground based DM - nucleon scattering  direct detection experiments.

\par On comparing the thermal averaged annihilation cross-sections for the indirect detection with the limits on the cross-section obtained from the  Fermi-LAT  and H.E.S.S. (2013) in the mass range less and greater than 500 GeV respectively through the observation of  monochromatic gamma-rays, we found that the model predicted annihilation cross-section with  the allowed lower bound of  scalar/ vector - scalar portal and fermionic  DM - pseudo-scalar portal couplings are being favoured by the limits obtained from  these experiments for the DM masses  400 GeV and above as shown in Figures \ref{fig:DMInDirectDetectionSV},  \ref{PSFindirectpp} and \ref{PSFindirectpz}.  
 However, the model prediction  of the annihilation cross-section for fermionic DM induced by the scalar portal lies far below the experimental upper limit and therefore favours the large model parameter space spanning the mass range upto 4 TeV as shown in the Figures \ref{SFindirectpp} and \ref{SFindirectpz}.

\par Comparing the constrained model prediction for spin-independent direct detection DM-nucleon elastic scattering experiments with the recent experimental results from  LUX (2016) \cite{Akerib:2016vxi}, XENON1T (2017) \cite{Aprile:2017iyp}, PANDA (2017) \cite{Cui:2017nnn} and Dark-Side50 (2016) \cite{Agnes:2015ftt} collaborations as shown in Figures \ref{ssdirect}, \ref{svdirect} and \ref{sfdirect} respectively, we find that almost the entire parameter space allowed by the LHC searches and the observed relic density is also allowed by the direct detection experiments for the fermionic dark matter. The case of real scalar and vector dark matter particles is different, where the parameter region is favoured only for the mass range greater than 300 GeV.

\par Finally,  it is important to comment upon the model restrictions, if it has to satisfy each and every  experimental results. This  helps in narrowing down the search for such DM candidates in the ongoing and upcoming collider based experiments.  To this end we translate the experimental results in terms of restriction on the DM mass and DM - portal coupling for the mass range varying between 10 GeV and 4 TeV. Therefore,  we summarise our analysis with the help of the three composite Figures \ref{DMcons270}, \ref{DMcons500} and \ref{DMcons750} corresponding to the three  choice of the scalar portal masses 270, 500 and 750 GeV respectively, where we have put all the  constraints  from the relic density, two indirect  detection experiments and two direct detection   experiments on the model  for a specific DM candidate in a plane spanned by its mass and  its coupling to the scalar and/ or pseudo-scalar portal.    These Figures spell out the implications of the experimental results on the model parameter space and depict   
\begin{itemize}
\item the contours drawn with the coordinates of the lower limit on the DM - portal coupling for a given DM mass   derived from the constant  relic density  \cite{WMAP1,Ade:2015xua} 
 \item the contours drawn with the coordinates of the  upper limit on the respective DM - portal coupling for a given DM mass by demanding such DM candidates  to satisfy the observed null results of the thermal averaged pair production cross-section for $\gamma\gamma $ final  state due to a DM pair annihilation from Fermi-LAT \cite{Ackermann:2015lka} (H.E.S.S. 2013 \cite{Abramowski:2013ax}) for DM mass range less (greater) than 500 GeV.
 \item the contours drawn with the coordinates of the  upper limit on the respective DM - portal coupling for a given DM mass by demanding such DM candidates  to satisfy  the observed null results of the sensitive DM - Nucleon scattering cross-section from the recent experiments XENON1T (2017) \cite{Aprile:2017iyp} and PANDA (2017) \cite{Cui:2017nnn}. However, since we do not have a stringent experimental constraints on the spin dependant fermionic DM - nucleon cross-sections, we do not show any direct detection contours for the pseudo-scalar induced fermionic DM.
\end{itemize}
\par It is remarkable to note that we have now shrunk the allowed parameter space which appear as white unshaded region in each of these figures. However, area of the allowed  unshaded white region  in 
 Figures \ref{DMcons270}-\ref{DMcons750}  for the DM masses greater than 1 TeV can be enhanced by  increasing the VLQ mass above 400 GeV. Thus this tightly constrained spin $0^\pm$ induced DM model has become challenging enough to be probed in the  colliders.
\vskip 5mm
\acknowledgments
 SD acknowledges the partial financial
 support from the CSIR grant No. 03(1340)/15/EMR-II. LKS acknowledges the UGC JRF fellowship for the partial financial support.   
\newpage 

\appendix
\begin{center}{\bf\Large Appendix}
\end{center}
\section{Partial decay-widths of the scalar/ pseudo-scalar portal with effective vertices} 
\label{Decaywidthcalculation}
Using the Lagrangian given in equations \eqref{LDMVLQ} -\eqref{LDMferm}, we calculate the effective strength of the interactions involving  SM neutral gauge Bosons and the scalar or pseudo-scalar portal. These effective interactions are computed by evaluating the triangle VLQ loop integrals  in the heavy fermion limit.   

\par The effective couplings  for the scalar portal are expressed as  
\begin{subequations}
\begin{eqnarray}
   \kappa_{g g} &=& \frac{y_{\phi^0}}{m_{\psi}} \,\frac{\alpha_{s}\,(m_{\phi^0})}{4\pi} \,I_{gg} \label{scal_eff_coup_gg} \\
   \kappa_{\gamma\gamma} &=&  \frac{y_{\phi^0}}{m_{ \psi  }} \frac{\alpha_{\rm em}(0)}{2\pi}\, N_{c}\, Q_{\psi}^{2} \,I_{\gamma\gamma} \label{scal_eff_coup_pp} \\
   \kappa_{\gamma Z} &=&  \frac{y_{\phi^0}}{m_{ \psi  }} \frac{\alpha_{\rm em}(0)}{\pi} \,N_{c}\, Q_{\psi}^{2}\, \tan \theta_{W} \,I_{\gamma Z} \label{scal_eff_coup_pz}\\
   \kappa_{ZZ} &=&  \frac{y_{\phi^0}}{m_{ \psi  }}\, \frac{\alpha_{\rm em}(0)}{2\,\pi} \,N_{c} Q_{\psi}^{2}\, \tan \theta^{2}_{W}\, I_{Z Z} \label{scal_eff_coup_zz}
   \end{eqnarray}
   \end{subequations}
   corresponding to the  $\phi^0\,g\,g$, $\phi^0\,\gamma\,\gamma$, $\phi^0\,\gamma\,Z$ and $\phi^0\, Z\,Z$ effective vertices. The respective loop integrals are given as
   \begin{subequations} 
   \begin{eqnarray}
  I_{gg} &=& I{\gamma\gamma}=\int_{0}^{1} dx \int_{0}^{1-x}  dy \frac{f(x,y)}{1 - \frac{m_{\phi^0}^{2}}{m_{ \psi  }^{2}}\,\, x\, y} \label{scalloopint1}\\
  I_{\gamma Z} &=& \int_{0}^{1} dx \int_{0}^{1-x} dy \frac{f(x,y)}{1 - \left(\frac{m_{\phi^0}^{2}}{m_{ \psi  }^{2}} - \frac{m_{Z}^{2}}{m_{ \psi  }^{2}}\right)\, x\, y - x\,\left(1-x\right)\,\frac{m_{Z}^{2}}{m_{ \psi  }^{2}}} \label{scalloopint2}\\ 
  I_{Z Z} &=& \int_{0}^{1} dx \int_{0}^{1-x}  dy \frac{f(x,y)}{1 - \left(\frac{m_{\phi^0}^{2}}{m_{ \psi  }^{2}} - 2 \frac{m_{Z}^{2}}{m_{ \psi  }^{2}}\right) x\, y - \left\{x\,\left(1-x\right) + y\,\left(1-y\right)\right\}\,\frac{m_{Z}^{2}}{m_{ \psi  }^{2}} }\label{scalloopint3}
 \end{eqnarray}
 \end{subequations}
 where $f(x,y)= (1-4\,x\,y)$.

\par The partial decay-widths of the  scalar portal to the SM neutral gauge Bosons are  computed and are given  as 
  \begin{subequations}
 \begin{eqnarray}
  \Gamma(\phi^0\to g\,g) &=& \frac{y_{\phi^0}^{2}}{8\pi} \,\,m_{\phi^0}\,\, \left(\frac{\alpha_{s}\left(m_{\phi^0}\right)}{\pi}\right)^{2} \frac{m_{\phi^0}^{2}}{m_{ \psi  }^{2}}\,\, \left\vert I_{gg} \right\vert^2\label{scaltogg}\\
  \Gamma(\phi^0\to \gamma\,\gamma) &=& \frac{y_{\phi^0}^{2}}{16\pi}\,\, m_{\phi^0}\,\, \left(\frac{\alpha_{\rm em}(0)}{\pi}\right)^{2} \frac{m_{\phi^0}^{2}}{m_{ \psi  }^{2}} \,\,N_{c}^{2} \,Q_{\psi}^{4} \,\,\left\vert I_{\gamma\gamma} \right\vert^2\label{scaltopp} \\
  \Gamma(\phi^0\to Z\,\gamma) &=& \frac{y_{\phi^0}^{2}}{8\pi}\,\, m_{\phi^0}\,\, \left(\frac{\alpha_{\rm em}(0)}{\pi}\right)^{2}\,\, \frac{m_{\phi^0}^{2}}{m_{ \psi  }^{2}}\, N_{c}^{2}\, Q_{\psi}^{4} \,\tan^{2}\theta_{W} \left(1 - \frac{m_{Z}^{2}}{m_{\phi^0}^{2}}\right)^{3}  \left\vert I_{\gamma Z} \right\vert^2\label{scaltozp}\\          
  \Gamma(\phi^0\to Z\,Z) &=& \frac{y_{\phi^0}^{2}}{16\pi} \,\, m_{\phi^0}\,\, \left(\frac{\alpha_{\rm em}(0)}{\pi}\right)^{2} \,\,\frac{m_{\phi^0}^{2}}{m_{ \psi  }^{2}}\,\, N_{c}^{2} \,Q_{\psi}^{4} \,\tan^{4}\theta_{W} \left(1 - 4\frac{m_{Z}^{2}}{m_{\phi^0}^{2}}\right)^{1/2} \nonumber\\
  &&\,\,\,\,\,\,\times\,\,\left(1 - 4\frac{m_{z}^{2}}{m_{\phi^0}^{2}} + 6\frac{m_{Z}^{4}}{m_{\phi^0}^{4}}\right)\,\,\, \left\vert I_{ZZ} \right\vert^2 \label{scaltozz}
 \end{eqnarray}
 \end{subequations}
 The expression for $\Gamma(\phi^0\to Z\, Z)$ is correct to $\sim 0.1 \%$.

 \par Similarly, the  effective couplings $\tilde\kappa_{VV}$ of the pseudo-scalar to the pair of $g\,g$, $\gamma\,\gamma$, $\gamma\, Z$ and $Z\,Z$  are obtained by replacing $y_{\phi^0}\to y_{A^0}$, $m_{\phi^0}\to m_{A^0}$ and $I_{VV}\to \tilde I_{VV}$ in equations \eqref{scal_eff_coup_gg}-\eqref{scal_eff_coup_zz} respectively.  The corresponding loop integrals $\tilde I_{VV}$ are obtained by substituting $m_{\phi^0}\to m_{A^0}$ and $f(x,y)$ by 1 in equations \eqref{scalloopint1}-\eqref{scalloopint3} respectively. Accordingly, the partial  decay-widths of the pseudo-scalar to a pair of SM neutral gauge Bosons can be obtained by replacing $y_{\phi^0}$ by $y_{A^0}$, $m_{\phi^0}$ by $m_{A^0}$ and $I_{VV}$ by $\tilde{I}_{VV}$ in equations \eqref{scaltogg}-\eqref{scaltozz}.
%\newpage
  
\section{ Thermal averaged DM pair annihilation Cross-Sections  } 
\label{ThermalAvCalculation}
\subsection{Scalar portal}
\label{ThermalAvCalculationscal}
The thermal averaged cross-sections for the annihilation of scalar DM to SM gauge Bosons are given as
\begin{eqnarray}
    \left\langle\sigma^{\phi^0}_{\eta\eta \rightarrow gg}\, v_{rel}\right\rangle &=& \frac{1}{\pi}\ \frac{\kappa_{\eta\phi^0}^2\ v_{\Phi}^2}{(4m_\eta^2-m_{\phi^0}^2)^2 + \Gamma_{\phi^0}^2 m_{\phi^0}^2}\ \frac{\alpha_s^2 y_{\phi^0}^2 I_{gg}^2}{\pi^2}\ \frac{m_\eta^2}{m_\psi^2} \label{thav:etaetatogg}\\
    \left\langle\sigma^{\phi^0}_{\eta\eta \rightarrow \gamma\gamma}\, v_{rel}\right\rangle &=& \frac{1}{2\, \pi}\ \frac{\kappa_{\eta\phi^0}^2\ v_{\Phi}^2}{(4m_\eta^2-m_{\phi^0}^2)^2 + \Gamma_{\phi^0}^2 m_{\phi^0}^2}\ \frac{\alpha_{\rm em}^2 y_{\phi^0}^2 I_{\gamma\gamma}^2}{\pi^2}\ \frac{m_\eta^2}{m_\psi^2}\ N_c^2 Q_\psi^4\label{thav:etaetatopp} \\
     \left\langle\sigma^{\phi^0}_{\eta\eta \rightarrow Z\gamma}\, v_{rel} \right\rangle &=& \frac{1}{\pi}\ \frac{\kappa_{\eta\phi^0}^2\ v_{\Phi}^2}{(4m_\eta^2-m_{\phi^0}^2)^2+ \Gamma_{\phi^0}^2 m_{\phi^0}^2}\ \frac{\alpha_{\rm em}^2 y_{\phi^0}^2 I_{\gamma Z}^2}{\pi^2}\ \frac{m_\eta^2}{m_\psi^2}\ N_c^2 Q_\psi^4 \tan^2\theta_W \bigg(1 - \frac{m_Z^2}{4m_\eta^2}\bigg)^3\ \hspace*{0.8cm} \label{thav:etaetatozp}\\
   \left\langle \sigma^{\phi^0}_{\eta\eta \rightarrow ZZ}\, v_{rel}\right\rangle &=& \frac{1}{2\, \pi}\ \frac{\kappa_{\eta\phi^0}^2\ v_{\Phi}^2}{(4m_\eta^2-m_{\phi^0}^2)^2+ \Gamma_{\phi^0}^2 m_{\phi^0}^2} \frac{\alpha_{\rm em}^2 y_{\phi^0}^2 I_{ZZ}^2}{\pi^2} \frac{m_\eta^2}{m_\psi^2} N_c^2 Q_\psi^4 \tan^4 \theta_W \nonumber \\
    && \hspace*{2cm} \bigg[\frac{3 m_Z^4}{8m_\eta^4}-\frac{m_Z^2}{m_\eta^2}+1\bigg]\, \bigg(1-\frac{m_Z^2}{m_\eta^2}\bigg)^{1/2}\,\hspace*{0.8cm}      \label{thav:etaetatozz}  \\
    \left\langle \sigma^{\phi^0}_{\eta\eta \rightarrow \psi\psi} v_{\rm rel} \right\rangle &=& \frac{y_{\phi^0}^2\, \kappa_{\eta\phi^0}^2\, v_{\Phi}^2}{4\pi\, m_\eta^3} \frac{\bigg(m_\eta^2 - m_\psi^2\bigg)^{3/2}}{16(m_\eta^2 - m_{\phi^0}^2)^2} \label{thav:etaetatopsipsi} \\   
    \left\langle \sigma_{\eta\eta \rightarrow \phi^0\phi^0} v_{\rm rel} \right\rangle &\simeq& \frac{\kappa_{\eta\phi^0}^4\, v_{\Phi}^4\, (1-\frac{m_{\phi^0}^2}{2m_\eta^2})}{16\, \pi\, m_\eta^6} \label{thav:etaetatophiphi}       
\end{eqnarray}

\par The thermal averaged cross-sections for the annihilation of the  vector DM pair via scalar portal to the di-Boson final states can be directly read out from the corresponding expressions for the scalar DM modulo the spin averaging of the initial states and the polarisation sum involved in  the matrix element  squared.   We obtain the pair annihilation cross-section to the respective final states directly by substituting $\kappa_{\eta\phi^0}$ by $\kappa_{V^0\phi^0}$/3 and $m_\eta$ by $m_{V^0}$ in equations \eqref{thav:etaetatogg} - \eqref{thav:etaetatozz}. The thermal averaged cross-section for the $s$ channel process $\left\langle\sigma \left(V^0V^0\to \bar\psi \psi\right)\,v\right\rangle$  is computed to be 1/6 of the $\left\langle\sigma \left(\eta\eta\to \bar\psi \psi\right)\,v\right\rangle$ given in \eqref{thav:etaetatopsipsi}.  The thermal averaged cross-section for $t$-channel pair is identical to that of the scalar DM annihilation to pair of scalar portals is given in \eqref{thav:etaetatophiphi}.

The thermal averaged cross-sections for the annihilation of fermionic DM via scalar portal are given as
\begin{eqnarray}
    \left\langle\sigma^{\phi^0}_{\chi\chi \rightarrow gg}\, v_{rel}\right\rangle &=&  \frac{\kappa_{\chi \phi^0}^2\, y_{\phi^0}^2\, \alpha_s^2}{2\, \pi^3 m_\psi^2}\,  \frac{m_\chi^4\, I_{gg}^2}{(4m_\chi^2 - m_{\phi^0}^2)^2 + \Gamma_{\phi^0}^2 m_{\phi^0}^2}\, \left(\frac{6}{x_F}\right) \label{thav:chichitogg} \\
    \left\langle\sigma^{\phi^0}_{\chi\chi \rightarrow \gamma\gamma}\, v_{\rm rel}\right\rangle &=& \frac{\kappa_{\chi \phi^0}^2\, y_{\phi^0}^2\, \alpha_{\rm em}^2}{4\, \pi^3\, m_\psi^2} \frac{m_\chi^4\, I_{\gamma\gamma}^2}{(4m_\chi^2 - m_{\phi^0}^2)^2 + \Gamma_{\phi^0}^2 m_{\phi^0}^2}\, N_c^2\, Q_\psi^4   \, \left(\frac{6}{x_F}\right)\label{thav:chichitopp}  \\
    \left\langle\sigma^{\phi^0}_{\chi\chi \rightarrow Z\gamma}\, v_{rel}\right\rangle &=&  \frac{\kappa_{\chi \phi^0}^2\, y_{\phi^0}^2\, \alpha_{\rm em}^2}{2\, \pi^3\, m_\psi^2}\, \frac{m_\chi^4\, I_{\gamma Z}^2}{(4m_\chi^2 - m_{\phi^0}^2)^2 + \Gamma_{\phi^0}^2 m_{\phi^0}^2}\ \left(1 - \frac{m_Z^2}{4m_V^2}\right)^3\, \tan^2\theta_W\, N_c^2\, Q_\psi^4 \, \left(\frac{6}{x_F}\right) \label{thav:chichitopz} \\   
    \left\langle\sigma^{\phi^0}_{\chi\chi \rightarrow ZZ}\, v_{rel}\right\rangle &=&  \frac{\kappa_{\chi \phi^0}^2\, y_{\phi^0}^2\, \alpha_{\rm em}^2}{4\, \pi^3\, m_\psi^2}\, \frac{m_\chi^4\, I_{ZZ}^2\, \tan^4\theta_W\, N_c^2\, Q_\psi^4}{(4m_\chi^2 - m_{\phi^0}^2)^2 + \Gamma_{\phi^0}^2 m_{\phi^0}^2}\, \sqrt{1-\frac{4m_Z^2}{s}}\, \left(1 - \frac{4\ m_Z^2}{m_{\phi^0}^2} + \frac{6\ m_Z^4}{m_{\phi^0}^4}\right)\,\left(\frac{6}{x_F}\right)\hspace{1cm}\\%\nonumber \\
%    && \hspace*{3cm} \tan^4\theta_W\, N_c^2\, Q_\psi^4 \hspace*{0.8cm}   \label{thav:chichitozz} \\   
    \left\langle \sigma^{\phi^0}_{\bar{\chi}\chi \rightarrow \psi\psi} v_{\rm rel} \right\rangle &=& \frac{y_{\phi^0}^2\, \kappa_{\chi \phi^0}^2}{8\pi m_\chi}\, \frac{\bigg(m_\chi^2 - m_\psi^2\bigg)^{3/2}}{16(m_\chi - m_{\phi^0}^2)^2}\, \left(\frac{6}{x_F}\right) \label{thav:chichitopsipsi} \\
    \left\langle \sigma_{\bar{\chi}\chi\rightarrow\phi^0\phi^0} v_{\rm rel} \right\rangle &\simeq&  \frac{3\, \kappa_{\chi \phi^0}^4}{128\, \pi\, m_\chi^2} \, \left(\frac{6}{x_F}\right)\label{thav:chichitophiphi}   
\end{eqnarray}

\subsection{Pseudo-Scalar portal }
\label{ThermalAvCalculationpscal}
The Thermal averaged cross-sections for the annihilation of fermionic DM via pseudo-scalar portal are given as
\begin{eqnarray}
\left\langle\sigma^{A^0}_{\bar{\chi}\chi \rightarrow g\,g}\, v_{rel}\right\rangle    &=& \frac{2\alpha_{s}^2\, \kappa_{\chi A^0}^2\, y_{A^0}^2}{\pi^3\, m_\psi^2}\ \frac{m_\chi^4}{(4\, m_\chi^2 - m_{A^0}^2)^2 + \Gamma_{A^0}^2 m_{A^0}^2}\  \tilde{I}_{gg}^2\, \, \left(1+\frac{15}{4\, x_F}\right) \label{pthav:chichitogg}\\
\left\langle\sigma^{A^0}_{\bar{\chi}\chi \rightarrow \gamma\gamma}\, v_{rel}\right\rangle    &=& \frac{\alpha_{\rm em}^2\, \kappa_{\chi A^0}^2\, y_{A^0}^2}{\pi^3\, m_\psi^2}\ \frac{m_\chi^4}{(4\, m_\chi^2 - m_{A^0}^2)^2 + \Gamma_{A^0}^2 m_{A^0}^2}\  \tilde{I}_{\gamma\gamma}^2\, N_c^2\, Q_\psi^4  \,\, \left(1+ \frac{15}{4\, x_F}\right)\label{pthav:chichitopp}\\
    \left\langle\sigma^{A^0}_{\bar{\chi}\chi \rightarrow Z\gamma}\, v_{rel}\right\rangle &=&  \frac{2\, \alpha_{\rm em}^2\, y_{A^0}^2\, \kappa_{\chi A^0}^2}{\pi^3\, m_\psi^2}\ \frac{m_\chi^4}{(4\, m_\chi^2 - m_{A^0}^2)^2 + \Gamma_{A^0}^2 m_{A^0}^2}\  \left(1 - \frac{m_Z^2}{4\, m_\chi^2}\right)^3\ \tilde{I}_{\gamma Z}^2\, N_c^2\ Q_\psi^4\, \tan^2\theta_W \hspace*{1cm}  \label{pthav:chichitopz}\\
    \left\langle\sigma^{A^0}_{\bar{\chi}\chi \rightarrow ZZ}\, v_{rel}\right\rangle &=&
 \frac{1}{8\, \pi}\ \sqrt{1-4\frac{m_{A^0}^2}{m_\chi^2}}\ \frac{y_{A^0}^2\, \kappa_{\chi A^0}^2\, \alpha_{\rm em}^2}{\pi^2}\ \frac{m_\chi^2}{m_\psi^2}\  \frac{ \tilde{I}_{ZZ}^2\, N_c^2\, Q_\psi^4\, \tan^4\theta_W}{(4 \, m_\chi^2 - m_{A^0}^2)^2 + \Gamma_{A^0}^2 m_{A^0}^2}\  \left(1 - \frac{4\ m_Z^2}{m_{A^0}^2}\right)^{3/2} \hspace*{1cm}\label{pthav:chichitozz} \\
    \left\langle \sigma^{A^0}_{\bar{\chi}\chi \rightarrow \psi\psi} v_{\rm rel} \right\rangle &=& \frac{y_{A^0}^2\, \kappa_{\chi A^0}^2}{2\pi}\, \frac{m_\chi \bigg(m_\chi^2 - m_\psi^2\bigg)^{1/2}}{16(m_\chi^2 - m_{A^0}^2)^2} \label{pthav:chichitopsipsi} \\
    \left\langle \sigma_{\bar{\chi}\chi\rightarrow A^0A^0} v_{\rm rel} \right\rangle &\simeq& \frac{\kappa_{\chi A^0}^4\, m_{A^0}^4}{320\, \pi\, m_\chi^6 }\, \bigg(1+\frac{3m_{A^0}^2}{10m_\chi^2}\bigg) \label{pthav:chichitoAA}
\end{eqnarray}

\newpage
%%%%%%%%%%%%%%%%%%%%%%%%%%%%%%%%%%%%%%%%%%%%%%%%%%%%%%%%%%%%%%%%%%%

\end{document}